\UseRawInputEncoding 
\documentclass[prl,twocolumn,superscriptaddress,noshowpacs,notitlepage,longbibliography]{revtex4-1}%
\usepackage{graphicx,bm,times}
\usepackage{amsmath}
\usepackage{amsfonts}
\usepackage{amssymb}
\usepackage{color}
\usepackage{hyperref}
\usepackage{xcolor}

\hypersetup{
	colorlinks = true,
	allcolors = {blue}
}

\begin{document}
	
	\title{FeSe and the missing electron pocket problem}
	
	\author{Luke C. Rhodes}
	
	\affiliation{School of Physics and Astronomy, University of St. Andrews, St. Andrews KY16 9SS, United Kingdom}	
	
	\author{Matthias Eschrig}
	\affiliation{Institute of Physics, University of Greifswald, Felix-Hausdorff-Strasse 6, 17489 Greifswald, Germany}
	
	\author{Timur K. Kim}
	\affiliation{Diamond Light Source, Harwell Campus, Didcot, OX11 0DE, United Kingdom}
	
	\author{Matthew D. Watson}
	\affiliation{Diamond Light Source, Harwell Campus, Didcot, OX11 0DE, United Kingdom}
	
	\begin{abstract}
		The nature and origin of electronic nematicity remains a significant challenge in our understanding of the iron-based superconductors. This is particularly evident in the iron chalcogenide, FeSe, where it is currently unclear how the experimentally determined Fermi surface near the M point evolves from having two electron pockets in the tetragonal state, to exhibiting just a single electron pocket in the nematic state. This has posed a major theoretical challenge, which has become known as the missing electron pocket problem of FeSe, and is of central importance if we wish to uncover the secrets behind nematicity and superconductivity in the wider iron-based superconductors. Here, we review the recent experimental work uncovering this nematic Fermi surface of FeSe from both ARPES and STM measurements, as well as current theoretical attempts to explain this missing electron pocket of FeSe, with a particular focus on the emerging importance of incorporating the $d_{xy}$ orbital into theoretical descriptions of the nematic state. Furthermore, we will discuss the consequence this missing electron pocket has on the theoretical understanding of superconductivity in this system and present several remaining open questions and avenues for future research.  
	\end{abstract}
	\date{\today}
	\maketitle

	\section{I. Introduction}
	One of the reasons for the huge interest in FeSe over the past decade has been the sense that it holds the key to the wider understanding of the whole Fe-based superconductor family \cite{Kreisel2020,Shibauchi2020, Fernandes2022}. With its minimalistic crystal structure and alluringly simple band structure in the tetragonal phase, alongside the prevalence of high-quality single crystals, it seemed like the ideal test bed to examine in detail the themes that were emerging in the field: strong orbital-dependent correlations \cite{Antoine2013,Medici2014,Nicola2013}, spin fluctuation pairing \cite{Mazin2008,Graser2009}, and most pertinently for this review, the so-called “nematic” phase \cite{Fernandes2014,Bohmer2017,Coldea2021}, where $C_4$ rotational symmetry is spontaneously broken below ~90 K. 
	
	The measurement of the momentum-dependence of the superconducting gap in FeSe, between 2016 and 2018, was a particular experimental triumph. The data from both scanning tunneling microscopy (STM) \cite{Sprau2017} and multiple angle-resolved photoemission spectroscopy (ARPES) measurements \cite{Xu2016,Hashimoto2018,Liu2018,Rhodes2018,Kushnirenko2018} revealed a clear conclusion: the gap structure is extremely anisotropic, and broadly follows the $d_{yz}$ orbital weight around the Fermi surface. While a twofold-symmetric gap is of course symmetry-allowed in an orthorhombic system, the fact that such a strong anisotropy was observed implied that the nematic state must also induce a profound anistropic effect on the Fermi surface of FeSe. However due to significant uncertainty as to the correct description of the low-temperature electronic structure, multiple theoretical explanations for the anisotropic gap structure were proposed \cite{Kreisel2017,Sprau2017,Benfatto2018,Rhodes2018,Kang2018,Yu2018}.
	
	A critical question required to understand this anisotropic superconducting gap is how does the nematic state influence the the low temperature Fermi surface and electronic structure of FeSe? Given that we have a second-order phase transition \cite{Bohmer2013}, and that the lattice distortion $\frac{|a-b|}{(a+b)}$ is only $\sim$0.2\%, the natural assumption, from an \textit{ab-initio} perspective \cite{Fedorov2016}, would be that nematicity should only weakly distort the established Fermi surface of the high-temperature tetragonal phase, which ARPES measurements have shown contains two hole pockets and two electron pockets \cite{Nakayama2014,Shimojima2014,Watson2015,Fanfarillo2016,Maletz2014,Fedorov2016,Watson2016,Coldea2018}. Yet ARPES measurements in the nematic state have revealed sizeable band shifts, of the order of 10-50~meV \cite{Coldea2018}, much larger that what would be predicted from ab-initio calculations \cite{Fedorov2016}. 
	
	\begin{figure*}
		\centering
		\includegraphics[width=\linewidth]{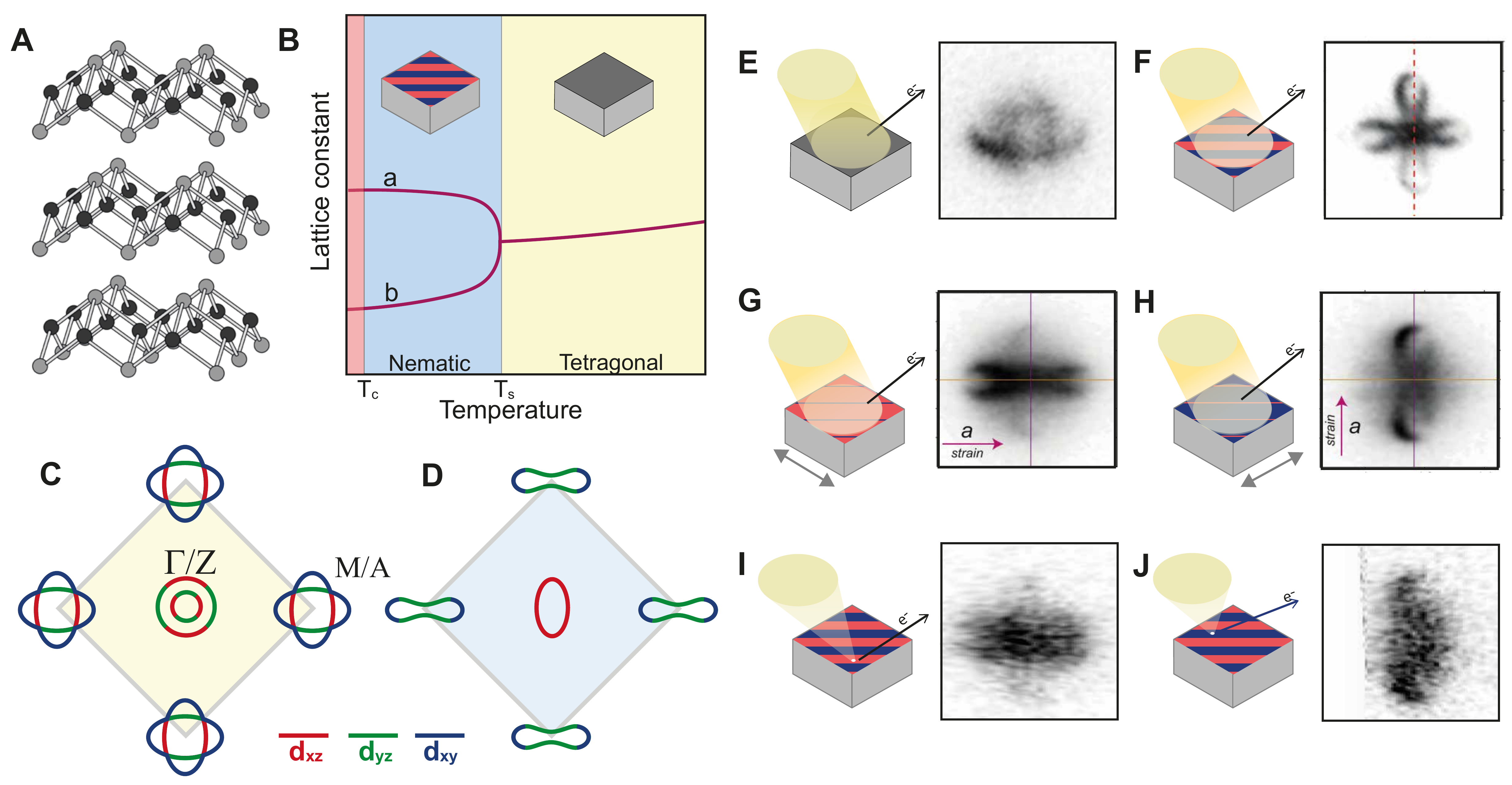}  
		\caption{\textbf{Summary of the Fermi surface of FeSe and photoemission measurements of a single electron pocket}. \textbf{(A)} Crystal structure of FeSe, Fe (black), Se (grey). \textbf{(B)} Sketch of the temperature evolution of the the lattice constants in FeSe, as described in Ref. \cite{Margadonna2008}, highlighting the evolution from a tetragonal (dark grey surface) to orthorhombic system with differently orientated domains (red and blue striped surface). \textbf{(C)} Sketch of the experimentally determined Fermi surface of FeSe in the tetragonal state and \textbf{(D)} in the nematic state.
			\textbf{(E-J)} Summary of the Fermi surface of FeSe measured around the M/A point via different photoemission techniques. \textbf{(E)} Measurement in the tetragonal state (100~K, LV $h_v=56$~eV \cite{Watson2016}) showing two electron pockets, \textbf{(F)} ARPES Measurement in the nematic state (10~K, LV, $h_v=56$~eV \cite{Watson2016})  arising from a superposition of two orthorhombic domain orientations (red and blue regions), referred to as a "twinned" measurement. \textbf{(G,H)} ARPES measurement of a detwinned crystal in the nematic state (10~K, $h_v=56$~eV \cite{Watson2017}, where strain is applied either along the $a$ or $b$ crystallographic axis and predominately probes a single domain orientation. \textbf{(I,J)} NanoARPES measurement using a photon beamspot of $<1$~$\mu$m (30~K, $h_v=56$~eV \cite{Rhodes2020}) in individual orthorhombic domains.}
		\label{fig:summary}
	\end{figure*}
	
	Unfortunately, the precise identification of specific parts of the band structure, the nematic energy scales and even the Fermi surface of FeSe has been complicated by the formation of orthorhombic domains upon entering the nematic state. In an orthorhombic crystal, conventional ARPES experiments measure a superposition of two perpendicularly orientated crystallographic domains, which doubles the number of bands observed in the experimental data and creates ambiguity about which bands arise from which domain. For this reason, a recent focal point of research has involved overcoming this technical challenge of orthorhombic domains, for example by applying uniaxial strain \cite{Shimojima2014,Watson2017,Yi2019,Huh2020,Cai2020,Cai2020b,Pfau2019,Pfau2021} or using NanoARPES \cite{Rhodes2020} or scanning tunneling microscopy \cite{Kasahara2014,Hanaguri2018,Sprau2017,Kostin2018}. The conclusion from these measurements have been unanimous, and have revealed that within the nematic state the Fermi surface of FeSe consists of one hole pocket and one electron pocket. 
	
	This finding, however, is very surprising and presents a fundamental theoretical conundrum that is at the heart of understanding the nematic and superconducting properties of FeSe. The bands that generate the two electron pockets observed in the tetragonal state form saddle points at the high symmetry M point close to the Fermi level. It is therefore not trivial to deform or shift these saddle points to lift one of these electron pockets away from the Fermi level upon entering the nematic state. This current theoretical challenge has become known as the ''missing electron pocket problem" of FeSe and resolving this problem promises deeper insight into the nematic state, and a wider understanding of superconductivity in the iron-based superconductors. 
	
	In this review we will overview the recent experimental and theoretical work uncovering the Fermi surface of FeSe in the nematic state and tackling the missing electron pocket problem. In section 2 we will briefly introduce the experimental electronic structure of FeSe in the tetragonal state, to use as the foundation for understanding the nematic electronic structure. In section 3 we will discuss the recent experimental data uncovering the electronic structure in the nematic state, in particular focusing on measurements which overcome the technical problems associated with orthorhombic crystals, including ARPES measurements under uniaxial strain, NanoARPES measurements and Scanning tunneling microscopy (STM) measurements. In Section 4 we will review the latest theoretical attempts to resolve this missing electron pocket problem, highlighting the necessity of considering the $d_{xy}$ orbital in the phenomenological description of the nematic state. And in section 5 we will discuss the consequence the updated Fermi surface has on the understanding of the superconducting properties of FeSe. A summary of the electronic structure and missing electron pocket problem of FeSe is presented in Fig. \ref{fig:summary}.

	\begin{figure}
		\centering
		\includegraphics[width=0.9\linewidth]{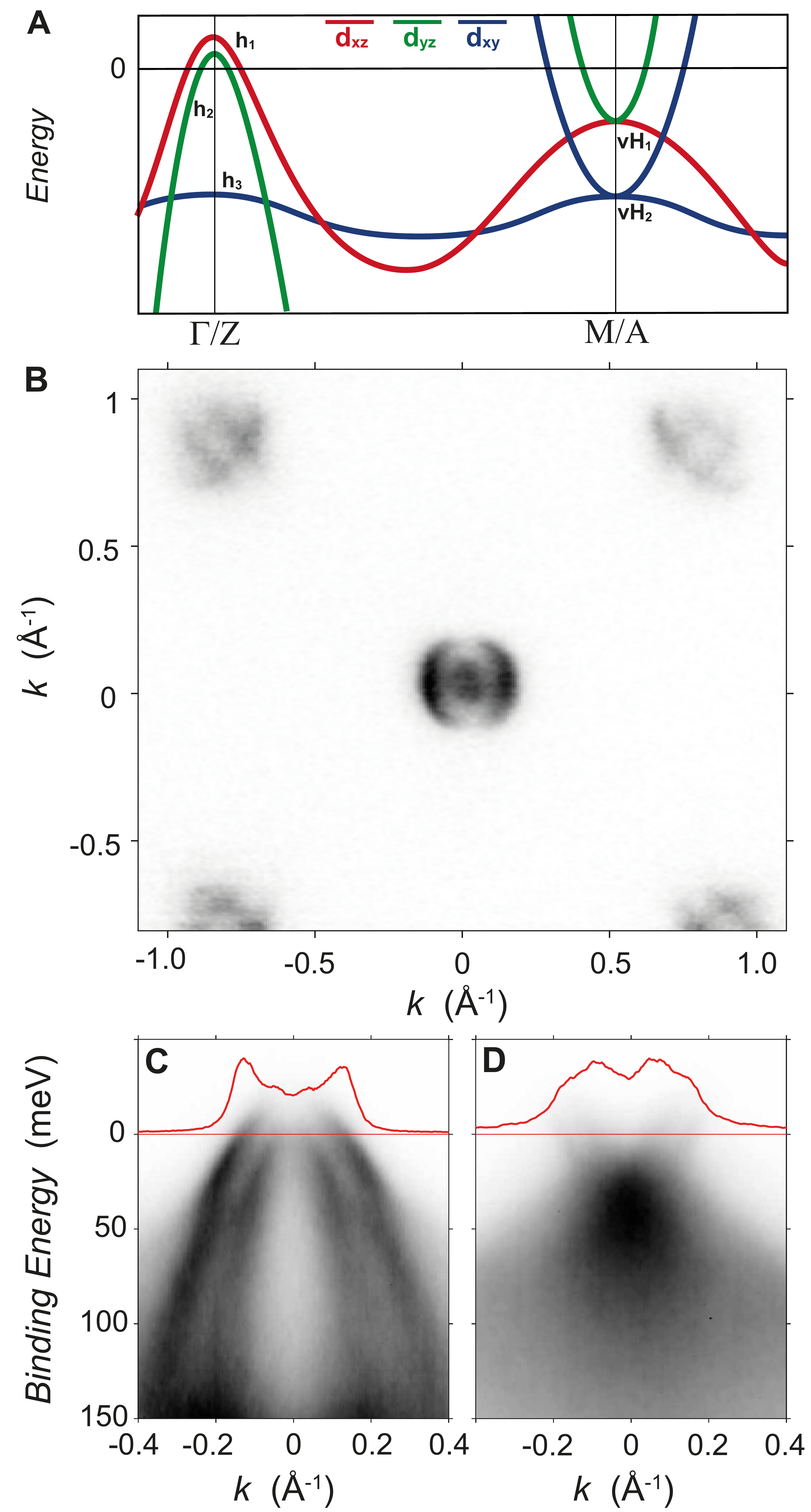}    \caption{\textbf{Electronic structure in the tetragonal state of FeSe}. \textbf{(A)} Sketch of the low-energy band structure of a typical $P4/nmm$ Fe-based superconductor along the $\Gamma-M$ ($k_z=0$) or $Z-A$ ($k_z = \pi$) high symmetry points. The colours indicate the dominant orbital character of the bands. \textbf{(B)} Fermi surface in the tetragonal state measured at 100~K close to $k_z = \pi$ ($h_\nu=56$~eV). \textbf{(C)} Cut along the $Z-A$ direction (equivalent to $\Gamma-M$ but at $k_z=\pi$) direction for the hole bands around the Z point for the same photon energy ($h_\nu=56$~eV). \textbf{(D)} Cut along the $Z-A$ direction for the electron bands around the A point. Figures adapted from \cite{Watson2015,Watson2016}.}
		\label{fig:Tetragonal}
	\end{figure}
	
	\section{II. Electronic structure in the Tetragonal state}
	
	From both a theoretical and experimental point of view, the electronic structure of the tetragonal state is relatively well understood. Prior to the onset of nematicity at $T_s=$ 90~K, FeSe exhibits tetragonal symmetry with a $P4/nmm$ crystal structure \cite{Margadonna2008}. This structure consists of layers of Fe atoms, in a 2D square lattice configuration, bridged by staggered out-of-plane Se atoms, giving rise to a crystallographic unit cell containing two Fe atoms and two Se atoms. The two Fe atoms are related by a glide-mirror symmetry, which can theoretically half the number of bands and allows for an unfolding to a 1-Fe Brillouin zone used by some authors \cite{AndersenBoeriGlideMirror}, but here we use the 2-Fe unit cell notation for comparison with ARPES measurements. 
	
	The low energy electronic properties are governed by the partially-filled $3d_{xz}$, $3d_{yz}$ and $3d_{xy}$ orbitals of the two Fe atoms, which in momentum space gives rise to three hole bands around the $\Gamma$ point and two symmetry-protected saddle point van-Hove singularities around the M point \cite{Eugenio2018} as shown in Fig. \ref{fig:Tetragonal}(a).
	
	Of the three hole bands, two exist as a $C_4$ symmetric pair exhibiting predominantly $d_{xz}$ and $d_{yz}$ orbital weight (labelled $h_1$ and $h_2$ in Fig.\ref{fig:Tetragonal}(a)) and the third is dominated by $d_{xy}$ orbital character ($h_3$). $h_1$ and $h_2$ would be energy degenerate at the high symmetry point, however spin orbit coupling lifts this degeneracy \cite{Borisenko2016}. As for the van-Hove singularities around the M point, one is a saddle point connecting bands of majority $d_{xz}$ and $d_{yz}$ weight (${vH_1}$) and the other is a saddle point connecting two $d_{xy}$ dominated bands (${vH_2}$). This general structure is broadly applicable to all $P4/nmm$ Fe-based superconductors (e.g. Fe(Te,Se,S), LiFeAs, NaFeAs, LaFeAsO), with some modifications for the 122 family due to the $I$-centering of the lattice. 
	
	The experimentally measured Fermi surface of FeSe at 100 K (or more precisely, a map of the experimental spectral function at the chemical potential) at approximately $k_z=\pi$ is shown in Fig. \ref{fig:Tetragonal}(b), revealing a two-hole pocket and two electron pocket Fermi surface.  Measurements around the center of the Brillouin zone show that both $h_1$ and $h_2$ cross the chemical potential at 100~K, as shown in Fig. \ref{fig:Tetragonal}(c). Their band maximas are separated by $\sim$20~meV due to spin-orbit coupling \cite{Watson2017b, Borisenko2016,Day2018}. At $k_z = 0$ these bands have a maxima at approximately $h_2=-13~meV$ and and $h_1=+7$~meV \cite{Watson2017b}, and at $k_z = \pi$ (shown in Fig. \ref{fig:Tetragonal}(c)) the bands have maxima of approximately $h_2=+5$~meV and $h_1=+30$~meV. The second smaller hole pocket of FeSe is thus only present at finite $k_z$, which highlights an important property of this system. Even though FeSe has a ``quasi-2D" structure, i.e the energy shift of the bands as a function of $k_z$ is only on the order of 20~meV, this energy scale is actually on the same order of magnitude as the total Fermi energy of this system, and therefore is non-negligible in quantitative descriptions of the physical properties of FeSe. We note in passing that, due to the small Fermi energy of this system, the electronic structure is subject to substantial temperature-dependence of the chemical potential, and the appearance of the ``Fermi surface" changes substantially between 100 and 300~K \cite{Rhodes2017}, although without any change of the symmetry. 
	
	The third $d_{xy}$ hole band, $h_3$, is observed to be much flatter and cross both $h_1$ and $h_2$ at an energy of approximately -50~meV. In most ARPES data sets, this band has a much lower intensity than the $h_1$ and $h_2$ bands, which is a consequence of photoemission-based matrix element effects, which ensures the intensity of photoelectrons originating from $d_{xy}$ states with momentum near $|\mathbf{k}| =  0$ will be suppressed \cite{Zhang2012}. Nevertheless, $h_3$ can be identified most clearly near where it hybridises with $h_1$ and $h_2$, and thus acquires some $d_{xz}$ and $d_{yz}$ orbital weight as shown in Fig. \ref{fig:Tetragonal}(c). 
	
	Near the corner of the Brillouin zone, both the $d_{xy}$ dominated electron band, connected to $vH_2$, and the $d_{xz}/d_{yz}$ electron band, connecting to $vH_1$, are observed to cross the Fermi level. Here the outer four-fold symmetric electron pocket is dominated by $d_{xy}$ orbital character while the inner pocket is dominated by $d_{xz}$ and $d_{yz}$ orbital weight \cite{Eschrig2009}. As this is a compensated system, the total Fermi volume of these electron pockets should be equal to that of the hole pockets \cite{Watson2015}.
	
	These two sets of electron bands connect to the saddle points which have an energy of approximately $vH_1=-20$~meV and $vH_2 = -40$~meV at the high symmetry point. The exact position of these stationary points, however, are masked by the presence of self-energy interactions which give rise to a broadening of the electronic states around the M point. This broadening is also captured in theoretical simulations involving spin and charge fluctuations \cite{Acharya2021}.
	
	The ARPES data presented in Fig. \ref{fig:Tetragonal} is taken from our own works \cite{Watson2016,Watson2015}, however multiple data sets are available in the literature and are all consistent with the interpretation presented here \cite{Watson2015b,Fedorov2016,Fanfarillo2016,Shimojima2014,Nakayama2014,Coldea2018}. Indeed, the electronic structure must be constrained by the symmetry based arguments of Fig. \ref{fig:Tetragonal}(a) \cite{Eschrig2009,Eugenio2018,Fernandes2014b} and each of the bands observed in the measurements can be mapped to corresponding bands calculated from \textit{ab-initio} techniques such as density functional theory (DFT) \cite{Eschrig2009,Watson2015,Fedorov2016} of the paramagnetic tetragonal phase. 
	
	There are, however, serious quantitative issues with DFT-based calculations, which severely limit its use in describing the low energy properties of FeSe. First, DFT-based calculations overestimate the bandwidth of the Fe $3d$-bands by a factor of $\sim$3\cite{Watson2015}. This is a generic finding across all Fe-based superconductors \cite{Yin2011}, and derives from the fact that electronic correlations are inadequately treated in DFT. It has been often argued that the correlation effects are orbital-dependent and particularly strong for the $d_{xy}$ orbital \cite{Watson2015,Yin2011,Yi2015}. More advanced theoretical simulations, such as DFT + DMFT \cite{Watson2017c} and QSGW + DMFT \cite{Acharya2021}, have had some success in capturing the global electronic structure on the eV scale \cite{Watson2017c,Evtushinsky2017_arXiv}, finding strongly incoherent spectral weight at 1-3~eV below $E_F$ and sharp quasiparticles only in the near vicinity of $E_F$. 
	However \textit{ab-initio} efforts still usually overestimate the size of the hole and electron Fermi surfaces, which are much smaller in experiment \cite{Watson2015,Watson2017c}. Most DFT-based simulations additionally predict that the $d_{xy}$ hole band also crosses the Fermi level, suggesting a three hole pocket and two electron pocket Fermi surface \cite{Eschrig2009,Watson2015}. 
	Finally, typical DFT-based calculations also suggest that a stripe or staggered-stripe antiferromagnetic ground state is the most stable configuration \cite{Yin2011,Glasbrenner2015}, when in reality FeSe remains paramagnetic (albeit with strong antiferromagnetic fluctuations \cite{He2018,Chen2019,Wang2020}). Current research is attempting to resolve this discrepancy from a pure \textit{ab-initio} perspective. Wang. \textit{et. al.} \cite{Wang2020} were able to reproduce the band structure around the Gamma point using a a polymorphus network of local structural distortions. The use of hybrid exchange correlation functionals and Hubbard-Hund correlations have recently been shown to also produce a substantial improvement on the tetragonal structure \cite{Gorni2021}.

	Due to the current limitations in \textit{ab-initio} modelling however, a substantial amount of work has gone into developing quantitatively accurate tight binding models of FeSe \cite{Eschrig2009, Graser2009,Mukherjee2015,Rhodes2017,Rhodes2021}. These models bypass the limitations in our current \textit{ab-initio} theories, allowing for an accurate, albeit phenomenological, description of the single-particle electronic structure to be defined, which we can compare with experimental measurements. Several hopping parameters sets have been developed, which have been obtained by directly comparing the numerical band dispersion with experimental ARPES data in the tetragonal state \cite{Mukherjee2015,Kreisel2017, Rhodes2017, Rhodes2021}. These models have been shown to reproduce the single-particle electronic properties of tetragonal FeSe much better than conventional DFT-based approaches \cite{Eschrig2009,Mukherjee2015,Rhodes2017}. In particular these models accurately capture the small Fermi energy of FeSe, which has been shown to lead to strong chemical potential renormalising effects as a function of temperature and nematic ordering \cite{Rhodes2017,Coldea2018,Kushnirenko2017,Pustovit2017,Pustovit2018}. By construction, such models allow for a quantitative description of the band positions of the hole and electron bands such that a comparison of the electronic structure in the nematic state can take place.

	\section{III. Experimental evidence for a missing electron pocket in the nematic state}
	We now focus on the electronic structure in the nematic state. Here experimental measurements encounter a major challenge. The nematic state is accompanied by a tetragonal to orthorhombic structural transition, at which point multiple orthorhombic domains form in the crystal. It has been identified that these domains are typically on the order of 1-5 $\mu$m in size \cite{Schwier2019,Rhodes2020,Shimojima2021,Tanatar2016}, which is much smaller than the cross section of the photon beam used in most high resolution synchrotron-based ARPES measurements ($>50\mu$m \cite{Hoesch2017}), as sketched in Fig. \ref{fig:detwinned}(a). Most of the initial photoemission data of FeSe in the nematic phase was collected on ``twinned" crystals. In such measurements, the band dispersion measured along the experimental $k_x$ axis contains contributions from domains with the orthorhombic $a$ axis both along, or perpendicular to, this direction, i.e. one measures a superposition of the spectral function arising from both domains. This creates an apparent $C_4$ symmetry in the measurements even at low temperatures (in the sense that the measured spectra are invariant under 90 degree rotation of the sample; the as-measured spectra are not generally fourfold-symmetric due to the ARPES matrix elements \cite{Brouet2012,Day2019}). This can lead to ambiguity about which band arises from which domain.

	\subsection{ARPES measurements on twinned crystals}
	
\begin{figure*}
		\centering
		\includegraphics[width=0.8\linewidth]{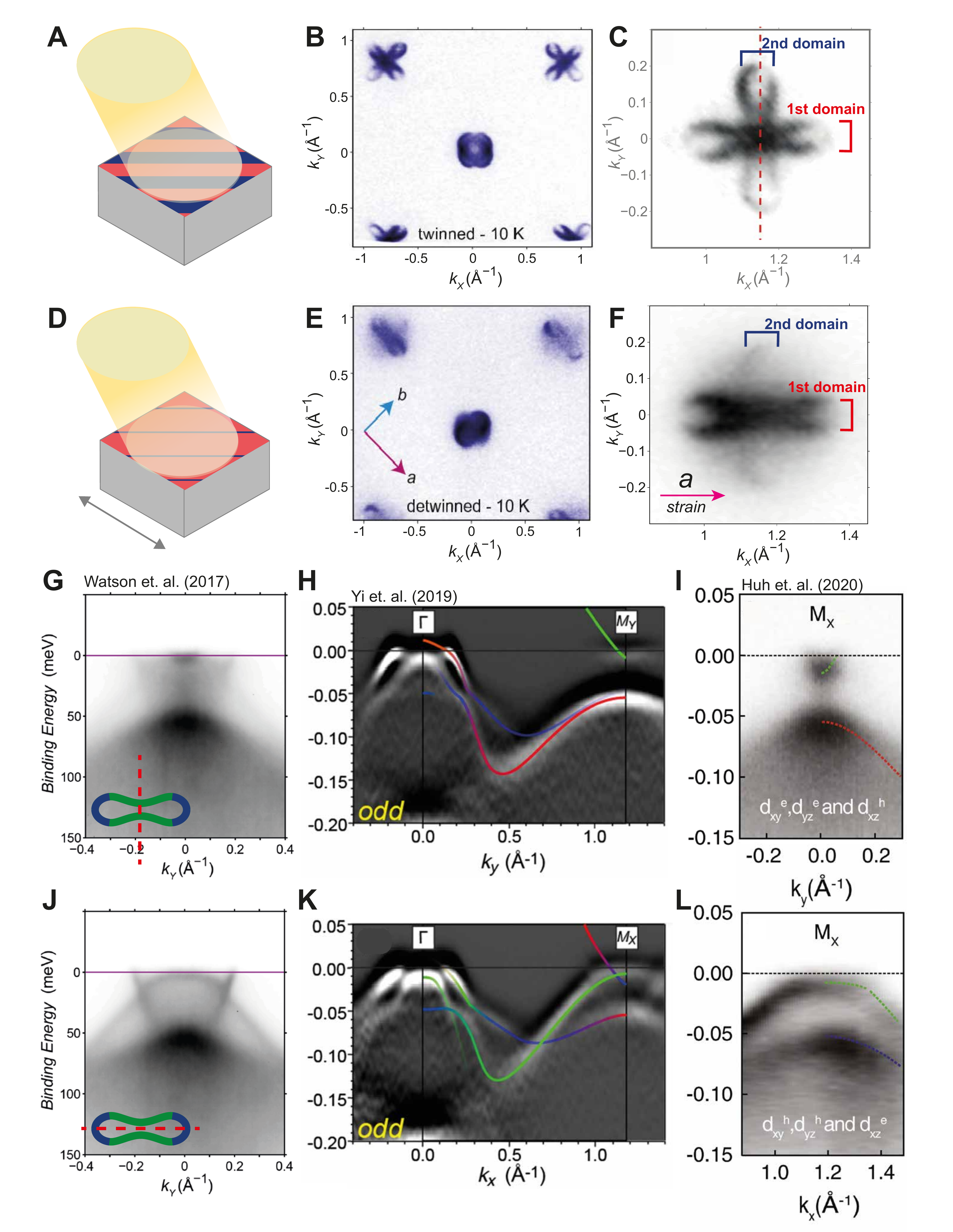}
		\caption{\textbf{Summary of ARPES measurements on detwinned crystals of FeSe} \textbf{(A)} sketch of a photoemission setup on a twinned crystal, showing equal coverage of both red and blue orthorhombic domains. \textbf{(B)} Fermi surface measured from a measurement on a twinned crystal ($h\nu$=56~eV). Taken from Watson \textit{et. al.} \cite{Watson2017}  \textbf{(C)} Close up of the electron pocket near the A point from Watson \textit{et. al.} \cite{Watson2016} ($h\nu$=56~eV).  \textbf{(D-F)} Equivalent sketch and measurement for a detwinned crystal of FeSe, which probes a majority of orthorhombic domains aligned in one direction. \textbf{(G)} Band dispersion of a detwinned crystal centered at the electron pocket. The insert shows the band path, from Watson \textit{et. al.} \cite{Watson2017}. h) Second derivative band dispersions of a detwinned crystal along the same path as \textbf{(G)} but extended from $Z$ to $A$, taken from Yi.\textit{ et. al.} \cite{Yi2019}. i) Band dispersion of a detwinned crystal along the same path as \textbf{(G)} from Huh. \textit{et. al.} \cite{Huh2020}. \textbf{(J-L)} Equivalent measurements but taken along the length of the electron pocket. \textbf{(H,K)} reproduced from Ref. \cite{Yi2019} under the Creative Commons Attribution 4.0 International License. \textbf{(I,J)} reproduced from Ref. \cite{Huh2020} under the Creative Commons Attribution 4.0 International License. }
		\label{fig:detwinned}
	\end{figure*}

	Multiple ARPES measurements on twinned crystals of FeSe have been reported \cite{Maletz2014,Nakayama2014,Shimojima2014,Watson2015,Watson2016,Fedorov2016,Fanfarillo2016,Watson2015b,Reiss2017,Kushnirenko2017,Kushnirenko2018,Rhodes2018} and have been extensively reviewed \cite{Pustovit2016,Coldea2018,Bohmer2017,Kreisel2020}. We present a representative Fermi surface obtained from a twinned crystal in Fig. \ref{fig:detwinned}(b) from Ref. \cite{Watson2016}. The hole pockets appear as two overlapping ellipses. Meanwhile, at the corner of the Brillouin zone, measurements reveal two electron pockets, which have been pinched in to produce what looks like two overlapping ``peanuts". 
	
	The challenge now lies in identifying which of these pockets, comes from which domain. The two hole pockets can be easily understood as one ellipse from each orthorhombic domain. Measurements of the band dispersion around the hole pocket reveal that the inner hole band ($h_2$) undergoes a Lifshitz transition as a function of temperature and resides below the Fermi level at 10~K, whilst the outer hole band ($h_1$) elongates into an elliptical shape. As all three hole bands can be tracked as a function of temperature from the tetragonal to nematic state, there is little ambiguity about the shape of the hole pocket Fermi surface at low temperatures. However, it is not possible to identify the orientation of the elliptical hole pocket from a single domain, i.e to identify whether it elongates along the orthorhombic $a$ or $b$ axis simply from these twinned measurements. 
	
	For the electron pocket, however, the understanding was less clear, and historically several distinct band structures have been interpreted from nearly identical data sets \cite{Fanfarillo2016,Watson2015,Watson2016,Kushnirenko2018,Rhodes2018}. As can be seen in Fig. \ref{fig:detwinned}(c), two electron pockets can be observed which look like overlapping ``peanuts'' in the twinned data. As the tetragonal state also exhibits two electron pockets, this may not appear that surprising. Indeed one interpretation was that the two oval shaped electron pockets in the tetragonal state simply pinched in at the sides, due to raising the binding energy of $vH_1$ \cite{Watson2016,Fedorov2016}. In other words, the electron pockets could retain approximate fourfold symmetry around the M point, and the pockets from each domain simply overlapped in twinned data sets \cite{Kushnirenko2018}.  However, other interpretations, particularly those attempting to understand the nematic band shifts from theoretical grounds, believed that the nematic state should have two differently shaped electron pockets \cite{Fanfarillo2016,Watson2015}. It was also equally plausible, experimentally at least, that only one electron pocket existed per domain \cite{Shimojima2014,Watson2017c,Rhodes2018}. Distinguishing between these scenarios was particularly challenging due to the broadness of the spectral weight around the M point in the tetragonal state (see Fig. \ref{fig:Tetragonal}(d)), which made a precise interpretation of the temperature evolution of the two van-Hove singularities ambiguous.

	\subsection{ARPES measurements on detwinned crystals}
	Compared to the measurements on twinned data, a much more preferable method to study the Fermi surface of FeSe would be to experimentally overcome the limitation imposed by these orthorhombic domains, and directly measure the electronic structure from a single crystallographic orientation. There are two strategies to overcome the twinning issue faced by ARPES measurements. Either 1) generate a sample with macroscopic ordering of the orthorhombic domains on length scales larger than the photon beam cross section, or 2) make the photon beam much smaller than the size of an orthorhombic domain. It has been known from earlier work on the 122 family of Fe-based superconductors that upon the application of ``uniaxial" strain along the Fe-Fe direction, it becomes energetically favourable for a majority of the orthorhombic domains to align along that axis \cite{Fisher2011}. While the resulting domain population is unlikely to be 100\% pure, measurements on strained, or ``detwinned", samples, as sketched in Fig. \ref{fig:detwinned}(d), allows one to distinguish between the intense spectral weight arising from the majority domain and the weak spectral weight arising from the 90 degree rotated minority domain.
	
	The first ARPES measurements on uniaxial strained samples of FeSe were performed in 2014 by Shimojima et. al. \cite{Shimojima2014}, where it was shown that the single hole pocket was elongated along the $k_y$ axis. Later, in 2017, Watson \textit{et. al.} \cite{Watson2017} was additionally able to resolve the detail of the electron pockets, as shown in Fig. \ref{fig:detwinned}(e). These measurements on detwinned crystals confirmed that the Fermi surface consisted of one elliptical hole-pocket, as expected from interpretation of the twinned measurements, but additionally revealed only one electron-pocket around the M point. This is shown in Fig. \ref{fig:detwinned}(f), where the majority of the spectral weight intensity now comes from one domain, and only a weak residual intensity comes from the minority domain. Unlike in the tetragonal state, at low temperatures, the electronic band structure around the M point produces sharp quasiparticle bands, a saddle point can be observed at $-5$~meV, which is electron like along the minor length of the electron pocket (as shown in Fig \ref{fig:detwinned}(g)), but hole-like when rotated by 90 degrees (Fig. \ref{fig:detwinned}(j)). Additionally, along the major length of the electron pocket, a deeper electron band and saddle point at $\sim-60$~meV can be observed that is dominated by $d_{xy}$ orbital weight. This gap between the upper and lower saddle points, is approximately 50~meV, and has been previously quoted as a ``nematic energy scale" \cite{Watson2015,Yi2019,Pfau2019}. However, as we will discuss in the theoretical section below, the exact energy scale of nematic shifts and splittings is slightly more complex and requires a linear combination of order parameters of different energy scales \cite{Rhodes2021}.
	
	This finding of only a single electron pocket at the Fermi level was not the expected theoretical result \cite{Mukherjee2015}, but nevertheless has now been reproduced by Yi \textit{et. al }\cite{Yi2019} (Fig. \ref{fig:detwinned}(h,k)) and Huh \textit{et. al.} \cite{Huh2020} (Fig. \ref{fig:detwinned}(i,l)). Further measurements on sulphur doped FeSe$_{1-x}$S$_{x}$ crystals under uniaxial strain by Cai \textit{et. al.} \cite{Cai2020, Cai2020b} have also reported very similar Fermi surfaces.
	
	\subsection{Temperature dependent detwinned ARPES measurements}
	A natural question when studying the evolution of the electronic structure of FeSe from the tetragonal to nematic state is to ask how do ARPES measurements evolve as a function of temperature. Many data sets on twinned samples exist (as discussed e.g by Coldea and Watson \cite{Coldea2018}), and recently Yi \textit{et. al.} and \cite{Yi2019}, Huh \textit{et. al.} \cite{Huh2020} have presented temperature dependent measurements on detwinned samples of FeSe. Similarly Cai \textit{et. al.} \cite{Cai2020b} have reported temperature dependent measurements on detwinned samples of 9\% doped FeSe$_{1-x}$S$_{x}$. 
	
	In twinnned crystals, the temperature dependence of the hole bands leaves little room for ambiguity, and can be neatly tracked as a function of temperature \cite{Watson2017c}. However the bands around the M point are a bit more ambiguous, due to the broad spectral feature of the M point in the tetragonal state (as shown in Fig. \ref{fig:Tetragonal}(d)). Whilst this broad spectral feature is observed to split as a function of temperature, precisely tracking the vHs from high to low temperatures requires a degree of interpretation and peak fitting, with multiple papers suggesting different evolution of the spectral weight \cite{Yi2015,Christensen2020,Watson2016,Fanfarillo2016,Fedorov2016}.
	
	Unfortunately, the additional complication of uniaxial strain in detwinned measurements makes temperature dependent analysis technically even more challenging. By changing the temperature of your system, you inevitably alter the amount of strain applied to the sample due to thermal expansion of the rig, which in turn may alter the relative population of orthorhombic domains you are probing, and moreover the energetics of domain formation may be temperature-dependent.  
	
	Cai \textit{et. al.} \cite{Cai2020b}, observed that as a function of decreasing temperature, the spectral weight of the second electron pocket simply decreases, which could be explained as a change in orthorhombic domain populations originating from a Fermi surface consisting of just one electron pocket per domain. However, Cai \textit{et. al.} argue that this is not the case and that the spectral weight loss is intrinsic to the nematic state \cite{Cai2020b}. 
	
	On the other hand, Yi \textit{et. al.} \cite{Yi2019} and Huh. \textit{et. al.} \cite{Huh2020} argue that they observe a band shifting above the Fermi level in their temperature dependent measurements. A temperature band shift would be independent of orthorhombic domain population, however it should also have been detected within twinned ARPES measurements, which so far has not been reported. 
	
	We conclude by noting that temperature-dependent ARPES measurements of the electron pocket are very challenging to perform, firstly because of the issue with orthorhombic domains, and secondly due to the intrinsic broadness of the van hove singularities measured by ARPES. One approach to overcome this limitation would be to systematically study the evolution of the low temperature electronic structure across the series of isoelectronic sulphur substituted FeSe$_{1-x}$S$_{x}$ crystals, extending the existing measurements on twinned samples \cite{Watson2015b,Reiss2017}. This will require future experimental investigation.
	
	\begin{figure*}
		\centering
		\includegraphics[width=\linewidth]{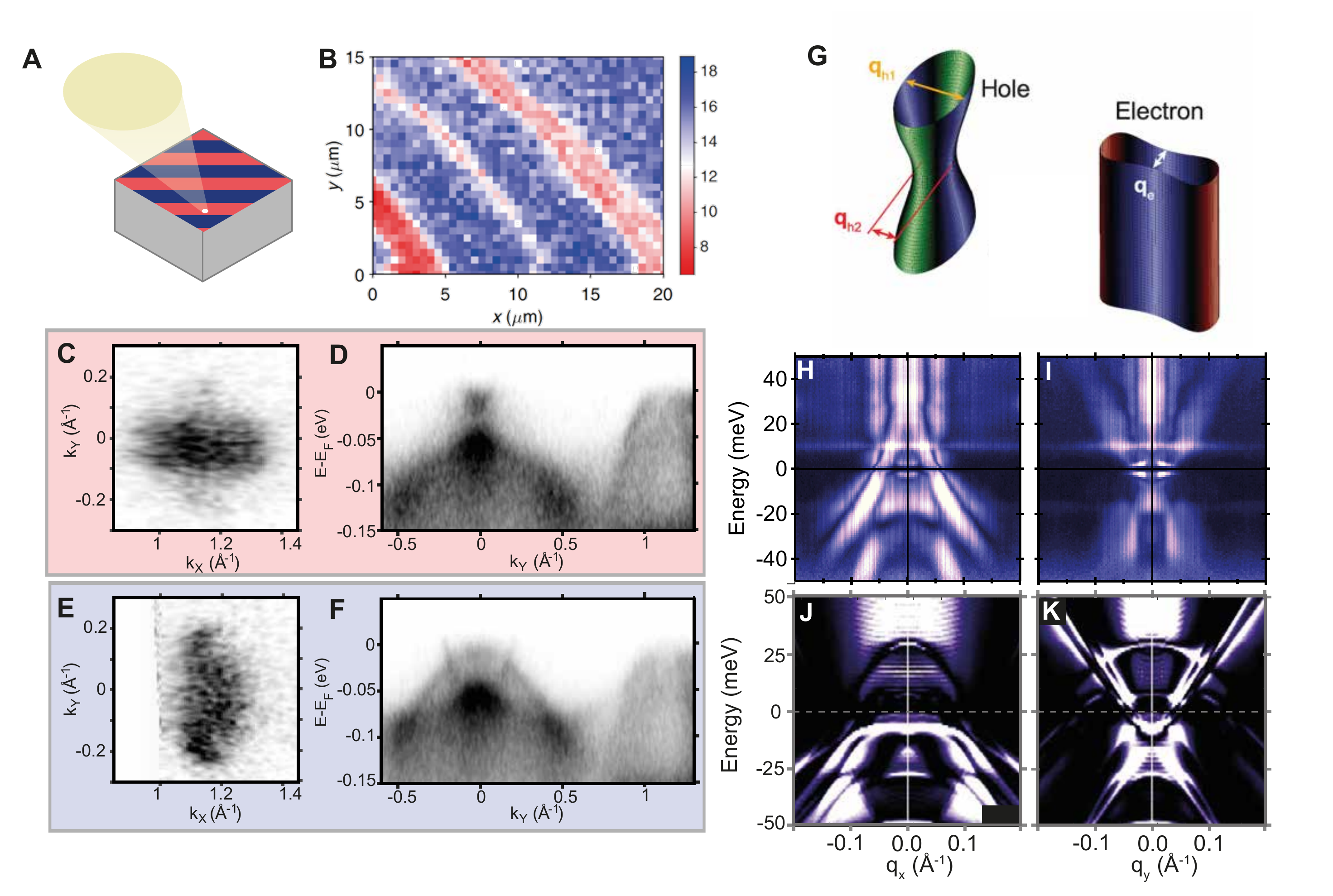}
		\caption{\textbf{Electronic structure within a single domain without the application of uniaxial strain}. \textbf{(A)} Sketch of a NanoARPES measurement, where the beam is focused to have a cross section $<1\mu$m. \textbf{(B)} Experimental spatial map of FeSe where the colour corresponds to the orthorhombic domains of FeSe, reproduced from Ref. \cite{Rhodes2020}. \textbf{(C,D)} Fermi surface around the electron pocket ($h_\nu=56$~eV, 30~K) and $A_y-Z$ cut taken within a single orthorhombic domain \cite{Rhodes2020}. \textbf{(E,F)} Fermi surface around the electron pocket and $A_y-Z$ cut taken in an adjacent orthorhombic domain \cite{Rhodes2020}. \textbf{(G)} Sketch of the Fermi surface scattering vectors inferred from STM measurements, as suggested by Ref. \cite{Hanaguri2018}.
			\textbf{(H,I)} STM measurements of the QPI scattering dispersions as a function of energy along the $q_x$ and $q_y$ high symmetry axes respectively, reproduced from Ref. \cite{Hanaguri2018}. \textbf{(J,K)} Simulated QPI scattering dispersion from a model of FeSe which described the band structure shown in Fig. \ref{fig:detwinned}, from Ref. \cite{Rhodes2019}. \textbf{(G-I)} are reproduced under the Creative Commons Attribution 4.0 International License. }
		\label{fig:NanoARPES}
	\end{figure*}
	
	\subsection{NanoARPES}
	There are experimental complications with performing ARPES measurements on uniaxially strained crystals, which may leave doubt as to the validity of the conclusions presented above. First, it is hard to fully exclude if the application of uniaxial strain has actually perturbed the underlying electronic structure of the crystal you are measuring. For example in the tetragonal material Sr$_2$RuO$_4$, uniaxial strain on the order of 1\% shifts the position of the vHs by nearly 20 meV \cite{Sunko2019}. In order to fully support the conclusions from these ARPES measurements on detwinned crystals, complementary techniques must be employed and their results compared. To this end, nanoARPES has also been performed on crystals of FeSe. In these technically demanding measurements, the photon beam is focused to sub-micrometer spatial resolution using a focusing optic close to the sample \cite{Iwasawa_2020}. The reduction of the spot size comes at the cost of dramatically reducing the photon flux, and thus the energy and angular resolutions are typically relaxed (compared to the earlier high-resolution results presented) in order to have a reasonable signal of photoelectrons. Nevertheless, the technique has been improved over the past 10 years to allow for energy resolution better than 20 meV \cite{Watson2019}.  This sub-micrometer spot size is smaller than a single orthorhombic domain, allowing for a spatial map of the sample from which the two orthorhombic domains can be distinguished by analysing their differing ARPES spectra, shown as red and blue stripes in Fig. \ref{fig:NanoARPES}(a,b). Measurements of the Fermi surface and band dispersion around the electron pocket in both domains (Fig. \ref{fig:NanoARPES}(c-f)) reveal an electronic structure totally consistent with that extracted from the ARPES measurements under uniaxial strain. In summary, the nanoARPES results fully support the conclusion of a Fermi surface in the nematic state consisting of a single hole pocket and a single electron pocket.

	\subsection{STM measurements}
	An entirely independent method to study the momentum resolved electronic structure within a single domain is to use scanning tunneling microscopy (STM). STM utilises quantum tunnelling, between the surface of a material and an atomically sharp tip, to study the electronic structure on the sub-nanometer scale. Information about the electronic structure can then be extracted in two ways. The first is by studying the differential conductance ($dI/dV$) to obtain a quantity proportional to the local density of states of the system. The second is to measure quasiparticle interference (QPI), to measure the perturbations to the local density of states generated by the presence of defects such as impurities or atomic vacancies. The wavelength associated with this perturbation contains direct information about the allowed momentum dependent scattering vectors associated with an electronic structure at a constant energy via $\mathbf{q} = \mathbf{k}-\mathbf{k'}$. 
	
	Multiple STM measurements have been reported for FeSe, and information regarding the nematic \cite{Kasahara2014,Kostin2018,Sprau2017} and superconducting state \cite{Kasahara2014,Jiao2017,Sprau2017,Hanaguri2018} have been determined, tetragonal state information has also been obtained from studies of isoelectronic sulphur doped crystals \cite{Hanaguri2018}. These measurements all contain a plethora of information regarding the local structure of the surface of FeSe, as well as information on defects \cite{Choubey2014,Bu2019}. Here, however, we focus on what the STM measurements can tell us about the low energy electronic structure in the nematic state, and whether this is consistent with the ARPES measurements discussed above. Although measuring QPI is an indirect method to measuring the electronic structure of a material, it is particularly powerful in determining band minimas and maximas, especially above the Fermi level, as well identifying whether bands have hole or electron scattering characteristics within a certain energy range.
	
	The scattering vector vs energy dispersion along the $q_x$ and $q_y$ directions, taken from Ref. \cite{Hanaguri2018}, are presented in Fig. \ref{fig:NanoARPES}(h,i). In agreement with other data sets \cite{Kasahara2014,Kostin2018}, several hole-like scattering vectors can be observed predominately along the $q_x$ axis, with a narrower hole-like dispersion along the $q_y$ direction. Also along the $q_y$ axis, one very clear electron-like scattering vector can be detected, which has a minima at $\sim-5$~meV, and has been identified as a scattering vector that connects the $d_{yz}$ parts of the electron pocket in FeSe (Fig. \ref{fig:NanoARPES}(g) \cite{Hanaguri2018,Kostin2018,Kasahara2014,Rhodes2019}. No corresponding electron-like dispersion can be observed along the $q_x$ direction, which should be the case in a two electron pocket scenario where all bands scatter equally. This was therefore interpreted as further evidence, from an independent technique to ARPES, that the Fermi surface of FeSe only consists of one hole pocket and one electron pocket, as sketched in Fig. \ref{fig:NanoARPES}(g).
	
	We note that due to the indirect nature of QPI measurements, there is a degree of interpretation and uncertainty about the assignment of the electronic states and often it is necessary to directly simulate the QPI dispersion from a theoretical assumption of the electronic structure and compare the agreement. Due to the intrinsic broadness of the experimentally measured scattering vectors, this can lead to differing conclusions based on initial assumptions. For example, Kostin \textit{et. al.} \cite{Kostin2018}, assuming that two electron pockets must be present at the Fermi level, interpreted a weak spectral feature as evidence for a second electron pocket, with a greatly reduced scattering intensity. Whereas Rhodes \textit{et. al.} \cite{Rhodes2019}, assuming that only one electron pocket was present at the Fermi level, interpreted this weak feature as an artifact of the Feenstra function, used in the experimental processing \cite{Macdonald2016}. Importantly however, both theoretical simulations agree that a Fermi surface consisting of one hole pocket and two electron pockets can not independently reproduce the observed data without some additional form of anisotropy, which implies that ARPES and STM are probing the same underlying electronic structure. We present the numerical simulations from Ref. \cite{Rhodes2019} in Fig. \ref{fig:NanoARPES}(j,k).
	
	As an aside, it is interesting to note that the hole band maxima in \ref{fig:NanoARPES}(h) extends to +25~meV \cite{Hanaguri2018}. It is known from ARPES that only one hole-like scattering vector at this energy can exist, and specifically must be generated by the $k_z=\pi$ states \cite{Coldea2018}. This reveals that QPI measurements are sensitive to states with different $k_z$. From arguments about the group velocity of electrons scattering off of defects \cite{Weismann2009,Lounis2011}, and the short range nature of quantum tunneling, it actually implies that QPI measurements will exhibit a $k_z$-selectivity rule \cite{Rhodes2019}, such that all stationary points along the $k_z$ axis will contribute to scattering vectors that will be detected by STM measurements, this has recently been realised in the fully 3D system, PbS \cite{Marques2021}. 
	
	\subsection{Points of contention}
	
	While we have so far presented a unified picture of the electronic structure of FeSe and have focused on points where broad agreement is found in the recent literature, historically there have been many points of disagreement surrounding the identification of bands and the nature of the Fermi surface, and there remain some points of contention.
	
	Regarding the hole pockets, an outlying report is a recent claim from laser-ARPES measurements that there is additional splitting, most prominently resulting in two hole pockets at the Fermi level instead of one \cite{Li2020PRXARPES}. The implication is that the Kramer’s degeneracy of the bands is lifted, i.e. that either time-reversal or inversion symmetry is broken. However, it is worth noting that at low photon energies used the $k_z$ is not well-defined as the final states are not free electron-like, and the two Fermi contours identified appear to be fairly close to the known Fermi contours at $k_z=0$ and $k_z=\pi$. Moreover, synchrotron-ARPES measurements with equally high energy resolution and better angular resolution (due to better definition of $k_z$) do not identify any additional splitting either in the $\Gamma$ or Z planes \cite{Rhodes2018}, and neither has any comparable splitting been observed for the electron pockets. Finally, there is no supporting evidence for time-reversal symmetry breaking from other techniques. Thus it remains our view that the Kramer’s degeneracy holds for all states and that there is only one hole pocket crossing EF, which is significantly warped along the $k_z$ axis.
	
	Regarding the electron pockets, while several groups have now coalesced around the one electron pocket scenario, it has previously been claimed that the ARPES data on twinned crystals is consistent with four features in the EDC at the M point \cite{Fedorov2016} such that there are two electron pockets per domain, with each domain contributing a pair of crossed peanuts with slightly differing shapes \cite{Kushnirenko2018}. This scenario is perhaps the most natural, as it is based on DFT predictions, and comes down to somewhat technical questions of whether asymmetric lineshapes at the M point contain one or two peaks, and whether the proposed small splittings can be resolved. Some of this groups data on twinned samples does indeed seem to show a splitting, which at face value would support their scenario. However, neither our group nor other groups have observed these claimed features and peak splittings in comparable data on twinned samples. Moreover, the detwinned data shows a complete absence of any spectral weight aside from the peanut along the $a$ direction, in multiple experimental geometries, which cannot easily be explained away by matrix element effects in ARPES (and similarly in QPI). We encourage all groups to continue to push for higher resolution data which could finally settle the controversy, especially on detwinned samples.

	\begin{figure*}
		\centering
		\includegraphics[width=0.7\linewidth]{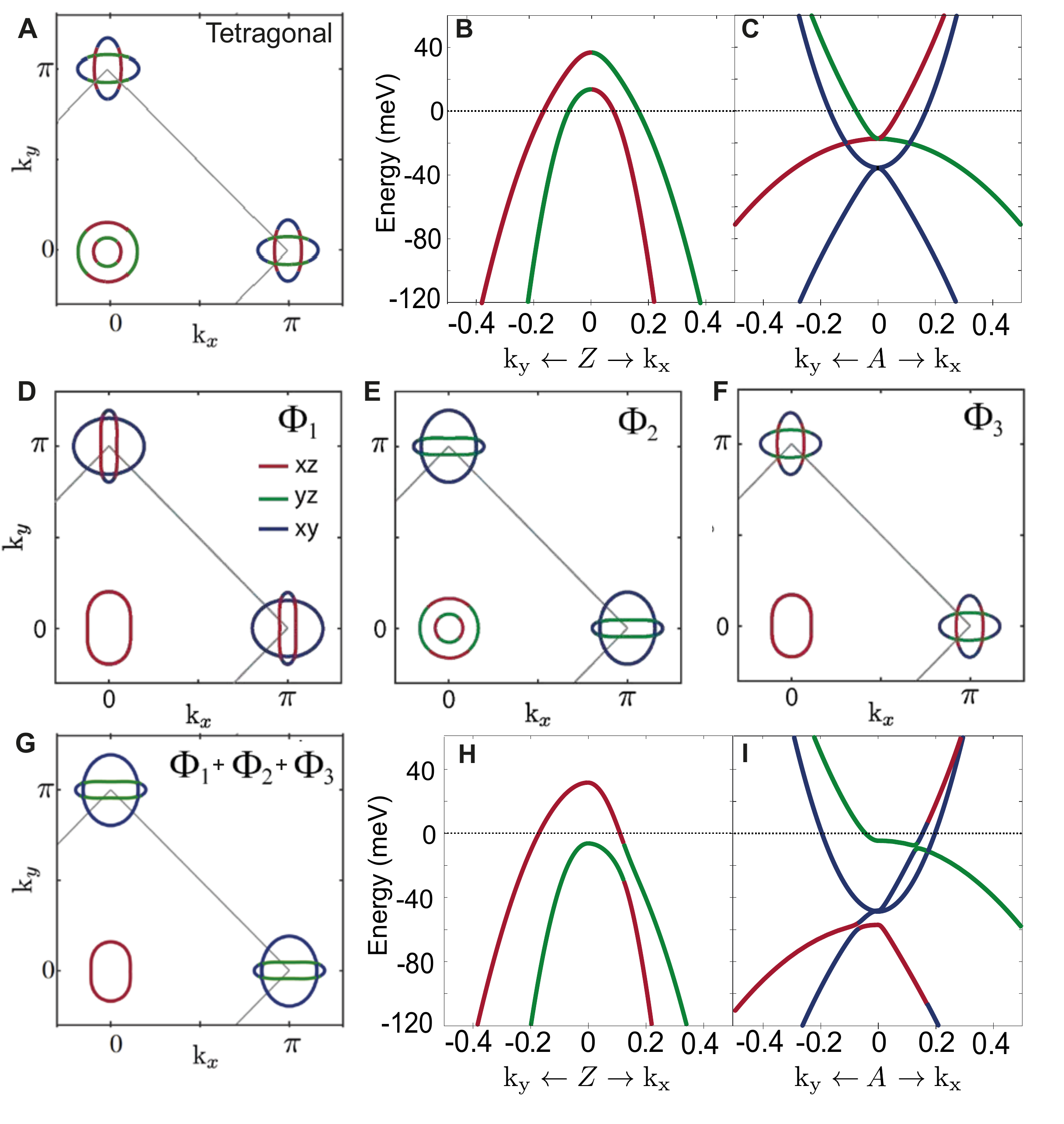}
		\caption{\textbf{Limitations of $d_{xz}/d_{yz}$ nematic ordering and origin of the missing electron pocket problem}. \textbf{(A,B,C)} Fermi surface and band dispersions around the Z and A point, for a tetragonal state model of the electronic structure from ref. \cite{Rhodes2021} in quantitative agreement with ARPES measurements. \textbf{(D,E,F)} The individual effect of the three symmetry breaking $d_{xz/yz}$ nematic order terms on the Fermi surface of the tetragonal state model. \textbf{(D)} Ferro orbital order ($\Phi_1 = 26$~meV) \textbf{(E)} d-wave bond order ($\Phi_2 = -26$~meV) \textbf{(F)} Extended s-wave bond order ($\Phi_3$=15~meV). \textbf{(G,H,I)} Fermi surface and band dispersions around the Z and A point, using a combination of $\Phi_1$, $\Phi_2$ and $\Phi_3$ as is often used in the literature. No matter what linear combination of these order parameters are used, a Fermi surface in agreement with the experimental data can not be produced.}
		\label{fig:summary_dxz_dyz_nematicity}
	\end{figure*}
	
	\section{IV.Theoretical explanations for the missing electron pocket}
	
	As we have discussed, the low energy electronic structure of the tetragonal state of FeSe can be qualitatively understood just from symmetry based arguments regarding the crystal structure and the $d_{xz}$, $d_{yz}$ and $d_{xy}$ orbitals of the Fe atoms. This band structure can be explained both from the framework of tight-binding modelling \cite{Eschrig2009,Graser2009,Mukherjee2015,Rhodes2017} as well as DFT-based simulations. All of this implies that, although a true quantitative explanation describing the renormalisation of the band structure from correlation effects may be missing, our understanding of the single-particle physics is complete.
	
	Within the nematic state, however, this is not the case. Following the previous logic, it would be assumed that the orthorhombic distortion produces a negligible change to the electronic structure, such that two hole pockets and two electron pockets should be present in the nematic state, which as the experimental data has revealed is clearly not the case. It is for this reason that the nematic state is believed to be of electronic or magnetic origin, yet the microscopic details still remain unclear. To address this, there has been a great deal of focus on trying to model how the nematic state evolves the electronic structure of a tetragonal-based model of FeSe, such as that shown in Fig. \ref{fig:summary_dxz_dyz_nematicity}(a-c) originally presented in Ref. \cite{Rhodes2021}. Specifically, theoretical research has attempted to develop a nematic order parameter which
	
	\begin{itemize}
		\item Lowers the symmetry from $C_4$ to $C_2$ whilst still preserving mirror symmetry.
		\item Generates an elliptical hole pocket dominated by $d_{xz}$ orbital weight.
		\item Removes one of the two electron pockets from the Fermi surface.
	\end{itemize}

	Historically, the first attempt to describe such a mechanism assumed that the $C_4$ symmetry breaking was governed by a lifting of the energy degeneracy of the $d_{xz}$ and $d_{yz}$ orbitals \cite{Lee2009}. 
	
	\begin{equation}
	\Phi_1(n_{xz} - n_{yz}),
	\label{Eq:Phi1}
	\end{equation}
	
	where $n_{xz/yz} = c^\dagger_{A,xz/yz}c_{A,xz/yz} + c^\dagger_{B,xz/yz}c_{B,xz/yz}$ is the number operator for the $xz$ or $yz$ orbital respectively on atom $A$ and $B$ in a two atom unit cell model of FeSe, and $\Phi_1$ is a scalar value used to describe the magnitude of the nematic order, which can in principle be fit to experiment.
	
	This term, referred to in the literature as ferro-orbital ordering, is the simplest form of $C_4$ symmetry breaking possible in this system. It acts in a momentum independent fashion to raise the binding energy of the $d_{xz}$ bands and lower the binding energy of the $d_{yz}$ band, similar to a Jahn-teller distortion \cite{Pradhan2021}. In this scenario, the electronic structure would evolve to produce a Fermi surface as shown in Fig \ref{fig:summary_dxz_dyz_nematicity}(d), which despite producing the correct elliptical hole pocket, does not generate the one-electron-pocket Fermi surface determined from experiment.  
	
	Following the train of thought that the phenomenology of the nematic state may be captured by a degeneracy breaking of the $d_{xz}$ and $d_{yz}$ states, it was also noted that there are two additional $B_{1g}$ symmetry breaking terms that can be defined and are equally valid in the nematic state \cite{Fernandes2014b,Mukherjee2015}
	
	\begin{equation}
	\Phi_2(n'_{xz} + n'_{yz})(\cos(k_x) - \cos(k_y))
	\label{Eq:Phi2}
	\end{equation}
	
	\begin{equation}
	\Phi_3(n'_{xz} - n'_{yz})(\cos(k_x) + \cos(k_y))
	\label{Eq:Phi3}
	\end{equation}
	
	Here, $n'_{xz/yz} = c^\dagger_{A,xz/yz}c_{B,xz/yz} + c^\dagger_{B,xz/yz}c_{A,xz/yz}$ describes a hopping from an $xz$ or $yz$ orbital on atom $A$ ($B$) to a $xz$ or $yz$ orbital on atom $B$ ($A$). These two terms, referred to as d-wave nematic bond order ($\Phi_2$) and extended-s wave bond order ($\Phi_3$) respectively, in combination with the ferro orbital order ($\Phi_1$) are the only possible nematic order parameters that can be defined for the $d_{xz}$ and $d_{yz}$ orbitals up to nearest neighbour hopping \cite{Fernandes2014b}, and have been extensively used in previous theoretical descriptions of the nematic state of FeSe \cite{Kreisel2015,Kreisel2018,Sprau2017,Kostin2018,Watson2016,Rhodes2018,Rhodes2019,Yu2018,Yu2021,Liu2018,Cercellier2019,Christensen2020,Kang2018,Jiang2016,Biswas2018,Xing2018,Kang2018b}. The individual consequences of these order parameters are shown in Fig. \ref{fig:summary_dxz_dyz_nematicity}(e) and \ref{fig:summary_dxz_dyz_nematicity}(f). 
	
	However, despite this vast amount of literature assuming these three $d_{xz}/d_{yz}$ nematic order parameters as the starting point for theoretical analysis, there lies one big problem. No matter what values of $\Phi_1$, $\Phi_2$ and $\Phi_3$ are chosen, a Fermi surface consisting of one hole pocket and a single electron pocket can not be produced, at least not starting from a quantitatively accurate ARPES-based model of FeSe in the tetragonal state \cite{Rhodes2021}. The best attempts to describe the ARPES data within this limitation result in a Fermi surface consisting of the correct elliptical hole pocket, a first electron pocket, of correct shape and size, and a second large electron pocket, dominated by $d_{xy}$ orbital character, as shown in Fig \ref{fig:summary_dxz_dyz_nematicity}(g-i).
	
	There is no experimental evidence for this large second electron pocket in the nematic state, and this discrepancy between theory and experiment has posed a major challenge for our theoretical understanding of nematicity. This is the central origin of the missing electron pocket problem. It has now become clear that a theory of nematicity only involving the physics captured in Eq. \eqref{Eq:Phi1}-\eqref{Eq:Phi3}, i.e nematicity derived solely from $d_{xz}$ and $d_{yz}$ orbital ordering, is insufficient to reproduce our experimental measurements, and additional explanations for this discrepancy have had to be developed.

	\subsection{Orbital selective quasiparticle weights}
	The earliest attempt to explain this discrepancy came from attempts to understand local spin fluctuations in tetragonal FeSe, such as those incorporated by DFT + dynamic mean field theory (DMFT). Within this framework it has been shown that the self-consistently determined quasiparticle weight ($Z$) of the $d_{xy}$ orbital was significantly smaller than the quasiparticle weight of the $d_{xz/yz}$ orbitals \cite{Yin2011,Mandal2017,Acharya2021}, approximately half. As the spectral function intensity measured by ARPES is directly proportional to the quasiparticle weight, the contribution of $d_{xy}$ dominated bands should be significantly reduced, compared to the $d_{xz}$ and $d_{yz}$ dominated bands in ARPES measurements. It was thus argued that ARPES measurements may not be able to observe the $d_{xy}$ orbital, and thus would not detect the second $d_{xy}$ dominated electron pocket in the nematic state, e.g in Ref. \cite{Christensen2020}, shown in Fig \ref{fig:theory_missingPocket}(a,b). 
	
	This argument however has not been supported by experimental measurements. Both in the tetragonal and nematic state, bands of $d_{xy}$ orbital character have been identified, particularly around the M point \cite{Coldea2018}. And although it is true that the $d_{xy}$ orbital appears to exhibit a larger effective mass renormalisation than the $d_{xz}$ and $d_{yz}$ orbitals \cite{Watson2015}, this extra renormalisation appears to not be enough to mask $d_{xy}$ spectral weight from ARPES-based measurements. 
	
	A similar, more phenomenological, approach was later employed by Kreisel \textit{et. al.} \cite{Kreisel2017} and popularised by Sprau \textit{et. al. }\cite{Sprau2017}. Here the values of the nematic order parameters ($\Phi_1 - \Phi_3$) were adjusted such that two similar shaped electron pockets were generated (Fig. \ref{fig:theory_missingPocket}(d)), one dominated by $d_{xz}$ orbital weight and one dominated by $d_{yz}$ orbital weight, with the tips retaining significant $d_{xy}$ orbital character. Specifically, starting from an ARPES-based tetragonal model of FeSe \cite{Kreisel2017} values of $\Phi_1 = 9.6$~meV, $\Phi_2=-8.9$~meV and $\Phi_3=0$~meV were used. It was then assumed that the nematic state could exhibit a significant reduction in the $d_{xz}$ quasiparticle weight compared to the $d_{yz}$ weight and, following the same argument as before, hidden from ARPES measurements of the spectral function. This is shown in Fig. \ref{fig:theory_missingPocket}(e). Following this logic, Sprau \textit{et. al.} attempted to determine which values of $Z$ by fitting them to experimental measurements of the angular dependence of the superconducting gap (discussed in Section 5) and the quasiparticle weight values chosen were $Z_{xy}=0.1$ $Z_{xz}=$0.2 and $Z_{yz}=$0.8, which in a later study was refined to $Z_{xy} = 0.073$, $Z_{xz} = 0.16$ and $Z_{yz} = 0.86$ \cite{Kostin2018}. In order to reproduce experimental data, it was also necessary to strongly suppress the quasiparticle weight of the $d_{xy}$ orbital, which as a consequence effectively fully suppressed one of the two electron pockets at the Fermi level. Slave-spin calculations, starting from a DFT-based tight binding model and varying the contributions of $\Phi_1-\Phi_3$ have also been performed and found that similar anisotropic ratios of the quasiparticle weights can be obtained \cite{Yu2018}, as shown in Fig. \ref{fig:theory_missingPocket}(c). A review of the slave-spin approach can be found in Ref. \cite{Yu2021}.
	
	This formalism of "orbital selective quasiparticle weights", i.e suppressing the contribution of electronic states with $d_{xz}$ and $d_{xy}$ orbital character in the nematic state, has received the most traction out of the potential theories of the missing electron pocket of FeSe. It has been claimed to be in agreement with STM and QPI measurements of the electronic structure \cite{Kostin2018}, the superconducting gap properties \cite{Sprau2017}, the spin susceptibility measured by inelastic neutron scattering \cite{Chen2019}, $\mu$SR measurements of spin relaxations \cite{Biswas2018} and thermodynamic based-measurements \cite{Cercellier2019}. A recent review on the topic can be found in Ref. \cite{Kreisel2020}.
	
	In our view, however, the success of this approach is due to accurately generating a Fermi surface of FeSe that has the correct one hole pocket and one electron pocket structure, and not necessarily due to the underlying assumptions behind the ansatz of highly anisotropic quasiparticle weights. Indeed, a change in spectral weight, on the order of magnitude as proposed by this theory, is something that should be directly observable with ARPES based measurements. In the tetragonal state, the quasiparticle weight of the $d_{xz}$ and $d_{yz}$ orbitals must be equivalent by symmetry, and thus, under this assumption, there would be a strong sudden suppression of the $d_{xz}$ dominated bands upon entering the nematic state. This is not what is observed in experimental measurements, bands of $d_{xz}$ dominated weight are detected at all temperatures within the nematic state, with no obvious reduction to the spectral intensity \cite{Rhodes2018,Hashimoto2018,Liu2018,Kushnirenko2018,Fanfarillo2016,Pfau2021}. Additionally, alternate explanations of the STM data and superconducting gap data, that do not rely on the assumption of orbital-selective quasiparticle weights, have been presented \cite{Hanaguri2018,Rhodes2018,Rhodes2019,Benfatto2018}. 
	
	\subsection{E-type order parameters}
\begin{figure*}
		\centering
		\includegraphics[width=\linewidth]{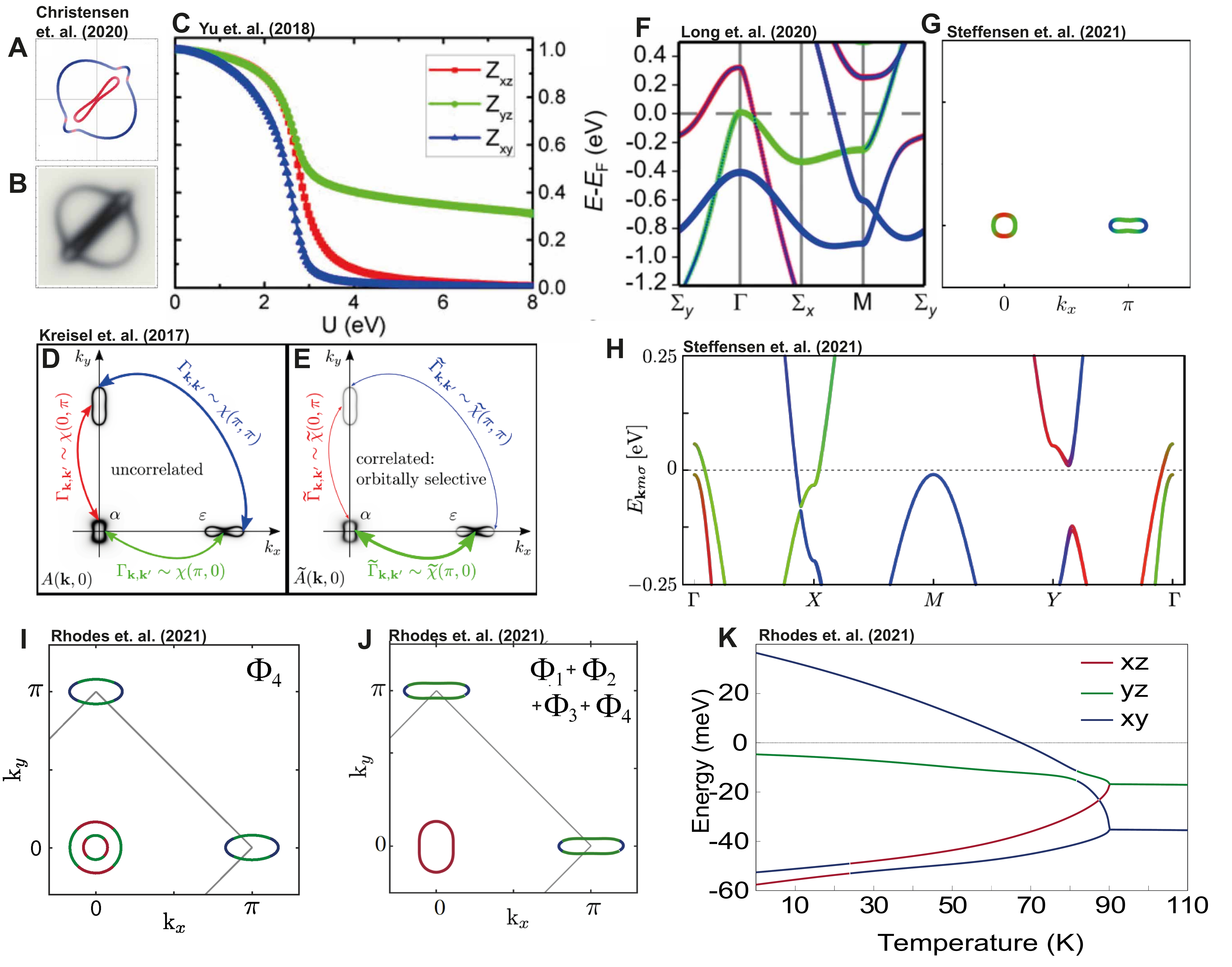}
		\caption{\textbf{Theoretical attempts to resolve the missing electron pocket problem}. \textbf{(A)} Fermi surface of the electron pockets in the nematic state proposed by Christensen \textit{et. al.} \cite{Christensen2020}. The spectral function is argued to have an increased decoherence of $d_{xy}$ weight, which is argued would not be observed by experiment and is simulated in \textbf{(B)}. \textbf{(C)} Slave-spin calculations from Yu \textit{et. al.} \cite{Yu2018}, revealing the possibility of highly anisotropic quasiparticle weights with local Coulomb repulsion. \textbf{(D,E)} Spectral function of the 1-Fe unit cell tight binding model from Kreisel \textit{et. al.} \cite{Kreisel2017}, with and without orbital-selective quasiparticle weights, highlighting the possible suppression of the second electron pocket via incoherent $d_{xz}$ and $d_{xy}$ spectral weight. f) Band dispersion of FeSe obtained from a DFT+U calculation with symmetry preconditioned wavefunctions from Long \textit{et. al.} \cite{Long2020}, highlighting the band hybridisation obtained if an E-type nematic order parameter is considered. \textbf{(G)} Fermi surface of the 1-Fe unit cell model from Steffensen \textit{et. al.} \cite{Steffensen2021} taking into account a self consistently obtained E-type nematic order parameter. \textbf{(H)} Band dispersion from the model used by Steffensen \textit{et. al.}\cite{Steffensen2021} showing a band hybridisation of the $d_{xz}$ (red) and $d_{xy}$ (blue) bands around the Y point (1-Fe unit cell), gapping out the second electron pocket. \textbf{(I)} Fermi surface obtained from the 2-Fe unit cell tetragonal model from Fig. \ref{fig:summary_dxz_dyz_nematicity}(a) assuming dominant $d_{xy}$ nematic ordering, as suggested by Rhodes \textit{et. al.}\cite{Rhodes2021}. \textbf{(J)} Equivalent Fermi surface including all four symmetry allowed nematic order parameters of FeSe and a symmetry allowed Hartree shift. \textbf{(L)} Mean-field temperature evolution of the electronic states at the high symmetry M point, highlighting a Lifshitz transition of the $d{xy}$ band and removal of the second electron pocket as proposed by Rhodes. \textit{et. al.} \cite{Rhodes2021}. \textbf{(A,B)} Reproduced from Ref. \cite{Christensen2020} under the Creative Commons Attribution 4.0 International License. \textbf{(C)} Reproduced from Ref. \cite{Yu2018} with permission from the American Physical Society. \textbf{(D,E)} Reproduced from Ref. \cite{Kreisel2017} with permission from the American Physical Society. \textbf{(F)} Reproduced from Ref. \cite{Long2020} under the Creative Commons Attribution 4.0 International License. }
		\label{fig:theory_missingPocket}
	\end{figure*}

	More recent attempts to explain the missing electron pocket have gone back to studying the single-particle physics of FeSe. A recent DFT + U calculation by Long \textit{et. al.}  \cite{Long2020}, involving symmetry preconditioned wavefunctions, found a lower energy configuration of FeSe by breaking the $E$ symmetry via a multipole nematic order, as shown in Fig \ref{fig:theory_missingPocket}(f). This has been further studied by Yamada \textit{et. al.} \cite{Yamada2021}. This symmetry breaking essentially generates a tetragonal to monolclinic distortion by generating an overlap between a $d_{xy}$ orbital and either $d_{xz}$ or $d_{yz}$ orbital, which as a bi-product also breaks $C_4$ symmetry. This consequentially generates a hybridisation between the $d_{xy}$ dominated electron band and either the $d_{xz}$ or $d_{yz}$ dominated electron band and was shown to produce a one-electron pocket Fermi surface within a certain parameter regime. 
	
	
	A stable E-type nematic order parameter was equally identified, within a tight-binding framework using parameters extracted from LDA-based calculations, by Steffensen \textit{et. al.} \cite{Steffensen2021}. Here it was shown that including nearest-neighbour Coloumbic repulsion, the self consistently calculated mean-field nematic order parameter that had the largest magnitude was an inter-orbital term hybridising the $d_{xz}$ and $d_{xy}$ orbitals (or $d_{yz}$ and $d_{xy}$). This order parameter was equally able to generate a one-electron pocket Fermi surface, via a similar hybridisation mechanism as the DFT-based calculation as shown in Fig. \ref{fig:theory_missingPocket}(g,h). 
	
	This appears to suggest that long-range Coulomb repulsion can stabilise a $C_4$ symmetry breaking ground state in FeSe. However, in this scenario, the $E$-type order parameter would also reduce the crystal symmetry of FeSe from tetragonal to monoclinic. Currently, the experimental evidence suggesting a tetragonal to monoclinic structural distortion in FeSe is lacking. However, upon $>85$\%  Te doping of the Se sites, a tetragonal to monoclinic transition has been realised \cite{Rodriquez2011}. This could hint that the known monoclinic structure of FeTe is actually stabilised by electron interactions \cite{Trainer2019}, however whether this mechanism can describe the physics of FeSe will require further experimental investigation.
	
	\subsection{non-local $d_{xy}$ nematic order parameter}
	When considering the relevant $d_{xz}$, $d_{yz}$ and $d_{xy}$ orbitals of tetragonal FeSe within a tight binding framework, there are only four order parameters that can be defined which break the $B_{1g}$ rotational symmetry of the material within a single unit cell. The first three, described in Eq. \eqref{Eq:Phi1} - \eqref{Eq:Phi3}, involve breaking the degeneracy of the $d_{xz}$ and $d{yz}$ orbitals. However, a fourth equally valid order parameter involving the $d{xy}$ orbital can also be defined as, 
	
	\begin{equation}
	\Phi_4(n'_{xy})(\cos(k_x) - \cos(k_y)).
	\label{Eq:Phi4}
	\end{equation}
	
	This term acts as a hopping anisotropy for nearest neighbour $d_{xy}$ orbitals, in a similar manner as \eqref{Eq:Phi2} for the $d_{xz}$ and $d_{yz}$ orbitals. It was initially defined by Fernandes \textit{et. al.} \cite{Fernandes2014b}, however in subsequent works it was assumed that this $d_{xy}$ nematic term would be much smaller, or negligible, compared to Eq. \eqref{Eq:Phi1} - \eqref{Eq:Phi3} \cite{Fernandes2014b}. Renormalisation group theory \cite{Chubukov2016,Xing2017,Classen2018} additionally found, that whilst Eq. \eqref{Eq:Phi4} was symmetry allowed, nematic symmetry breaking only had stable RG flow in either the $d_{xz}/d_{yz}$ channel or the $d_{xy}$ channel, implying that finite $\Phi_1-\Phi_3$ and $\Phi_4$ would not both be present simultaneously \cite{Chubukov2016}. However a weakly unstable trajectory suggested that this may not be the case \cite{Xing2017}. 
	
	In Ref. \cite{Rhodes2021} Rhodes \textit{et. al.} looked at the qualitative effect $\Phi_4$ has on the electronic structure. They showed that a one-electron pocket Fermi surface could be generated from a ARPES-based tight binding model of FeSe solely using the $\Phi_4$ term, as shown in Fig. \ref{fig:theory_missingPocket}(i). It was shown that $\Phi_4$ has the effect of breaking the degeneracy of the $d_{xy}$ vHs ($vH_2$ in Fig. \ref{fig:Tetragonal}(a)), which if made large enough ($\sim$50~meV) would induce a Lifshitz transition of the $d_{xy}$ band, and thus reduce the total number of electron pockets crossing the Fermi level to one. This is shown in Fig. \ref{fig:theory_missingPocket}(k). in combination with $\Phi_1$ to $\Phi_3$, the addition of $\Phi_4$ made it possible to generate a Fermi surface in agreement with the ARPES measurements, as shown in Fig. \ref{fig:theory_missingPocket}(j). A recent study has also found this order to be consistent with specific heat measurements \cite{Islam2021}. 
	
	However, in order to get quantitative agreement with the Fermi surface and low-energy electronic structure using Eq. \eqref{Eq:Phi1} - \eqref{Eq:Phi4}, it was observed that the the splitting of the $d_{xy}$ van-Hove singularity must be asymmetric. Specifically, ARPES measurements as a function of temperature find that the lower part of the $d_{xy}$ vHs around the M point remains approximately at the same energy \cite{Watson2016,Fanfarillo2016,Yi2019}. This is not captured by the $\Phi_4$ term that assumes a symmetric splitting of the bands. To account for this, Rhodes \textit{et. al.} \cite{Rhodes2021} included a $d_{xy}$-specific Hartree shift, a constant energy shift of the $d_{xy}$ orbital at the $M$ point, that although allowed by symmetry, did not have an obvious origin. Additionally, in order to generate a Lifshitz transition of the electron pocket, and obtain quantitative agreement with experimental data as a function of temperature both $d_{xy}$ terms, $\Phi_4$ and the Hartree shift, had to be significantly larger than the the $d_{xz}/d_{yz}$ terms ($\Phi_1 - \Phi_3$). Specifically, in order to reproduce the ARPES measurements $\Phi_1 + \Phi_3=15$~meV, $\Phi_1 + \Phi_2 = -26$~meV and $\Phi_4 = \Delta_{Hartree}$ = 45 meV \cite{Rhodes2021}. It is also worth noting that the mean-field analysis by Steffensen \textit{et. al.} \cite{Steffensen2021} equally found that the $\Phi_4$ nematic order parameter should be finite, but found it to be of approximately equal magnitude as $\Phi_1$-$\Phi_3$ rather than twice as large, as suggested by Rhodes \textit{et. al.} \cite{Rhodes2021}.

	\subsection{Importance of the $d_{xy}$ orbital in theories of nematicity}
	Each theory proposed to describe the low-energy electronic structure of the nematic state of FeSe has it's relative strengths and weaknesses. Nevertheless a common theme in these different attempts has begun to emerge. In all methods used to theoretically remove an electron pocket from the Fermi level, it has been necessary to modify the $d_{xy}$ orbital in some way. Whether that's suppressing its contribution via quasiparticle weights, gapping out the $d_{xy}$ band via hybridisation, or rigidly shifting the $d_{xy}$ band above the Fermi level. What we can gleam from this analysis therefore, is that we should view the nematic state in a new light, not originating from a specific orbital ordering mechanism of $d_{xz}$ and $d_{yz}$ states, but rather as a symmetry breaking phenomena which couples to every orbital at the Fermi level. Further theoretical investigations are required in order to elucidate the origin of the nematic state. The importance of the $d_{xy}$ orbital has also been recently noted from NMR measurements \cite{Li2020NMR} and angular dependent magnetoresistance \cite{Liu2021}.
	
	\section{V. Consequences for the superconducting gap symmetry}
	One of the most striking properties of FeSe is it's highly tuneable superconducting transition temperature, ranging from 8~K in bulk crystals \cite{Margadonna2008}, 36.7~K under pressure \cite{Medvedev2009}, and up to 65~K when a monolayer is placed on SrTiO$_3$ \cite{Huang2017}, and hence the nature of superconductivity in FeSe is an important question that attracted a lot of attention.
	
	From an experimental point of view, the momentum dependence of the superconducting gap of bulk FeSe, has been extensively determined from ARPES \cite{Xu2016,Hashimoto2018,Liu2018,Kushnirenko2018,Rhodes2018}, STM \cite{Jiao2017, Sprau2017,Hanaguri2018}, Spectific heat \cite{Hardy2019,Sun2017} and muSR measurements \cite{Biswas2018}, with surprisingly near unanimous agreement as to the angular dependence of the gap structure around both the hole and electron pocket. This achievement provided the perfect opportunity to directly compare theories of superconductivity with experimental measurements.
	
	In this section, we will review the experimental data of the momentum dependence of the superconducting gap, particularly from ARPES measurements, and discuss the theoretical consequence the updated Fermi surface topology has on the theoretical understanding of superconductivity in FeSe.

	\begin{figure*}
		\centering
		\includegraphics[width=0.7\linewidth]{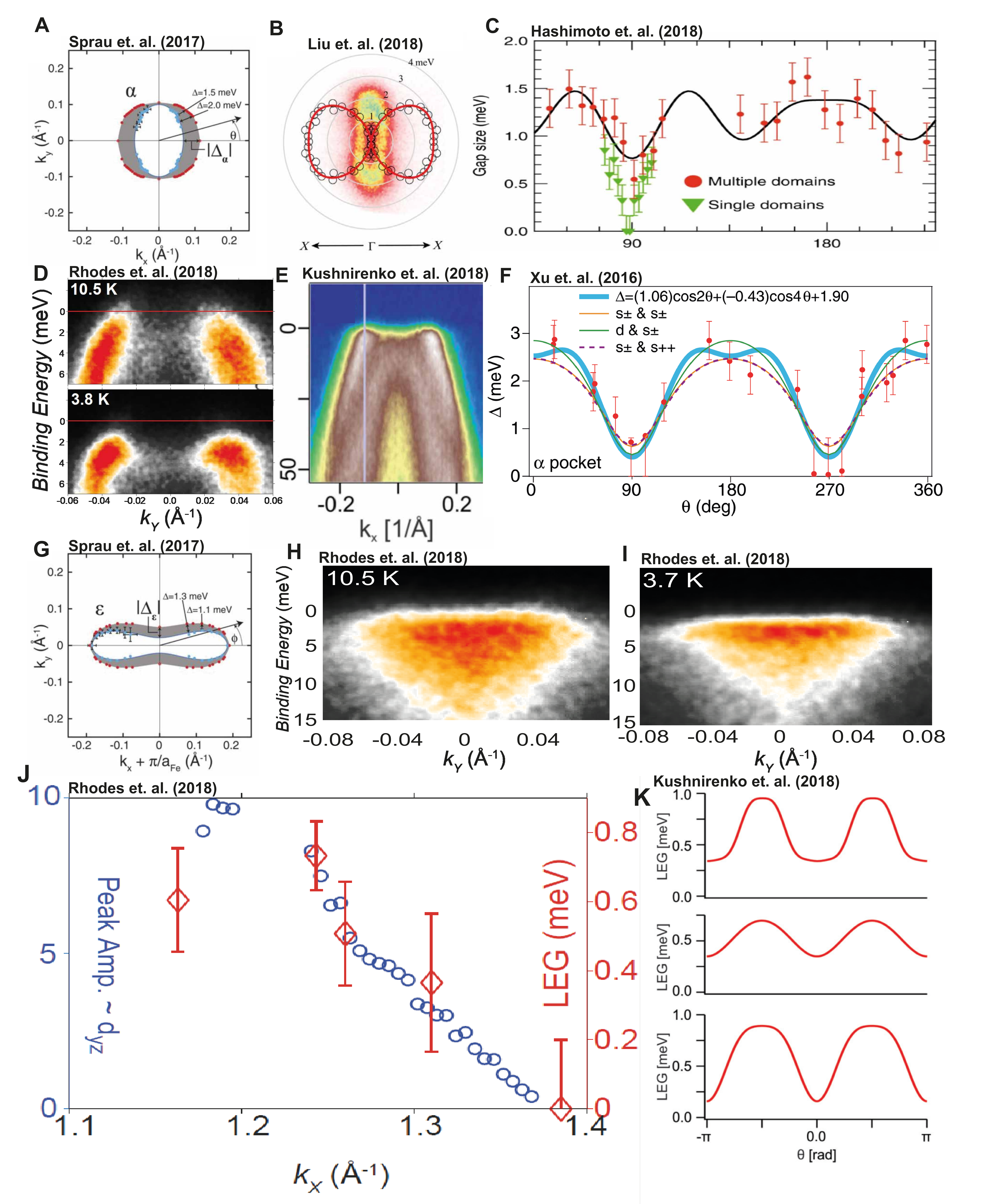}
		\caption{\textbf{Experimental measurements of the superconducting gap of FeSe.} \textbf{(A)} Angular dependence of the gap around the hole pocket as extracted from BQPI measurements from Sprau \textit{et. al.} \cite{Sprau2017}. \textbf{(B)} Angular dependence of the gap around the hole pocket as extracted from Laser ARPES measurements from Liu et. al. \cite{Liu2018}. \textbf{(C)} Angular dependence of the gap around the hole pocket from Hashimoto \textit{et. al.} \cite{Hashimoto2018}. Red dots are data from a twinned sample, whereas green data was measured on an accidentally strained sample. \textbf{(D)} Band dispersion of the $k_z=0$ hole band ($h\nu = 37$~eV) from Rhodes \textit{et. al.} \cite{Rhodes2018} taken along the direction where the hole band gap is largest, above and below $T_c$. \textbf{(E)} Equivalent band dispersion of the $k_z =\pi$ hole band ($h\nu = 21$~eV) below $T_c$ from Kushnirenko \textit{et. al.} \cite{Kushnirenko2018}. \textbf{(F)} Angular dependence of the hole band of FeSe$_{0.93}$S$_{0.07}$ from Xu \textit{et. al.} \cite{Xu2016}, showing equivalent momentum dependence as the undoped sample. \textbf{(G)} Angular dependence of the gap around the electron pocket as extracted from BQPI measurements from Sprau \textit{et. al.} \cite{Sprau2017}. \textbf{(H,I)} Band dispersion along the minor length of the electron pocket above and below $T_c$, along the high symmetry axis from Rhodes \textit{et. al.}\cite{Rhodes2018}. \textbf{(J)} Comparison of the gap magnitude (Leading Edge Gap - LEG) and the intensity of the spectral weight from Linear Vertical polarised light as a function of $k_x$, which is directly correlated to the amplitude of $d_{yz}$ orbital weight. The gap is observed to decrease with decreasing $d_{yz}$ weight. Taken from Rhodes \textit{et. al.} \cite{Rhodes2018}. \textbf{(K)} Sketch of the angular dependence of the electron pocket at $k_z=0$ (bottom) $k_z=\frac{\pi}{2}$ (middle) and $k_z=\pi$ (top) from Kushirenko \textit{et. al.} \cite{Kushnirenko2018}. \textbf{(A,G)} Reproduced from Ref. \cite{Sprau2017} with permission from the AAAS. \textbf{(B)} Reproduced from Ref \cite{Liu2018} under the Creative Commons Attribution 4.0 International License. \textbf{(C)} b) Reproduced from Ref \cite{Hashimoto2018} under the Creative Commons Attribution 4.0 International License. \textbf{(F)} Reproduced from Ref. \cite{Xu2016} with permission from the American Physical Society. \textbf{(E,K)} Reproduced from Ref. \cite{Kushnirenko2018} with permission from the American Physical Society.}
		\label{fig:SCgap_Experiment}
	\end{figure*}

	\subsection{Experimental measurements of the superconducting gap}
	The key findings from the multiple ARPES and QPI measurements are presented in Fig. \ref{fig:SCgap_Experiment}. For the gap situated on the hole pocket, a highly two-fold anisotropic momentum dependence of the gap was measured, as shown from QPI analysis by Sprau \textit{et. al.} in Fig. \ref{fig:SCgap_Experiment}(a). The angular dependence of the hole pocket using ARPES was first reported in 2016 by Xu \textit{et. al.} \cite{Xu2016} on 7\% sulphur doped FeSe measured at 6.3~K, as shown in Fig. \ref{fig:SCgap_Experiment}(f). It was found that the angular dependence at both  $k_z=0$ (using a photon energy of $h\nu=37$~eV) and $k_z=\pi$ ($h\nu=21$~eV) produced near identical momentum distributions. This sulphur doped system has a very similar electronic structure to undoped FeSe, albeit with a slightly reduced nematic transition temperature \cite{Matsuura2017} and slightly higher superconducting transition temperature (9.8~K \cite{Xu2016}). Later, in 2018,  Liu \textit{et. al.} \cite{Liu2018} and Hashimoto \textit{et. al.} \cite{Hashimoto2018} used laser ARPES, with $h\nu=6.994$~eV, on FeSe at 1.6~K and observed the same highly anistropic angular dependence of the gap at the hole pocket, as shown in Fig. \ref{fig:SCgap_Experiment}(b,c). By using such a low photon energy and temperature these authors ensured the greatest possible energy resolution for resolving the gap of the hole pocket. However the trade-off here is that information about states with large angular momentum, e.g the electron pockets, as well as the $k_z$-dependence of the hole pocket, can not be obtained. Kushnirenko \textit{et. al.} \cite{Kushnirenko2018}, as well as Rhodes \textit{et. al.} \cite{Rhodes2018}, were able to resolve the three dimensional gap structure of both the hole and electron pockets using synchrotron radiation, as shown in Fig. \ref{fig:SCgap_Experiment}(d,e). In these manuscripts, it was again confirmed that the gap structure of the hole pocket at both $k_z=0$ and $k_z=\pi$ exhibited the same highly anisotropic two-fold angular dependence of the gap as determined in the Sulphur doped sample of Xu. \textit{et. al.} \cite{Xu2016}. Kushnirenko \textit{et. al.} claimed that the superconducting gap that was larger at $k_z = \pi$ and smaller at $k_z = 0$, however Rhodes \textit{et. al.} suggested the opposite: the gap was observed to be larger at $k_z =0$ and smaller at $k_z = \pi$. We note that in order to reach the $k_z=0$ hole pocket, a higher photon energy of 37~eV is required, which makes the measurement of the gap at the $\Gamma$ point exceedingly challenging, and the measurements are at the cutting edge of what is currently achievable by synchrotron-based ARPES measurements. 
	
	Hashimoto \textit{et. al.} additionally claimed that the gap structure produced a  different behaviour with and without the presence of uniaxial strain. Without strain, they observed a $\cos(8\theta)$ behaviour \cite{Hashimoto2018}, which when accidentally detwinned via uniaxial strain, yielded a gap structure that is consistent with the other measurements. So far this $\cos(8\theta)$ dependence of the gap has not been reproduced.
	
	As for the electron pocket, the angular dependence of the gap from QPI measurements is presented in Fig. \ref{fig:SCgap_Experiment}(g). Revealing a particularly constant gap magnitude across the length of the ellipse, which quickly decays towards zero at the tips of the pocket. This is where the orbital character of the pocket transforms from predominantly $d_{yz}$ weight to $d_{xy}$ weight. ARPES measurements by Kushnirenko \textit{et. al} \cite{Kushnirenko2018}, and Rhodes \textit{et. al.} \cite{Rhodes2018}, were also able to resolve the angular dependence of the superconducting gap at the electron pocket. ARPES measurements along the minor length of the electron pocket, above and below $T_c$, are shown in Fig. \ref{fig:SCgap_Experiment}(h,i). Thanks to the orbital sensitivity of ARPES-based measurements, Rhodes \textit{et. al.}  found a direct correlation between the intensity of $d_{yz}$ orbital weight and the size of the superconducting gap, establishing a direct link between orbital character and gap magnitude. Kushnirenko \textit{et. al.} \cite{Kushnirenko2018} also observed that the rate that the gap decreased as a function of momentum was slightly different for intermediate $k_z$ values (Fig. \ref{fig:SCgap_Experiment}(m). 
	
	This extremely aniostropic gap structure for both the hole and electron pocket raises a question as to whether FeSe is a nodal or nodeless superconductor, which could have a profound effect on our understanding of the gap symmetry in this system. For example, neglecting the electron pocket, it was argued by Hashimoto \textit{et. al.} that a nodal gap structure of the hole pocket would be consistent with p-wave superconductivity \cite{Hashimoto2018} (This is not consistent once the gap structure of the electron pocket is additionally taken into account). It is not possible to clearly distinguish between a nodal gap or a very small gap in ARPES measurements, due to the limitations of energy resolution arising from thermal broadening and the choice of photon energy. Alternate techniques, such as STM and specific heat measurements, do have sufficient energy and thermal resolution to tackle this issue, but here STM measurements of the density of states by Sprau \textit{et. al.} \cite{Sprau2017} suggest a fully gapped, nodeless, superconducting ground state, whereas specific heat measurements have argued that the measured data is consistent with a nodal superconducting gap \cite{Hardy2019}. It is still unclear whether FeSe exhibits nodes or very small superconducting gaps, however as we will discuss below, theoretical arguments appear to suggest that if any nodes do exist, they would be accidental in nature.

	\begin{figure*}
		\centering
		\includegraphics[width=\linewidth]{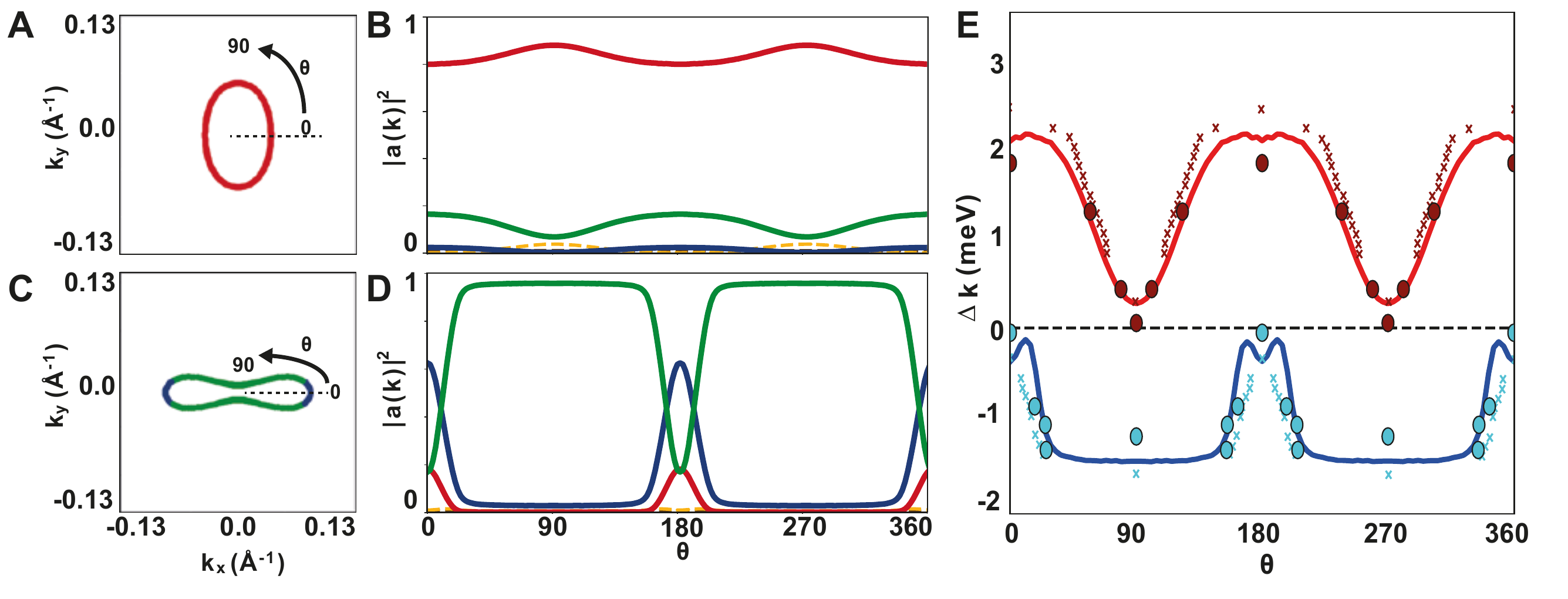}    \caption{\textbf{Theoretical simulation of the momentum dependence of the superconducting gap} from Rhodes \textit{et. al.} \cite{Rhodes2018}. Here, a Fermi surface consisting of one hole pocket a single electron pocket were considered and a spin fluctuation pairing mechanism was assumed. \textbf{(A,B)} Fermi surface of the hole pocket and angular dependence of the orbital content of the hole pocket. \textbf{(C,D)} Fermi surface of the one electron pocket and angular dependence of the orbital content of the electron pocket. The colour labels are red - $d_{xz}$, green $d_{yz}$ and blue $d_{xy}$. \textbf{(E)} Simulated angular dependence of the superconducting gap for the hole pocket (red) and electron pocket (blue), revealing a direct correlation with the $d_{yz}$ weight shown in \textbf{(B)} and \textbf{(D)}. The crosses and dots are experimental data extracted from STM \cite{Sprau2017} and ARPES \cite{Rhodes2018} measurements respectively.}
		\label{fig:SCgap_Theory}
	\end{figure*}
	
	\subsection{Theoretical understanding of the superconducting gap}
	
	The most striking result from the experimentally determined gap structure of FeSe, is the clear realisation that the size of the superconducting gap at the Fermi level is correlated with the magnitude of $d_{yz}$ orbital weight. This tells us that the superconducting pairing mechanism is sensitive to orbital character, and is evidence for superconductivity mediated by Coulomb interactions, such as via a spin-fluctuation mechanism of superconductivity. 
	
	Although the idea that spin fluctuations govern the Cooper pairing in the iron-based superconductors, was originally proposed back when superconductivity in these materials were first discovered \cite{Mazin2008}, the evidence for this has often been inferred from gap symmetry arguments, such as a sign-changing $s^\pm$ order parameter \cite{Sprau2017}, or from the general argument that FeSe is close to a magnetic instability. FeSe, being  such a clean system, has enabled a direct comparison between theoretical simulations and experimental data.

	Indeed many theoretical simulations of the angular dependence of the superconducting gap in FeSe have been performed \cite{Sprau2017,Kreisel2017,Rhodes2018,Kang2018,Yu2018,Benfatto2018,Steffensen2021,Rhodes2021}. However, as the formation of Cooper pairs are directly sensitive to the states at the Fermi level, the starting model used to describe FeSe is very important. Numerical simulations have shown that models of FeSe which do not account for the missing electron pocket of the nematic state, i.e a model Fermi surface which describes two electron pockets around the M point, can not reproduce the experimentally observed gap structure \cite{Sprau2017,Rhodes2018,Benfatto2018,Kang2018}. 
	
	Initially, this was a confusing result, but with hindsight it is not that surprising. The presence of an extra electron pocket in the simulations would naturally influence the superconducting pairing. Due to the local nature of Coulomb repulsion, the pairing between electrons in real space will be largest for electrons located on the same atom in the same orbital. It follows from this argument, that the pairing of electrons in momentum space would be favoured if a spin scattering process occurs which couples electronic states of the same orbital character. In the nematic state of FeSe, spin-fluctuations are strongest when connecting the hole and electron pocket \cite{Wang2016,Wang2016b, Chen2019}. In a one-electron pocket scenario, the only common orbital content between the two pockets are the $d_{yz}$ orbital weight, as shown in Fig. \ref{fig:SCgap_Theory}(a-d), and thus this would dominate the superconducting gap magnitude. This would not be the case in a two-electron pocket scenario, where scattering with $d_{xz}$ electrons between the hole and electron pocket would also contribute. 
	
	It has now been shown that irregardless of the theoretical mechanism employed to remove this second electron pocket from the superconducting calculation, whether that's orbital selective quasiparticle weights \cite{Sprau2017,Kreisel2017,Yu2018}, orbital selective spin fluctuations \cite{Benfatto2018},  E-type nematic ordering \cite{Steffensen2021}, a non-local $d_{xy}$ nematic order parameter \cite{Rhodes2021} or simply ignoring it from simulations of the superconducting pairing outright \cite{Rhodes2018} (as shown in Fig. \ref{fig:SCgap_Theory}), the correct momentum dependence of the gap structure can be naturally captured assuming weak-coupling spin fluctuation mediated pairing.   
	
	This is a remarkable finding, not only does it further support the theory of spin-fluctuation mediated superconductivity in the iron-based superconductors, but it provides another independent piece of evidence for a single electron pocket around the M point in the nematic state of FeSe. This result highlights the incredible importance of correctly accounting for the missing electron pocket in the nematic state, as without it we can not begin to understand the superconducting properties of this material. 
	
	\section{Discussion}
	This review has been wholly focused on what at first glance might appear to be an esoteric point of discussion, namely, the characterisation and modelling of the Fermi surface of FeSe in the nematic state. However, we propose that after the hundreds of papers and many years of debate and controversy on the subject, that there are very important conclusions to be drawn, which have wider implications for our understanding of both nematic ordering and superconductivity across the wider family of Fe-based superconductors. 
	
	The first conclusions surround nematic ordering, where the results establish
	\begin{itemize}
		\item That nematic ordering affects all bands at the Fermi level, with the $d_{xy}$ derived bands playing as significant a role as the $d_{xz}$ and $d_{yz}$ derived bands.
		\item That nematic order manifests in the band structure through a combination of all allowed symmetry-breaking terms, primarily anisotropic hopping terms, and cannot be exclusively treated by on-site orbital ordering. 
		\item That nematic ordering does not cause a minor perturbation of the electronic structure, but can lift an entire electron pocket away from the Fermi level. 
	\end{itemize}
	
	We believe that these conclusions should be widely applicable across other Fe-based superconductors. While these conclusions do not yet constitute a self-consistent microscopic mechanism of nematic order, they do present strong constraints to any proposed microscopic models. 
	
	The second set of conclusions relate to the superconductivity:     
	
	\begin{itemize}
		\item The superconducting gap of FeSe is remarkably anisotropic.
		\item The fact that the gap follows the $d_{yz}$ orbital character is strong experimental evidence that the pairing mechanism is sensitive to local orbital degrees of freedom, i.e. for spin-fluctuation pairing.
		\item The superconducting gap of FeSe can be naturally reproduced by spin-fluctuation calculations assuming only one electron pocket at the Fermi level.
	\end{itemize}
	
	There has long been a consensus that the superconductivity in the Fe-based systems is mediated by spin-fluctuation pairing, but we argue that FeSe provides some of the most direct experimental support for this. As long as one starts with the one-electron pocket Fermi surface, the further details of the calculation are not critical, because in this scenario the only orbital component which is present on both the hole and electron pockets is the $d_{yz}$ character, and so this channel dominates the structure of the gap. The success of this result justifies the use of similar spin-fluctuation pairing calculations on other Fe-based superconductors, although we emphasize the importance of starting with an experimentally accurate Fermi surface. 
	
	Importantly this insight has only been unlocked once we understand that the true Fermi surface of FeSe consists of one hole pocket and a single electron pocket, rather than one hole pocket and two electron pockets as was initially believed. However, despite us emphasizing how the one electron pocket scenario is key to the understanding of the unusual properties of FeSe, we believe it is still an open question as to what mechanism really drives this modification of the electronic structure. The models of describing the electronic structure in the nematic state have grown more accurate and more sophisticated, yet there is a lack of intuition about what is the real driving force for the evolution of the electronic structure that we observe. In our opinion it remains a delicate and important open question, but solving it in the case of FeSe could unlock a wider understanding of nematicity in the iron-based superconductors.

	Additionally, whilst the experimental challenge imposed by measuring the electronic structure of orthorhombic crystals has always been present, the focus on an answer to the origin of nematicity in FeSe has particularly emphasised the continued development of detwinning methods in ARPES \cite{Shimojima2014,Watson2017,Yi2019,Huh2020,Pfau2019,Pfau2021,Cai2020,Cai2020b, Watson2018,Watson2019}, as well as showcasing the potential of NanoARPES for strongly correlated materials with local domain structures \cite{Watson2019,Rhodes2021}.

	\section{Outlook and Conclusion}
	
	With an outlook to the future, there are still multiple open questions regarding the missing electron pocket problem, nematicity and superconductivity in FeSe. Firstly, can we experimentally identify the exact conditions when one of the electron pockets in the tetragaonal state appears or disappears from the Fermi level? So far, this has remained slightly ambiguous, with some experiments claiming a gradual disappearance of the electron pocket \cite{Cai2020} and others claiming a Lifshitz transition around 70~K \cite{Yi2019,Huh2020,Rhodes2021}. 
	
	Another open question is how the missing electron pocket scenario can be reconciled with the QPI measurements as a function of sulphur doping \cite{Hanaguri2018} or Tellurium doping \cite{Shibauchi2020}, each providing an isoelectronic tuning parameter to control the evolution of the Fermi surface. The systematic evolution of the Fermi surface has been studied by Quantum Oscillations \cite{Coldea2019}, however due to the tiny size of the Fermi energy in this system, the unambiguous assignment of the quantum oscillation frequencies is challenging \cite{Rhodes2021}. Equally, twinned ARPES measurements on sulphur doped FeSe have already been performed \cite{Watson2015b,Reiss2017}, as well as several studies on detwinned crystals for 9\% sulphur doping \cite{Cai2020,Cai2020b}. So far it is unclear when the missing electron pocket reappears, and so further measurements of detwinned FeSe$_{1-x}$S$_{x}$ are desirable, although by 18\% the system is tetragonal once more and two electron pockets are certainly observed \cite{Reiss2017}. 
	
	Finally, an important avenue of research is how does the momentum-dependence of the superconducting gap change as nematicity is supressed, e.g as a function of sulphur doping. The momentum dependence of the superconducting gap for undoped FeSe has now been extensively characterised, and theoretical predictions of how the gap should evolve as nematicity is suppressed have been proposed \cite{Rhodes2021}. This much needed experimental data would again place important constraints on our theories of nematicity and superconductivity in these systems. 
	
	As the study of the Fe-based superconductors has matured since they exploded onto the scene in 2008, the emphasis has shifted from basic characterisation of a wide variety of superconducting families, to detailed examination of particular cases. FeSe has been the subject of particularly focused attention, and the effort has been worthwhile, with two remarkable results emerging: the one electron pocket Fermi surface, and the highly anisotropic superconducting gap structure. We have argued that these two, taken together, provide strong evidence for spin-fluctuation pairing in FeSe, which is presumably applicable to the wider family of Fe-based superconductors. However, the extent to which the one electron pocket phenomenology may be applicable to the nematic phase of other material systems is a large open question; as well as FeSe$_{1-x}$S$_{x}$ and FeSe$_{1-x}$Te$_{x}$, we propose NaFeAs \cite{Watson2018} as a candidate worthy of re-examination. Thus as this review of FeSe concludes, we propose it is time to take the experimental and theoretical tools developed for case of FeSe, and apply them with renewed vigour to the wider field of Fe-based superconductors.

	\section*{Conflict of Interest Statement}
	The authors declare that the research was conducted in the absence of any commercial or financial relationships that could be construed as a potential conflict of interest.
	
	\section*{Author Contributions}
	All authors contributed to the development of this review article. 
	
	\section*{Funding}
	LCR acknowledges funding from the royal commission for the exhibition for the 1851. 
	

\begin{thebibliography}{120}%
		\makeatletter
		\providecommand \@ifxundefined [1]{%
			\@ifx{#1\undefined}
		}%
		\providecommand \@ifnum [1]{%
			\ifnum #1\expandafter \@firstoftwo
			\else \expandafter \@secondoftwo
			\fi
		}%
		\providecommand \@ifx [1]{%
			\ifx #1\expandafter \@firstoftwo
			\else \expandafter \@secondoftwo
			\fi
		}%
		\providecommand \natexlab [1]{#1}%
		\providecommand \enquote  [1]{``#1''}%
		\providecommand \bibnamefont  [1]{#1}%
		\providecommand \bibfnamefont [1]{#1}%
		\providecommand \citenamefont [1]{#1}%
		\providecommand \href@noop [0]{\@secondoftwo}%
		\providecommand \href [0]{\begingroup \@sanitize@url \@href}%
		\providecommand \@href[1]{\@@startlink{#1}\@@href}%
		\providecommand \@@href[1]{\endgroup#1\@@endlink}%
		\providecommand \@sanitize@url [0]{\catcode `\\12\catcode `\$12\catcode
			`\&12\catcode `\#12\catcode `\^12\catcode `\_12\catcode `\%12\relax}%
		\providecommand \@@startlink[1]{}%
		\providecommand \@@endlink[0]{}%
		\providecommand \url  [0]{\begingroup\@sanitize@url \@url }%
		\providecommand \@url [1]{\endgroup\@href {#1}{\urlprefix }}%
		\providecommand \urlprefix  [0]{URL }%
		\providecommand \Eprint [0]{\href }%
		\providecommand \doibase [0]{http://dx.doi.org/}%
		\providecommand \selectlanguage [0]{\@gobble}%
		\providecommand \bibinfo  [0]{\@secondoftwo}%
		\providecommand \bibfield  [0]{\@secondoftwo}%
		\providecommand \translation [1]{[#1]}%
		\providecommand \BibitemOpen [0]{}%
		\providecommand \bibitemStop [0]{}%
		\providecommand \bibitemNoStop [0]{.\EOS\space}%
		\providecommand \EOS [0]{\spacefactor3000\relax}%
		\providecommand \BibitemShut  [1]{\csname bibitem#1\endcsname}%
		\let\auto@bib@innerbib\@empty
		\bibitem [{\citenamefont {Kreisel}\ \emph {et~al.}(2020)\citenamefont
			{Kreisel}, \citenamefont {Hirschfeld},\ and\ \citenamefont
			{Andersen}}]{Kreisel2020}%
		\BibitemOpen
		\bibfield  {author} {\bibinfo {author} {\bibfnamefont {A.}~\bibnamefont
				{Kreisel}}, \bibinfo {author} {\bibfnamefont {P.~J.}\ \bibnamefont
				{Hirschfeld}}, \ and\ \bibinfo {author} {\bibfnamefont {B.~M.}\ \bibnamefont
				{Andersen}},\ }\bibfield  {title} {\enquote {\bibinfo {title} {{On the
						Remarkable Superconductivity of FeSe and Its Close Cousins}},}\ }\href
		{\doibase 10.3390/sym12091402} {\bibfield  {journal} {\bibinfo  {journal}
				{Symmetry}\ }\textbf {\bibinfo {volume} {12}},\ \bibinfo {pages} {1402}
			(\bibinfo {year} {2020})}\BibitemShut {NoStop}%
		\bibitem [{\citenamefont {Shibauchi}\ \emph {et~al.}(2020)\citenamefont
			{Shibauchi}, \citenamefont {Hanaguri},\ and\ \citenamefont
			{Matsuda}}]{Shibauchi2020}%
		\BibitemOpen
		\bibfield  {author} {\bibinfo {author} {\bibfnamefont {T.}~\bibnamefont
				{Shibauchi}}, \bibinfo {author} {\bibfnamefont {T.}~\bibnamefont {Hanaguri}},
			\ and\ \bibinfo {author} {\bibfnamefont {Y.}~\bibnamefont {Matsuda}},\
		}\bibfield  {title} {\enquote {\bibinfo {title} {{Exotic Superconducting
						States in FeSe-based Materials}},}\ }\href {\doibase 10.7566/JPSJ.89.102002}
		{\bibfield  {journal} {\bibinfo  {journal} {Journal of the Physical Society
					of Japan}\ }\textbf {\bibinfo {volume} {89}},\ \bibinfo {pages} {102002}
			(\bibinfo {year} {2020})}\BibitemShut {NoStop}%
		\bibitem [{\citenamefont {Fernandes}\ \emph {et~al.}(2022)\citenamefont
			{Fernandes}, \citenamefont {Coldea}, \citenamefont {Ding}, \citenamefont
			{Fisher}, \citenamefont {Hirschfeld},\ and\ \citenamefont
			{Kotliar}}]{Fernandes2022}%
		\BibitemOpen
		\bibfield  {author} {\bibinfo {author} {\bibfnamefont {R.~M.}\ \bibnamefont
				{Fernandes}}, \bibinfo {author} {\bibfnamefont {A.~I.}\ \bibnamefont
				{Coldea}}, \bibinfo {author} {\bibfnamefont {H.}~\bibnamefont {Ding}},
			\bibinfo {author} {\bibfnamefont {I.~R.}\ \bibnamefont {Fisher}}, \bibinfo
			{author} {\bibfnamefont {P.~J.}\ \bibnamefont {Hirschfeld}}, \ and\ \bibinfo
			{author} {\bibfnamefont {G.}~\bibnamefont {Kotliar}},\ }\bibfield  {title}
		{\enquote {\bibinfo {title} {{Iron pnictides and chalcogenides: a
						new paradigm for superconductivity}},}\ }\href {\doibase
			10.1038/s41586-021-04073-2} {\bibfield  {journal} {\bibinfo  {journal}
				{Nature}\ }\textbf {\bibinfo {volume} {601}},\ \bibinfo {pages} {35--44}
			(\bibinfo {year} {2022})}\BibitemShut {NoStop}%
		\bibitem [{\citenamefont {Georges}\ \emph {et~al.}(2013)\citenamefont
			{Georges}, \citenamefont {Medici},\ and\ \citenamefont
			{Mravlje}}]{Antoine2013}%
		\BibitemOpen
		\bibfield  {author} {\bibinfo {author} {\bibfnamefont {A.}~\bibnamefont
				{Georges}}, \bibinfo {author} {\bibfnamefont {L.~de'}\ \bibnamefont
				{Medici}}, \ and\ \bibinfo {author} {\bibfnamefont {J.}~\bibnamefont
				{Mravlje}},\ }\bibfield  {title} {\enquote {\bibinfo {title} {{Strong
						Correlations from Hund’s Coupling}},}\ }\href {\doibase
			10.1146/annurev-conmatphys-020911-125045} {\bibfield  {journal} {\bibinfo
				{journal} {Annual Review of Condensed Matter Physics}\ }\textbf {\bibinfo
				{volume} {4}},\ \bibinfo {pages} {137--178} (\bibinfo {year}
			{2013})}\BibitemShut {NoStop}%
		\bibitem [{\citenamefont {de' Medici}\ \emph {et~al.}(2014)\citenamefont {de'
				Medici}, \citenamefont {Giovannetti},\ and\ \citenamefont
			{Capone}}]{Medici2014}%
		\BibitemOpen
		\bibfield  {author} {\bibinfo {author} {\bibfnamefont {L}~\bibnamefont {de'
					Medici}}, \bibinfo {author} {\bibfnamefont {G.}~\bibnamefont {Giovannetti}},
			\ and\ \bibinfo {author} {\bibfnamefont {M.}~\bibnamefont {Capone}},\
		}\bibfield  {title} {\enquote {\bibinfo {title} {{Selective Mott Physics as a
						Key to Iron Superconductors}},}\ }\href {\doibase
			10.1103/PhysRevLett.112.177001} {\bibfield  {journal} {\bibinfo  {journal}
				{Phys. Rev. Lett.}\ }\textbf {\bibinfo {volume} {112}},\ \bibinfo {pages}
			{177001} (\bibinfo {year} {2014})}\BibitemShut {NoStop}%
		\bibitem [{\citenamefont {Lanat\`a}\ \emph {et~al.}(2013)\citenamefont
			{Lanat\`a}, \citenamefont {Strand}, \citenamefont {Giovannetti},
			\citenamefont {Hellsing}, \citenamefont {de' Medici},\ and\ \citenamefont
			{Capone}}]{Nicola2013}%
		\BibitemOpen
		\bibfield  {author} {\bibinfo {author} {\bibfnamefont {N.}~\bibnamefont
				{Lanat\`a}}, \bibinfo {author} {\bibfnamefont {H.~U.~R.}\ \bibnamefont
				{Strand}}, \bibinfo {author} {\bibfnamefont {G.}~\bibnamefont {Giovannetti}},
			\bibinfo {author} {\bibfnamefont {B.}~\bibnamefont {Hellsing}}, \bibinfo
			{author} {\bibfnamefont {L.}~\bibnamefont {de' Medici}}, \ and\ \bibinfo
			{author} {\bibfnamefont {M.}~\bibnamefont {Capone}},\ }\bibfield  {title}
		{\enquote {\bibinfo {title} {{Orbital selectivity in Hund's metals: The iron
						chalcogenides}},}\ }\href {\doibase 10.1103/PhysRevB.87.045122} {\bibfield
			{journal} {\bibinfo  {journal} {Phys. Rev. B}\ }\textbf {\bibinfo {volume}
				{87}},\ \bibinfo {pages} {045122} (\bibinfo {year} {2013})}\BibitemShut
		{NoStop}%
		\bibitem [{\citenamefont {Mazin}\ \emph {et~al.}(2008)\citenamefont {Mazin},
			\citenamefont {Singh}, \citenamefont {Johannes},\ and\ \citenamefont
			{Du}}]{Mazin2008}%
		\BibitemOpen
		\bibfield  {author} {\bibinfo {author} {\bibfnamefont {I.~I.}\ \bibnamefont
				{Mazin}}, \bibinfo {author} {\bibfnamefont {D.~J.}\ \bibnamefont {Singh}},
			\bibinfo {author} {\bibfnamefont {M.~D.}\ \bibnamefont {Johannes}}, \ and\
			\bibinfo {author} {\bibfnamefont {M.~H.}\ \bibnamefont {Du}},\ }\bibfield
		{title} {\enquote {\bibinfo {title} {{Unconventional Superconductivity with a
						Sign Reversal in the Order Parameter of
						${\mathrm{LaFeAsO}}_{1\ensuremath{-}x}{\mathrm{F}}_{x}$}},}\ }\href {\doibase
			10.1103/PhysRevLett.101.057003} {\bibfield  {journal} {\bibinfo  {journal}
				{Phys. Rev. Lett.}\ }\textbf {\bibinfo {volume} {101}},\ \bibinfo {pages}
			{057003} (\bibinfo {year} {2008})}\BibitemShut {NoStop}%
		\bibitem [{\citenamefont {Graser}\ \emph {et~al.}(2009)\citenamefont {Graser},
			\citenamefont {Maier}, \citenamefont {Hirschfeld},\ and\ \citenamefont
			{Scalapino}}]{Graser2009}%
		\BibitemOpen
		\bibfield  {author} {\bibinfo {author} {\bibfnamefont {S.}~\bibnamefont
				{Graser}}, \bibinfo {author} {\bibfnamefont {T.~A.}\ \bibnamefont {Maier}},
			\bibinfo {author} {\bibfnamefont {P.~J.}\ \bibnamefont {Hirschfeld}}, \ and\
			\bibinfo {author} {\bibfnamefont {D.~J.}\ \bibnamefont {Scalapino}},\
		}\bibfield  {title} {\enquote {\bibinfo {title} {{Near-degeneracy of several
						pairing channels in multiorbital models for the Fe pnictides}},}\ }\href
		{\doibase 10.1088/1367-2630/11/2/025016} {\bibfield  {journal} {\bibinfo
				{journal} {New J. Phys.}\ }\textbf {\bibinfo {volume} {11}},\ \bibinfo
			{pages} {025016} (\bibinfo {year} {2009})}\BibitemShut {NoStop}%
		\bibitem [{\citenamefont {Fernandes}\ \emph {et~al.}(2014)\citenamefont
			{Fernandes}, \citenamefont {Chubukov},\ and\ \citenamefont
			{Schmalian}}]{Fernandes2014}%
		\BibitemOpen
		\bibfield  {author} {\bibinfo {author} {\bibfnamefont {R.~M.}\ \bibnamefont
				{Fernandes}}, \bibinfo {author} {\bibfnamefont {A.~V.}\ \bibnamefont
				{Chubukov}}, \ and\ \bibinfo {author} {\bibfnamefont {J.}~\bibnamefont
				{Schmalian}},\ }\bibfield  {title} {\enquote {\bibinfo {title} {{What drives
						nematic order in iron-based superconductors?}}}\ }\href {\doibase
			10.1038/nphys2877} {\bibfield  {journal} {\bibinfo  {journal} {Nat. Phys.}\
			}\textbf {\bibinfo {volume} {10}},\ \bibinfo {pages} {97--104} (\bibinfo
			{year} {2014})}\BibitemShut {NoStop}%
		\bibitem [{\citenamefont {Böhmer}\ and\ \citenamefont
			{Kreisel}(2017)}]{Bohmer2017}%
		\BibitemOpen
		\bibfield  {author} {\bibinfo {author} {\bibfnamefont {A.~E.}\ \bibnamefont
				{Böhmer}}\ and\ \bibinfo {author} {\bibfnamefont {A.}~\bibnamefont
				{Kreisel}},\ }\bibfield  {title} {\enquote {\bibinfo {title} {{Nematicity,
						magnetism and superconductivity in FeSe}},}\ }\href {\doibase
			10.1088/1361-648x/aa9caa} {\bibfield  {journal} {\bibinfo  {journal} {J.
					Phys.: Condens. Matter}\ }\textbf {\bibinfo {volume} {30}},\ \bibinfo {pages}
			{023001} (\bibinfo {year} {2017})}\BibitemShut {NoStop}%
		\bibitem [{\citenamefont {Coldea}(2021)}]{Coldea2021}%
		\BibitemOpen
		\bibfield  {author} {\bibinfo {author} {\bibfnamefont {A.~I.}\ \bibnamefont
				{Coldea}},\ }\bibfield  {title} {\enquote {\bibinfo {title} {{Electronic
						Nematic States Tuned by Isoelectronic Substitution in Bulk
						${\mathrm{FeSe}}_{1-x}{\mathrm{S}}_{x}$}},}\ }\href {\doibase
			10.3389/fphy.2020.594500} {\bibfield  {journal} {\bibinfo  {journal}
				{Frontiers in Physics}\ }\textbf {\bibinfo {volume} {8}},\ \bibinfo {pages}
			{528} (\bibinfo {year} {2021})}\BibitemShut {NoStop}%
		\bibitem [{\citenamefont {Sprau}\ \emph {et~al.}(2017)\citenamefont {Sprau},
			\citenamefont {Kostin}, \citenamefont {Kreisel}, \citenamefont
			{B{\"{o}}hmer}, \citenamefont {Taufour}, \citenamefont {Canfield},
			\citenamefont {Mukherjee}, \citenamefont {Hirschfeld}, \citenamefont
			{Andersen},\ and\ \citenamefont {S{\'{e}}amus~Davis}}]{Sprau2017}%
		\BibitemOpen
		\bibfield  {author} {\bibinfo {author} {\bibfnamefont {P.~O.}\ \bibnamefont
				{Sprau}}, \bibinfo {author} {\bibfnamefont {A.}~\bibnamefont {Kostin}},
			\bibinfo {author} {\bibfnamefont {A.}~\bibnamefont {Kreisel}}, \bibinfo
			{author} {\bibfnamefont {A.~E.}\ \bibnamefont {B{\"{o}}hmer}}, \bibinfo
			{author} {\bibfnamefont {V.}~\bibnamefont {Taufour}}, \bibinfo {author}
			{\bibfnamefont {P.~C.}\ \bibnamefont {Canfield}}, \bibinfo {author}
			{\bibfnamefont {S.}~\bibnamefont {Mukherjee}}, \bibinfo {author}
			{\bibfnamefont {P.~J.}\ \bibnamefont {Hirschfeld}}, \bibinfo {author}
			{\bibfnamefont {B.~M.}\ \bibnamefont {Andersen}}, \ and\ \bibinfo {author}
			{\bibfnamefont {J.~C.}\ \bibnamefont {S{\'{e}}amus~Davis}},\ }\bibfield
		{title} {\enquote {\bibinfo {title} {{Discovery of Orbital-Selective Cooper
						Pairing in FeSe}},}\ }\href {\doibase 10.1126/science.aal1575} {\bibfield
			{journal} {\bibinfo  {journal} {Science}\ }\textbf {\bibinfo {volume}
				{357}},\ \bibinfo {pages} {75--80} (\bibinfo {year} {2017})}\BibitemShut
		{NoStop}%
		\bibitem [{\citenamefont {Xu}\ \emph {et~al.}(2016)\citenamefont {Xu},
			\citenamefont {Niu}, \citenamefont {Xu}, \citenamefont {Jiang}, \citenamefont
			{Yao}, \citenamefont {Chen}, \citenamefont {Song}, \citenamefont
			{Abdel-Hafiez}, \citenamefont {Chareev}, \citenamefont {Vasiliev},
			\citenamefont {Wang}, \citenamefont {Wo}, \citenamefont {Zhao}, \citenamefont
			{Peng},\ and\ \citenamefont {Feng}}]{Xu2016}%
		\BibitemOpen
		\bibfield  {author} {\bibinfo {author} {\bibfnamefont {H.~C.}\ \bibnamefont
				{Xu}}, \bibinfo {author} {\bibfnamefont {X.~H.}\ \bibnamefont {Niu}},
			\bibinfo {author} {\bibfnamefont {D.~F.}\ \bibnamefont {Xu}}, \bibinfo
			{author} {\bibfnamefont {J.}~\bibnamefont {Jiang}}, \bibinfo {author}
			{\bibfnamefont {Q.}~\bibnamefont {Yao}}, \bibinfo {author} {\bibfnamefont
				{Q.~Y.}\ \bibnamefont {Chen}}, \bibinfo {author} {\bibfnamefont
				{Q.}~\bibnamefont {Song}}, \bibinfo {author} {\bibfnamefont {M.}~\bibnamefont
				{Abdel-Hafiez}}, \bibinfo {author} {\bibfnamefont {D.~A.}\ \bibnamefont
				{Chareev}}, \bibinfo {author} {\bibfnamefont {A.~N.}\ \bibnamefont
				{Vasiliev}}, \bibinfo {author} {\bibfnamefont {Q.~S.}\ \bibnamefont {Wang}},
			\bibinfo {author} {\bibfnamefont {H.~L.}\ \bibnamefont {Wo}}, \bibinfo
			{author} {\bibfnamefont {J.}~\bibnamefont {Zhao}}, \bibinfo {author}
			{\bibfnamefont {R.}~\bibnamefont {Peng}}, \ and\ \bibinfo {author}
			{\bibfnamefont {D.~L.}\ \bibnamefont {Feng}},\ }\bibfield  {title} {\enquote
			{\bibinfo {title} {{Highly Anisotropic and Twofold Symmetric Superconducting
						Gap in Nematically Ordered ${\mathrm{FeSe}}_{0.93}{\mathrm{S}}_{0.07}$}},}\
		}\href {\doibase 10.1103/PhysRevLett.117.157003} {\bibfield  {journal}
			{\bibinfo  {journal} {Phys. Rev. Lett.}\ }\textbf {\bibinfo {volume} {117}},\
			\bibinfo {pages} {157003} (\bibinfo {year} {2016})}\BibitemShut {NoStop}%
		\bibitem [{\citenamefont {Hashimoto}\ \emph {et~al.}(2018)\citenamefont
			{Hashimoto}, \citenamefont {Ota}, \citenamefont {Yamamoto}, \citenamefont
			{Suzuki}, \citenamefont {Shimojima}, \citenamefont {Watanabe}, \citenamefont
			{Chen}, \citenamefont {Kasahara}, \citenamefont {Matsuda}, \citenamefont
			{Shibauchi}, \citenamefont {Okazaki},\ and\ \citenamefont
			{Shin}}]{Hashimoto2018}%
		\BibitemOpen
		\bibfield  {author} {\bibinfo {author} {\bibfnamefont {T.}~\bibnamefont
				{Hashimoto}}, \bibinfo {author} {\bibfnamefont {Y.}~\bibnamefont {Ota}},
			\bibinfo {author} {\bibfnamefont {H.~Q.}\ \bibnamefont {Yamamoto}}, \bibinfo
			{author} {\bibfnamefont {Y.}~\bibnamefont {Suzuki}}, \bibinfo {author}
			{\bibfnamefont {T.}~\bibnamefont {Shimojima}}, \bibinfo {author}
			{\bibfnamefont {S.}~\bibnamefont {Watanabe}}, \bibinfo {author}
			{\bibfnamefont {C.}~\bibnamefont {Chen}}, \bibinfo {author} {\bibfnamefont
				{S.}~\bibnamefont {Kasahara}}, \bibinfo {author} {\bibfnamefont
				{Y.}~\bibnamefont {Matsuda}}, \bibinfo {author} {\bibfnamefont
				{T.}~\bibnamefont {Shibauchi}}, \bibinfo {author} {\bibfnamefont
				{K.}~\bibnamefont {Okazaki}}, \ and\ \bibinfo {author} {\bibfnamefont
				{S.}~\bibnamefont {Shin}},\ }\bibfield  {title} {\enquote {\bibinfo {title}
				{{Superconducting gap anisotropy sensitive to nematic domains in FeSe}},}\
		}\href {\doibase 10.1038/s41467-017-02739-y} {\bibfield  {journal} {\bibinfo
				{journal} {Nat. Commun.}\ }\textbf {\bibinfo {volume} {9}},\ \bibinfo {pages}
			{282} (\bibinfo {year} {2018})}\BibitemShut {NoStop}%
		\bibitem [{\citenamefont {Liu}\ \emph {et~al.}(2018)\citenamefont {Liu},
			\citenamefont {Li}, \citenamefont {Huang}, \citenamefont {Lei}, \citenamefont
			{Wang}, \citenamefont {Wu}, \citenamefont {Shen}, \citenamefont {Gao},
			\citenamefont {Zhang}, \citenamefont {Liu}, \citenamefont {Hu}, \citenamefont
			{Xu}, \citenamefont {Liang}, \citenamefont {Liu}, \citenamefont {Ai},
			\citenamefont {Zhao}, \citenamefont {He}, \citenamefont {Yu}, \citenamefont
			{Liu}, \citenamefont {Mao}, \citenamefont {Dong}, \citenamefont {Jia},
			\citenamefont {Zhang}, \citenamefont {Zhang}, \citenamefont {Yang},
			\citenamefont {Wang}, \citenamefont {Peng}, \citenamefont {Shi},
			\citenamefont {Hu}, \citenamefont {Xiang}, \citenamefont {Chen},
			\citenamefont {Xu}, \citenamefont {Chen},\ and\ \citenamefont
			{Zhou}}]{Liu2018}%
		\BibitemOpen
		\bibfield  {author} {\bibinfo {author} {\bibfnamefont {D.}~\bibnamefont
				{Liu}}, \bibinfo {author} {\bibfnamefont {C.}~\bibnamefont {Li}}, \bibinfo
			{author} {\bibfnamefont {J.}~\bibnamefont {Huang}}, \bibinfo {author}
			{\bibfnamefont {B.}~\bibnamefont {Lei}}, \bibinfo {author} {\bibfnamefont
				{L.}~\bibnamefont {Wang}}, \bibinfo {author} {\bibfnamefont {X.}~\bibnamefont
				{Wu}}, \bibinfo {author} {\bibfnamefont {B.}~\bibnamefont {Shen}}, \bibinfo
			{author} {\bibfnamefont {Q.}~\bibnamefont {Gao}}, \bibinfo {author}
			{\bibfnamefont {Y.}~\bibnamefont {Zhang}}, \bibinfo {author} {\bibfnamefont
				{X.}~\bibnamefont {Liu}}, \bibinfo {author} {\bibfnamefont {Y.}~\bibnamefont
				{Hu}}, \bibinfo {author} {\bibfnamefont {Y.}~\bibnamefont {Xu}}, \bibinfo
			{author} {\bibfnamefont {A.}~\bibnamefont {Liang}}, \bibinfo {author}
			{\bibfnamefont {J.}~\bibnamefont {Liu}}, \bibinfo {author} {\bibfnamefont
				{P.}~\bibnamefont {Ai}}, \bibinfo {author} {\bibfnamefont {L.}~\bibnamefont
				{Zhao}}, \bibinfo {author} {\bibfnamefont {S.}~\bibnamefont {He}}, \bibinfo
			{author} {\bibfnamefont {L.}~\bibnamefont {Yu}}, \bibinfo {author}
			{\bibfnamefont {G.}~\bibnamefont {Liu}}, \bibinfo {author} {\bibfnamefont
				{Y.}~\bibnamefont {Mao}}, \bibinfo {author} {\bibfnamefont {X.}~\bibnamefont
				{Dong}}, \bibinfo {author} {\bibfnamefont {X.}~\bibnamefont {Jia}}, \bibinfo
			{author} {\bibfnamefont {F.}~\bibnamefont {Zhang}}, \bibinfo {author}
			{\bibfnamefont {S.}~\bibnamefont {Zhang}}, \bibinfo {author} {\bibfnamefont
				{F.}~\bibnamefont {Yang}}, \bibinfo {author} {\bibfnamefont {Z.}~\bibnamefont
				{Wang}}, \bibinfo {author} {\bibfnamefont {Q.}~\bibnamefont {Peng}}, \bibinfo
			{author} {\bibfnamefont {Y.}~\bibnamefont {Shi}}, \bibinfo {author}
			{\bibfnamefont {J.}~\bibnamefont {Hu}}, \bibinfo {author} {\bibfnamefont
				{T.}~\bibnamefont {Xiang}}, \bibinfo {author} {\bibfnamefont
				{X.}~\bibnamefont {Chen}}, \bibinfo {author} {\bibfnamefont {Z.}~\bibnamefont
				{Xu}}, \bibinfo {author} {\bibfnamefont {C.}~\bibnamefont {Chen}}, \ and\
			\bibinfo {author} {\bibfnamefont {X.~J.}\ \bibnamefont {Zhou}},\ }\bibfield
		{title} {\enquote {\bibinfo {title} {{Orbital Origin of Extremely Anisotropic
						Superconducting Gap in Nematic Phase of FeSe Superconductor}},}\ }\href
		{\doibase 10.1103/PhysRevX.8.031033} {\bibfield  {journal} {\bibinfo
				{journal} {Phys. Rev. X}\ }\textbf {\bibinfo {volume} {8}},\ \bibinfo {pages}
			{031033} (\bibinfo {year} {2018})}\BibitemShut {NoStop}%
		\bibitem [{\citenamefont {Rhodes}\ \emph {et~al.}(2018)\citenamefont {Rhodes},
			\citenamefont {Watson}, \citenamefont {Haghighirad}, \citenamefont
			{Evtushinsky}, \citenamefont {Eschrig},\ and\ \citenamefont
			{Kim}}]{Rhodes2018}%
		\BibitemOpen
		\bibfield  {author} {\bibinfo {author} {\bibfnamefont {L.~C.}\ \bibnamefont
				{Rhodes}}, \bibinfo {author} {\bibfnamefont {M.~D.}\ \bibnamefont {Watson}},
			\bibinfo {author} {\bibfnamefont {A.~A.}\ \bibnamefont {Haghighirad}},
			\bibinfo {author} {\bibfnamefont {D.~V.}\ \bibnamefont {Evtushinsky}},
			\bibinfo {author} {\bibfnamefont {M.}~\bibnamefont {Eschrig}}, \ and\
			\bibinfo {author} {\bibfnamefont {T.~K.}\ \bibnamefont {Kim}},\ }\bibfield
		{title} {\enquote {\bibinfo {title} {{Scaling of the superconducting gap with
						orbital character in FeSe}},}\ }\href {\doibase 10.1103/PhysRevB.98.180503}
		{\bibfield  {journal} {\bibinfo  {journal} {Phys. Rev. B}\ }\textbf {\bibinfo
				{volume} {98}},\ \bibinfo {pages} {180503} (\bibinfo {year}
			{2018})}\BibitemShut {NoStop}%
		\bibitem [{\citenamefont {Kushnirenko}\ \emph {et~al.}(2018)\citenamefont
			{Kushnirenko}, \citenamefont {Fedorov}, \citenamefont {Haubold},
			\citenamefont {Thirupathaiah}, \citenamefont {Wolf}, \citenamefont
			{Aswartham}, \citenamefont {Morozov}, \citenamefont {Kim}, \citenamefont
			{B\"uchner},\ and\ \citenamefont {Borisenko}}]{Kushnirenko2018}%
		\BibitemOpen
		\bibfield  {author} {\bibinfo {author} {\bibfnamefont {Y.~S.}\ \bibnamefont
				{Kushnirenko}}, \bibinfo {author} {\bibfnamefont {A.~V.}\ \bibnamefont
				{Fedorov}}, \bibinfo {author} {\bibfnamefont {E.}~\bibnamefont {Haubold}},
			\bibinfo {author} {\bibfnamefont {S.}~\bibnamefont {Thirupathaiah}}, \bibinfo
			{author} {\bibfnamefont {T.}~\bibnamefont {Wolf}}, \bibinfo {author}
			{\bibfnamefont {S.}~\bibnamefont {Aswartham}}, \bibinfo {author}
			{\bibfnamefont {I.}~\bibnamefont {Morozov}}, \bibinfo {author} {\bibfnamefont
				{T.~K.}\ \bibnamefont {Kim}}, \bibinfo {author} {\bibfnamefont
				{B.}~\bibnamefont {B\"uchner}}, \ and\ \bibinfo {author} {\bibfnamefont
				{S.~V.}\ \bibnamefont {Borisenko}},\ }\bibfield  {title} {\enquote {\bibinfo
				{title} {{Three-dimensional superconducting gap in FeSe from angle-resolved
						photoemission spectroscopy}},}\ }\href {\doibase 10.1103/PhysRevB.97.180501}
		{\bibfield  {journal} {\bibinfo  {journal} {Phys. Rev. B}\ }\textbf {\bibinfo
				{volume} {97}},\ \bibinfo {pages} {180501} (\bibinfo {year}
			{2018})}\BibitemShut {NoStop}%
		\bibitem [{\citenamefont {Kreisel}\ \emph {et~al.}(2017)\citenamefont
			{Kreisel}, \citenamefont {Andersen}, \citenamefont {Sprau}, \citenamefont
			{Kostin}, \citenamefont {Davis},\ and\ \citenamefont
			{Hirschfeld}}]{Kreisel2017}%
		\BibitemOpen
		\bibfield  {author} {\bibinfo {author} {\bibfnamefont {A.}~\bibnamefont
				{Kreisel}}, \bibinfo {author} {\bibfnamefont {B.~M.}\ \bibnamefont
				{Andersen}}, \bibinfo {author} {\bibfnamefont {P.~O.}\ \bibnamefont {Sprau}},
			\bibinfo {author} {\bibfnamefont {A.}~\bibnamefont {Kostin}}, \bibinfo
			{author} {\bibfnamefont {J.~C.~S\'eamus}\ \bibnamefont {Davis}}, \ and\
			\bibinfo {author} {\bibfnamefont {P.~J.}\ \bibnamefont {Hirschfeld}},\
		}\bibfield  {title} {\enquote {\bibinfo {title} {Orbital selective pairing
					and gap structures of iron-based superconductors},}\ }\href {\doibase
			10.1103/PhysRevB.95.174504} {\bibfield  {journal} {\bibinfo  {journal} {Phys.
					Rev. B}\ }\textbf {\bibinfo {volume} {95}},\ \bibinfo {pages} {174504}
			(\bibinfo {year} {2017})}\BibitemShut {NoStop}%
		\bibitem [{\citenamefont {Benfatto}\ \emph {et~al.}(2018)\citenamefont
			{Benfatto}, \citenamefont {Valenzuela},\ and\ \citenamefont
			{Fanfarillo}}]{Benfatto2018}%
		\BibitemOpen
		\bibfield  {author} {\bibinfo {author} {\bibfnamefont {L.}~\bibnamefont
				{Benfatto}}, \bibinfo {author} {\bibfnamefont {B.}~\bibnamefont
				{Valenzuela}}, \ and\ \bibinfo {author} {\bibfnamefont {L.}~\bibnamefont
				{Fanfarillo}},\ }\bibfield  {title} {\enquote {\bibinfo {title} {{Nematic
						pairing from orbital-selective spin fluctuations in FeSe}},}\ }\href
		{\doibase 10.1038/s41535-018-0129-9} {\bibfield  {journal} {\bibinfo
				{journal} {npj Quantum Materials}\ }\textbf {\bibinfo {volume} {3}},\
			\bibinfo {pages} {56} (\bibinfo {year} {2018})}\BibitemShut {NoStop}%
		\bibitem [{\citenamefont {Kang}\ \emph
			{et~al.}(2018{\natexlab{a}})\citenamefont {Kang}, \citenamefont {Fernandes},\
			and\ \citenamefont {Chubukov}}]{Kang2018}%
		\BibitemOpen
		\bibfield  {author} {\bibinfo {author} {\bibfnamefont {J.}~\bibnamefont
				{Kang}}, \bibinfo {author} {\bibfnamefont {R.~M.}\ \bibnamefont {Fernandes}},
			\ and\ \bibinfo {author} {\bibfnamefont {A.}~\bibnamefont {Chubukov}},\
		}\bibfield  {title} {\enquote {\bibinfo {title} {Superconductivity in fese:
					The role of nematic order},}\ }\href {\doibase
			10.1103/PhysRevLett.120.267001} {\bibfield  {journal} {\bibinfo  {journal}
				{Phys. Rev. Lett.}\ }\textbf {\bibinfo {volume} {120}},\ \bibinfo {pages}
			{267001} (\bibinfo {year} {2018}{\natexlab{a}})}\BibitemShut {NoStop}%
		\bibitem [{\citenamefont {Yu}\ \emph {et~al.}(2018)\citenamefont {Yu},
			\citenamefont {Zhu},\ and\ \citenamefont {Si}}]{Yu2018}%
		\BibitemOpen
		\bibfield  {author} {\bibinfo {author} {\bibfnamefont {R.}~\bibnamefont
				{Yu}}, \bibinfo {author} {\bibfnamefont {J-X.}\ \bibnamefont {Zhu}}, \ and\
			\bibinfo {author} {\bibfnamefont {Q.}~\bibnamefont {Si}},\ }\bibfield
		{title} {\enquote {\bibinfo {title} {{Orbital Selectivity Enhanced by Nematic
						Order in FeSe}},}\ }\href {\doibase 10.1103/PhysRevLett.121.227003}
		{\bibfield  {journal} {\bibinfo  {journal} {Phys. Rev. Lett.}\ }\textbf
			{\bibinfo {volume} {121}},\ \bibinfo {pages} {227003} (\bibinfo {year}
			{2018})}\BibitemShut {NoStop}%
		\bibitem [{\citenamefont {B{\"{o}}hmer}\ \emph {et~al.}(2013)\citenamefont
			{B{\"{o}}hmer}, \citenamefont {Hardy}, \citenamefont {Eilers}, \citenamefont
			{Ernst}, \citenamefont {Adelmann}, \citenamefont {Schweiss}, \citenamefont
			{Wolf},\ and\ \citenamefont {Meingast}}]{Bohmer2013}%
		\BibitemOpen
		\bibfield  {author} {\bibinfo {author} {\bibfnamefont {A.~E.}\ \bibnamefont
				{B{\"{o}}hmer}}, \bibinfo {author} {\bibfnamefont {F.}~\bibnamefont {Hardy}},
			\bibinfo {author} {\bibfnamefont {F.}~\bibnamefont {Eilers}}, \bibinfo
			{author} {\bibfnamefont {D.}~\bibnamefont {Ernst}}, \bibinfo {author}
			{\bibfnamefont {P.}~\bibnamefont {Adelmann}}, \bibinfo {author}
			{\bibfnamefont {P.}~\bibnamefont {Schweiss}}, \bibinfo {author}
			{\bibfnamefont {T.}~\bibnamefont {Wolf}}, \ and\ \bibinfo {author}
			{\bibfnamefont {C.}~\bibnamefont {Meingast}},\ }\bibfield  {title} {\enquote
			{\bibinfo {title} {{Lack of coupling between superconductivity and
						orthorhombic distortion in stoichiometric single-crystalline FeSe}},}\ }\href
		{\doibase 10.1103/PhysRevB.87.180505} {\bibfield  {journal} {\bibinfo
				{journal} {Phys. Rev. B}\ }\textbf {\bibinfo {volume} {87}},\ \bibinfo
			{pages} {180505(R)} (\bibinfo {year} {2013})}\BibitemShut {NoStop}%
		\bibitem [{\citenamefont {Fedorov}\ \emph {et~al.}(2016)\citenamefont
			{Fedorov}, \citenamefont {Yaresko}, \citenamefont {Kim}, \citenamefont
			{Kushnirenko}, \citenamefont {Haubold}, \citenamefont {Wolf}, \citenamefont
			{Hoesch}, \citenamefont {Grueneis}, \citenamefont {Buechner},\ and\
			\citenamefont {Borisenko}}]{Fedorov2016}%
		\BibitemOpen
		\bibfield  {author} {\bibinfo {author} {\bibfnamefont {A.}~\bibnamefont
				{Fedorov}}, \bibinfo {author} {\bibfnamefont {A.}~\bibnamefont {Yaresko}},
			\bibinfo {author} {\bibfnamefont {T.~K.}\ \bibnamefont {Kim}}, \bibinfo
			{author} {\bibfnamefont {E.}~\bibnamefont {Kushnirenko}}, \bibinfo {author}
			{\bibfnamefont {E.}~\bibnamefont {Haubold}}, \bibinfo {author} {\bibfnamefont
				{T.}~\bibnamefont {Wolf}}, \bibinfo {author} {\bibfnamefont {M.}~\bibnamefont
				{Hoesch}}, \bibinfo {author} {\bibfnamefont {A.}~\bibnamefont {Grueneis}},
			\bibinfo {author} {\bibfnamefont {B.}~\bibnamefont {Buechner}}, \ and\
			\bibinfo {author} {\bibfnamefont {S.~V.}\ \bibnamefont {Borisenko}},\
		}\bibfield  {title} {\enquote {\bibinfo {title} {{Effect of nematic ordering
						on electronic structure of FeSe}},}\ }\href {\doibase 10.1038/srep36834}
		{\bibfield  {journal} {\bibinfo  {journal} {Sci. Rep.}\ }\textbf {\bibinfo
				{volume} {6}},\ \bibinfo {pages} {36834} (\bibinfo {year}
			{2016})}\BibitemShut {NoStop}%
		\bibitem [{\citenamefont {Nakayama}\ \emph {et~al.}(2014)\citenamefont
			{Nakayama}, \citenamefont {Miyata}, \citenamefont {Phan}, \citenamefont
			{Sato}, \citenamefont {Tanabe}, \citenamefont {Urata}, \citenamefont
			{Tanigaki},\ and\ \citenamefont {Takahashi}}]{Nakayama2014}%
		\BibitemOpen
		\bibfield  {author} {\bibinfo {author} {\bibfnamefont {K.}~\bibnamefont
				{Nakayama}}, \bibinfo {author} {\bibfnamefont {Y.}~\bibnamefont {Miyata}},
			\bibinfo {author} {\bibfnamefont {G.~N.}\ \bibnamefont {Phan}}, \bibinfo
			{author} {\bibfnamefont {T.}~\bibnamefont {Sato}}, \bibinfo {author}
			{\bibfnamefont {Y.}~\bibnamefont {Tanabe}}, \bibinfo {author} {\bibfnamefont
				{T.}~\bibnamefont {Urata}}, \bibinfo {author} {\bibfnamefont
				{K.}~\bibnamefont {Tanigaki}}, \ and\ \bibinfo {author} {\bibfnamefont
				{T.}~\bibnamefont {Takahashi}},\ }\bibfield  {title} {\enquote {\bibinfo
				{title} {{Reconstruction of band structure induced by electronic nematicity
						in an FeSe superconductor}},}\ }\href {\doibase
			10.1103/PhysRevLett.113.237001} {\bibfield  {journal} {\bibinfo  {journal}
				{Phys. Rev. Lett.}\ }\textbf {\bibinfo {volume} {113}},\ \bibinfo {pages}
			{237001} (\bibinfo {year} {2014})}\BibitemShut {NoStop}%
		\bibitem [{\citenamefont {Shimojima}\ \emph {et~al.}(2014)\citenamefont
			{Shimojima}, \citenamefont {Suzuki}, \citenamefont {Sonobe}, \citenamefont
			{Nakamura}, \citenamefont {Sakano}, \citenamefont {Omachi}, \citenamefont
			{Yoshioka}, \citenamefont {Kuwata-Gonokami}, \citenamefont {Ono},
			\citenamefont {Kumigashira}, \citenamefont {B{\"{o}}hmer}, \citenamefont
			{Hardy}, \citenamefont {Wolf}, \citenamefont {Meingast}, \citenamefont
			{L{\"{o}}hneysen}, \citenamefont {Ikeda},\ and\ \citenamefont
			{Ishizaka}}]{Shimojima2014}%
		\BibitemOpen
		\bibfield  {author} {\bibinfo {author} {\bibfnamefont {T.}~\bibnamefont
				{Shimojima}}, \bibinfo {author} {\bibfnamefont {Y.}~\bibnamefont {Suzuki}},
			\bibinfo {author} {\bibfnamefont {T.}~\bibnamefont {Sonobe}}, \bibinfo
			{author} {\bibfnamefont {A.}~\bibnamefont {Nakamura}}, \bibinfo {author}
			{\bibfnamefont {M.}~\bibnamefont {Sakano}}, \bibinfo {author} {\bibfnamefont
				{J.}~\bibnamefont {Omachi}}, \bibinfo {author} {\bibfnamefont
				{K.}~\bibnamefont {Yoshioka}}, \bibinfo {author} {\bibfnamefont
				{M.}~\bibnamefont {Kuwata-Gonokami}}, \bibinfo {author} {\bibfnamefont
				{K.}~\bibnamefont {Ono}}, \bibinfo {author} {\bibfnamefont {H.}~\bibnamefont
				{Kumigashira}}, \bibinfo {author} {\bibfnamefont {A.~E.}\ \bibnamefont
				{B{\"{o}}hmer}}, \bibinfo {author} {\bibfnamefont {F.}~\bibnamefont {Hardy}},
			\bibinfo {author} {\bibfnamefont {T.}~\bibnamefont {Wolf}}, \bibinfo {author}
			{\bibfnamefont {C.}~\bibnamefont {Meingast}}, \bibinfo {author}
			{\bibfnamefont {H.~V.}\ \bibnamefont {L{\"{o}}hneysen}}, \bibinfo {author}
			{\bibfnamefont {H.}~\bibnamefont {Ikeda}}, \ and\ \bibinfo {author}
			{\bibfnamefont {K.}~\bibnamefont {Ishizaka}},\ }\bibfield  {title} {\enquote
			{\bibinfo {title} {{Lifting of xz / yz orbital degeneracy at the structural
						transition in detwinned FeSe}},}\ }\href {\doibase
			10.1103/PhysRevB.90.121111} {\bibfield  {journal} {\bibinfo  {journal} {Phys.
					Rev. B}\ }\textbf {\bibinfo {volume} {90}},\ \bibinfo {pages} {121111(R)}
			(\bibinfo {year} {2014})}\BibitemShut {NoStop}%
		\bibitem [{\citenamefont {Watson}\ \emph
			{et~al.}(2015{\natexlab{a}})\citenamefont {Watson}, \citenamefont {Kim},
			\citenamefont {Haghighirad}, \citenamefont {Davies}, \citenamefont
			{McCollam}, \citenamefont {Narayanan}, \citenamefont {Blake}, \citenamefont
			{Chen}, \citenamefont {Ghannadzadeh}, \citenamefont {Schofield},
			\citenamefont {Hoesch}, \citenamefont {Meingast}, \citenamefont {Wolf},\ and\
			\citenamefont {Coldea}}]{Watson2015}%
		\BibitemOpen
		\bibfield  {author} {\bibinfo {author} {\bibfnamefont {M.~D.}\ \bibnamefont
				{Watson}}, \bibinfo {author} {\bibfnamefont {T.~K.}\ \bibnamefont {Kim}},
			\bibinfo {author} {\bibfnamefont {A.~A.}\ \bibnamefont {Haghighirad}},
			\bibinfo {author} {\bibfnamefont {N.~R.}\ \bibnamefont {Davies}}, \bibinfo
			{author} {\bibfnamefont {A.}~\bibnamefont {McCollam}}, \bibinfo {author}
			{\bibfnamefont {A.}~\bibnamefont {Narayanan}}, \bibinfo {author}
			{\bibfnamefont {S.~F.}\ \bibnamefont {Blake}}, \bibinfo {author}
			{\bibfnamefont {Y.~L.}\ \bibnamefont {Chen}}, \bibinfo {author}
			{\bibfnamefont {S.}~\bibnamefont {Ghannadzadeh}}, \bibinfo {author}
			{\bibfnamefont {A.~J.}\ \bibnamefont {Schofield}}, \bibinfo {author}
			{\bibfnamefont {M.}~\bibnamefont {Hoesch}}, \bibinfo {author} {\bibfnamefont
				{C.}~\bibnamefont {Meingast}}, \bibinfo {author} {\bibfnamefont
				{T.}~\bibnamefont {Wolf}}, \ and\ \bibinfo {author} {\bibfnamefont {A.~I.}\
				\bibnamefont {Coldea}},\ }\bibfield  {title} {\enquote {\bibinfo {title}
				{{Emergence of the nematic electronic state in FeSe}},}\ }\href {\doibase
			http://dx.doi.org/10.1103/PhysRevB.91.155106} {\bibfield  {journal} {\bibinfo
				{journal} {Phys. Rev. B}\ }\textbf {\bibinfo {volume} {91}},\ \bibinfo
			{pages} {155106} (\bibinfo {year} {2015}{\natexlab{a}})}\BibitemShut
		{NoStop}%
		\bibitem [{\citenamefont {Fanfarillo}\ \emph {et~al.}(2016)\citenamefont
			{Fanfarillo}, \citenamefont {Mansart}, \citenamefont {Toulemonde},
			\citenamefont {Cercellier}, \citenamefont {Le~F\`evre}, \citenamefont
			{Bertran}, \citenamefont {Valenzuela}, \citenamefont {Benfatto},\ and\
			\citenamefont {Brouet}}]{Fanfarillo2016}%
		\BibitemOpen
		\bibfield  {author} {\bibinfo {author} {\bibfnamefont {L.}~\bibnamefont
				{Fanfarillo}}, \bibinfo {author} {\bibfnamefont {J.}~\bibnamefont {Mansart}},
			\bibinfo {author} {\bibfnamefont {P.}~\bibnamefont {Toulemonde}}, \bibinfo
			{author} {\bibfnamefont {H.}~\bibnamefont {Cercellier}}, \bibinfo {author}
			{\bibfnamefont {P.}~\bibnamefont {Le~F\`evre}}, \bibinfo {author}
			{\bibfnamefont {F.}~\bibnamefont {Bertran}}, \bibinfo {author} {\bibfnamefont
				{B.}~\bibnamefont {Valenzuela}}, \bibinfo {author} {\bibfnamefont
				{L.}~\bibnamefont {Benfatto}}, \ and\ \bibinfo {author} {\bibfnamefont
				{V.}~\bibnamefont {Brouet}},\ }\bibfield  {title} {\enquote {\bibinfo {title}
				{{Orbital-dependent Fermi surface shrinking as a fingerprint of nematicity in
						FeSe}},}\ }\href {\doibase 10.1103/PhysRevB.94.155138} {\bibfield  {journal}
			{\bibinfo  {journal} {Phys. Rev. B}\ }\textbf {\bibinfo {volume} {94}},\
			\bibinfo {pages} {155138} (\bibinfo {year} {2016})}\BibitemShut {NoStop}%
		\bibitem [{\citenamefont {Maletz}\ \emph {et~al.}(2014)\citenamefont {Maletz},
			\citenamefont {Zabolotnyy}, \citenamefont {Evtushinsky}, \citenamefont
			{Thirupathaiah}, \citenamefont {Wolter}, \citenamefont {Harnagea},
			\citenamefont {Yaresko}, \citenamefont {Vasiliev}, \citenamefont {Chareev},
			\citenamefont {B\"ohmer}, \citenamefont {Hardy}, \citenamefont {Wolf},
			\citenamefont {Meingast}, \citenamefont {Rienks}, \citenamefont {B\"uchner},\
			and\ \citenamefont {Borisenko}}]{Maletz2014}%
		\BibitemOpen
		\bibfield  {author} {\bibinfo {author} {\bibfnamefont {J.}~\bibnamefont
				{Maletz}}, \bibinfo {author} {\bibfnamefont {V.~B.}\ \bibnamefont
				{Zabolotnyy}}, \bibinfo {author} {\bibfnamefont {D.~V.}\ \bibnamefont
				{Evtushinsky}}, \bibinfo {author} {\bibfnamefont {S.}~\bibnamefont
				{Thirupathaiah}}, \bibinfo {author} {\bibfnamefont {A.~U.~B.}\ \bibnamefont
				{Wolter}}, \bibinfo {author} {\bibfnamefont {L.}~\bibnamefont {Harnagea}},
			\bibinfo {author} {\bibfnamefont {A.~N.}\ \bibnamefont {Yaresko}}, \bibinfo
			{author} {\bibfnamefont {A.~N.}\ \bibnamefont {Vasiliev}}, \bibinfo {author}
			{\bibfnamefont {D.~A.}\ \bibnamefont {Chareev}}, \bibinfo {author}
			{\bibfnamefont {A.~E.}\ \bibnamefont {B\"ohmer}}, \bibinfo {author}
			{\bibfnamefont {F.}~\bibnamefont {Hardy}}, \bibinfo {author} {\bibfnamefont
				{T.}~\bibnamefont {Wolf}}, \bibinfo {author} {\bibfnamefont {C.}~\bibnamefont
				{Meingast}}, \bibinfo {author} {\bibfnamefont {E.~D.~L.}\ \bibnamefont
				{Rienks}}, \bibinfo {author} {\bibfnamefont {B.}~\bibnamefont {B\"uchner}}, \
			and\ \bibinfo {author} {\bibfnamefont {S.~V.}\ \bibnamefont {Borisenko}},\
		}\bibfield  {title} {\enquote {\bibinfo {title} {{Unusual band
						renormalization in the simplest iron-based superconductor
						${\text{FeSe}}_{1\ensuremath{-}x}$}},}\ }\href {\doibase
			10.1103/PhysRevB.89.220506} {\bibfield  {journal} {\bibinfo  {journal} {Phys.
					Rev. B}\ }\textbf {\bibinfo {volume} {89}},\ \bibinfo {pages} {220506}
			(\bibinfo {year} {2014})}\BibitemShut {NoStop}%
		\bibitem [{\citenamefont {Watson}\ \emph {et~al.}(2016)\citenamefont {Watson},
			\citenamefont {Kim}, \citenamefont {Rhodes}, \citenamefont {Eschrig},
			\citenamefont {Hoesch}, \citenamefont {Haghighirad},\ and\ \citenamefont
			{Coldea}}]{Watson2016}%
		\BibitemOpen
		\bibfield  {author} {\bibinfo {author} {\bibfnamefont {M.~D.}\ \bibnamefont
				{Watson}}, \bibinfo {author} {\bibfnamefont {T.~K.}\ \bibnamefont {Kim}},
			\bibinfo {author} {\bibfnamefont {L.~C.}\ \bibnamefont {Rhodes}}, \bibinfo
			{author} {\bibfnamefont {M.}~\bibnamefont {Eschrig}}, \bibinfo {author}
			{\bibfnamefont {M.}~\bibnamefont {Hoesch}}, \bibinfo {author} {\bibfnamefont
				{A.~A.}\ \bibnamefont {Haghighirad}}, \ and\ \bibinfo {author} {\bibfnamefont
				{A.~I.}\ \bibnamefont {Coldea}},\ }\bibfield  {title} {\enquote {\bibinfo
				{title} {{Evidence for unidirectional nematic bond ordering in FeSe}},}\
		}\href {\doibase 10.1103/PhysRevB.94.201107} {\bibfield  {journal} {\bibinfo
				{journal} {Phys. Rev. B}\ }\textbf {\bibinfo {volume} {94}},\ \bibinfo
			{pages} {201107} (\bibinfo {year} {2016})}\BibitemShut {NoStop}%
		\bibitem [{\citenamefont {Coldea}\ and\ \citenamefont
			{Watson}(2018)}]{Coldea2018}%
		\BibitemOpen
		\bibfield  {author} {\bibinfo {author} {\bibfnamefont {A.~I.}\ \bibnamefont
				{Coldea}}\ and\ \bibinfo {author} {\bibfnamefont {M.~D.}\ \bibnamefont
				{Watson}},\ }\bibfield  {title} {\enquote {\bibinfo {title} {{The Key
						Ingredients of the Electronic Structure of FeSe}},}\ }\href {\doibase
			10.1146/annurev-conmatphys-033117-054137} {\bibfield  {journal} {\bibinfo
				{journal} {Annual Review of Condensed Matter Physics}\ }\textbf {\bibinfo
				{volume} {9}},\ \bibinfo {pages} {125--146} (\bibinfo {year}
			{2018})}\BibitemShut {NoStop}%
		\bibitem [{\citenamefont {Watson}\ \emph
			{et~al.}(2017{\natexlab{a}})\citenamefont {Watson}, \citenamefont
			{Haghighirad}, \citenamefont {Rhodes}, \citenamefont {Hoesch},\ and\
			\citenamefont {Kim}}]{Watson2017}%
		\BibitemOpen
		\bibfield  {author} {\bibinfo {author} {\bibfnamefont {M.~D.}\ \bibnamefont
				{Watson}}, \bibinfo {author} {\bibfnamefont {A.~A.}\ \bibnamefont
				{Haghighirad}}, \bibinfo {author} {\bibfnamefont {L.~C.}\ \bibnamefont
				{Rhodes}}, \bibinfo {author} {\bibfnamefont {M.}~\bibnamefont {Hoesch}}, \
			and\ \bibinfo {author} {\bibfnamefont {T.~K.}\ \bibnamefont {Kim}},\
		}\bibfield  {title} {\enquote {\bibinfo {title} {{Electronic anisotropies
						revealed by detwinned angle-resolved photo-emission spectroscopy measurements
						of FeSe}},}\ }\href
		{https://iopscience.iop.org/article/10.1088/1367-2630/aa8a04} {\bibfield
			{journal} {\bibinfo  {journal} {New J. Phys.}\ }\textbf {\bibinfo {volume}
				{19}},\ \bibinfo {pages} {103021} (\bibinfo {year}
			{2017}{\natexlab{a}})}\BibitemShut {NoStop}%
		\bibitem [{\citenamefont {Yi}\ \emph {et~al.}(2019)\citenamefont {Yi},
			\citenamefont {Pfau}, \citenamefont {Zhang}, \citenamefont {He},
			\citenamefont {Wu}, \citenamefont {Chen}, \citenamefont {Ye}, \citenamefont
			{Hashimoto}, \citenamefont {Yu}, \citenamefont {Si}, \citenamefont {Lee},
			\citenamefont {Dai}, \citenamefont {Shen}, \citenamefont {Lu},\ and\
			\citenamefont {Birgeneau}}]{Yi2019}%
		\BibitemOpen
		\bibfield  {author} {\bibinfo {author} {\bibfnamefont {M.}~\bibnamefont
				{Yi}}, \bibinfo {author} {\bibfnamefont {H.}~\bibnamefont {Pfau}}, \bibinfo
			{author} {\bibfnamefont {Y.}~\bibnamefont {Zhang}}, \bibinfo {author}
			{\bibfnamefont {Y.}~\bibnamefont {He}}, \bibinfo {author} {\bibfnamefont
				{H.}~\bibnamefont {Wu}}, \bibinfo {author} {\bibfnamefont {T.}~\bibnamefont
				{Chen}}, \bibinfo {author} {\bibfnamefont {Z.~R.}\ \bibnamefont {Ye}},
			\bibinfo {author} {\bibfnamefont {M.}~\bibnamefont {Hashimoto}}, \bibinfo
			{author} {\bibfnamefont {R.}~\bibnamefont {Yu}}, \bibinfo {author}
			{\bibfnamefont {Q.}~\bibnamefont {Si}}, \bibinfo {author} {\bibfnamefont
				{D.-H.}\ \bibnamefont {Lee}}, \bibinfo {author} {\bibfnamefont {Pengcheng}\
				\bibnamefont {Dai}}, \bibinfo {author} {\bibfnamefont {Z.-X.}\ \bibnamefont
				{Shen}}, \bibinfo {author} {\bibfnamefont {D.~H.}\ \bibnamefont {Lu}}, \ and\
			\bibinfo {author} {\bibfnamefont {R.~J.}\ \bibnamefont {Birgeneau}},\
		}\bibfield  {title} {\enquote {\bibinfo {title} {{Nematic Energy Scale and
						the Missing Electron Pocket in FeSe}},}\ }\href {\doibase
			10.1103/PhysRevX.9.041049} {\bibfield  {journal} {\bibinfo  {journal} {Phys.
					Rev. X}\ }\textbf {\bibinfo {volume} {9}},\ \bibinfo {pages} {041049}
			(\bibinfo {year} {2019})}\BibitemShut {NoStop}%
		\bibitem [{\citenamefont {Huh}\ \emph {et~al.}(2020)\citenamefont {Huh},
			\citenamefont {Seo}, \citenamefont {Kim}, \citenamefont {Cho}, \citenamefont
			{Jung}, \citenamefont {Kim}, \citenamefont {Koh}, \citenamefont {Kwon},
			\citenamefont {Kim}, \citenamefont {Kyung}, \citenamefont {Denlinger},
			\citenamefont {Kim}, \citenamefont {Chae}, \citenamefont {Kim}, \citenamefont
			{Kim},\ and\ \citenamefont {Kim}}]{Huh2020}%
		\BibitemOpen
		\bibfield  {author} {\bibinfo {author} {\bibfnamefont {S.~S.}\ \bibnamefont
				{Huh}}, \bibinfo {author} {\bibfnamefont {J.~J.}\ \bibnamefont {Seo}},
			\bibinfo {author} {\bibfnamefont {B.~S.}\ \bibnamefont {Kim}}, \bibinfo
			{author} {\bibfnamefont {S.~H.}\ \bibnamefont {Cho}}, \bibinfo {author}
			{\bibfnamefont {J.~K.}\ \bibnamefont {Jung}}, \bibinfo {author}
			{\bibfnamefont {S.}~\bibnamefont {Kim}}, \bibinfo {author} {\bibfnamefont
				{Y.~Y.}\ \bibnamefont {Koh}}, \bibinfo {author} {\bibfnamefont {C.~I.}\
				\bibnamefont {Kwon}}, \bibinfo {author} {\bibfnamefont {J.~S.}\ \bibnamefont
				{Kim}}, \bibinfo {author} {\bibfnamefont {W.~S.}\ \bibnamefont {Kyung}},
			\bibinfo {author} {\bibfnamefont {J.~D.}\ \bibnamefont {Denlinger}}, \bibinfo
			{author} {\bibfnamefont {Y.~H.}\ \bibnamefont {Kim}}, \bibinfo {author}
			{\bibfnamefont {B.~N.}\ \bibnamefont {Chae}}, \bibinfo {author}
			{\bibfnamefont {N.~D.}\ \bibnamefont {Kim}}, \bibinfo {author} {\bibfnamefont
				{Y.~K}\ \bibnamefont {Kim}}, \ and\ \bibinfo {author} {\bibfnamefont
				{C.}~\bibnamefont {Kim}},\ }\bibfield  {title} {\enquote {\bibinfo {title}
				{{Absence of Y-pocket in 1-Fe Brillouin zone and reversed orbital occupation
						imbalance in FeSe}},}\ }\href
		{https://www.nature.com/articles/s42005-020-0319-1} {\bibfield  {journal}
			{\bibinfo  {journal} {Commun. Phys.}\ }\textbf {\bibinfo {volume} {3}},\
			\bibinfo {pages} {52} (\bibinfo {year} {2020})}\BibitemShut {NoStop}%
		\bibitem [{\citenamefont {Cai}\ \emph {et~al.}(2020{\natexlab{a}})\citenamefont
			{Cai}, \citenamefont {Han}, \citenamefont {Wang}, \citenamefont {Chen},
			\citenamefont {Wang}, \citenamefont {Xin}, \citenamefont {Ma}, \citenamefont
			{Li},\ and\ \citenamefont {Zhang}}]{Cai2020}%
		\BibitemOpen
		\bibfield  {author} {\bibinfo {author} {\bibfnamefont {C.}~\bibnamefont
				{Cai}}, \bibinfo {author} {\bibfnamefont {T.~T.}\ \bibnamefont {Han}},
			\bibinfo {author} {\bibfnamefont {Z.~G.}\ \bibnamefont {Wang}}, \bibinfo
			{author} {\bibfnamefont {L.}~\bibnamefont {Chen}}, \bibinfo {author}
			{\bibfnamefont {Y.~D.}\ \bibnamefont {Wang}}, \bibinfo {author}
			{\bibfnamefont {Z.~M.}\ \bibnamefont {Xin}}, \bibinfo {author} {\bibfnamefont
				{M.~W.}\ \bibnamefont {Ma}}, \bibinfo {author} {\bibfnamefont {Yuan}\
				\bibnamefont {Li}}, \ and\ \bibinfo {author} {\bibfnamefont {Y.}~\bibnamefont
				{Zhang}},\ }\bibfield  {title} {\enquote {\bibinfo {title}
				{{Momentum-resolved measurement of electronic nematic susceptibility in the
						${\mathrm{FeSe}}_{0.9}{\mathrm{S}}_{0.1}$ superconductor}},}\ }\href
		{\doibase 10.1103/PhysRevB.101.180501} {\bibfield  {journal} {\bibinfo
				{journal} {Phys. Rev. B}\ }\textbf {\bibinfo {volume} {101}},\ \bibinfo
			{pages} {180501} (\bibinfo {year} {2020}{\natexlab{a}})}\BibitemShut
		{NoStop}%
		\bibitem [{\citenamefont {Cai}\ \emph {et~al.}(2020{\natexlab{b}})\citenamefont
			{Cai}, \citenamefont {Han}, \citenamefont {Wang}, \citenamefont {Chen},
			\citenamefont {Wang}, \citenamefont {Xin}, \citenamefont {Ma}, \citenamefont
			{Li},\ and\ \citenamefont {Zhang}}]{Cai2020b}%
		\BibitemOpen
		\bibfield  {author} {\bibinfo {author} {\bibfnamefont {C.}~\bibnamefont
				{Cai}}, \bibinfo {author} {\bibfnamefont {T.~T.}\ \bibnamefont {Han}},
			\bibinfo {author} {\bibfnamefont {Z.~G.}\ \bibnamefont {Wang}}, \bibinfo
			{author} {\bibfnamefont {L.}~\bibnamefont {Chen}}, \bibinfo {author}
			{\bibfnamefont {Y.~D.}\ \bibnamefont {Wang}}, \bibinfo {author}
			{\bibfnamefont {Z.~M.}\ \bibnamefont {Xin}}, \bibinfo {author} {\bibfnamefont
				{M.~W.}\ \bibnamefont {Ma}}, \bibinfo {author} {\bibfnamefont {Yuan}\
				\bibnamefont {Li}}, \ and\ \bibinfo {author} {\bibfnamefont {Y.}~\bibnamefont
				{Zhang}},\ }\bibfield  {title} {\enquote {\bibinfo {title} {{Anomalous
						spectral weight transfer in the nematic state of iron-selenide
						superconductor}},}\ }\href
		{http://cpb.iphy.ac.cn/article/2020/2034/cpb_29_7_077401.html} {\bibfield
			{journal} {\bibinfo  {journal} {Chin. Phys. B}\ }\textbf {\bibinfo {volume}
				{29}},\ \bibinfo {pages} {077401} (\bibinfo {year}
			{2020}{\natexlab{b}})}\BibitemShut {NoStop}%
		\bibitem [{\citenamefont {Pfau}\ \emph {et~al.}(2019)\citenamefont {Pfau},
			\citenamefont {Chen}, \citenamefont {Yi}, \citenamefont {Hashimoto},
			\citenamefont {Rotundu}, \citenamefont {Palmstrom}, \citenamefont {Chen},
			\citenamefont {Dai}, \citenamefont {Straquadine}, \citenamefont {Hristov},
			\citenamefont {Birgeneau}, \citenamefont {Fisher}, \citenamefont {Lu},\ and\
			\citenamefont {Shen}}]{Pfau2019}%
		\BibitemOpen
		\bibfield  {author} {\bibinfo {author} {\bibfnamefont {H.}~\bibnamefont
				{Pfau}}, \bibinfo {author} {\bibfnamefont {S.~D.}\ \bibnamefont {Chen}},
			\bibinfo {author} {\bibfnamefont {M.}~\bibnamefont {Yi}}, \bibinfo {author}
			{\bibfnamefont {M.}~\bibnamefont {Hashimoto}}, \bibinfo {author}
			{\bibfnamefont {C.~R.}\ \bibnamefont {Rotundu}}, \bibinfo {author}
			{\bibfnamefont {J.~C.}\ \bibnamefont {Palmstrom}}, \bibinfo {author}
			{\bibfnamefont {T.}~\bibnamefont {Chen}}, \bibinfo {author} {\bibfnamefont
				{P.-C.}\ \bibnamefont {Dai}}, \bibinfo {author} {\bibfnamefont
				{J.}~\bibnamefont {Straquadine}}, \bibinfo {author} {\bibfnamefont
				{A.}~\bibnamefont {Hristov}}, \bibinfo {author} {\bibfnamefont {R.~J.}\
				\bibnamefont {Birgeneau}}, \bibinfo {author} {\bibfnamefont {I.~R.}\
				\bibnamefont {Fisher}}, \bibinfo {author} {\bibfnamefont {D.}~\bibnamefont
				{Lu}}, \ and\ \bibinfo {author} {\bibfnamefont {Z.-X.}\ \bibnamefont
				{Shen}},\ }\bibfield  {title} {\enquote {\bibinfo {title} {{Momentum
						Dependence of the Nematic Order Parameter in Iron-Based Superconductors}},}\
		}\href {https://journals.aps.org/prl/abstract/10.1103/PhysRevLett.123.066402}
		{\bibfield  {journal} {\bibinfo  {journal} {Phys. Rev. Lett.}\ }\textbf
			{\bibinfo {volume} {123}},\ \bibinfo {pages} {066402} (\bibinfo {year}
			{2019})}\BibitemShut {NoStop}%
		\bibitem [{\citenamefont {Pfau}\ \emph {et~al.}(2021)\citenamefont {Pfau},
			\citenamefont {Yi}, \citenamefont {Hashimoto}, \citenamefont {Chen},
			\citenamefont {Dai}, \citenamefont {Shen}, \citenamefont {Mo},\ and\
			\citenamefont {Lu}}]{Pfau2021}%
		\BibitemOpen
		\bibfield  {author} {\bibinfo {author} {\bibfnamefont {H.}~\bibnamefont
				{Pfau}}, \bibinfo {author} {\bibfnamefont {M.}~\bibnamefont {Yi}}, \bibinfo
			{author} {\bibfnamefont {M.}~\bibnamefont {Hashimoto}}, \bibinfo {author}
			{\bibfnamefont {T.}~\bibnamefont {Chen}}, \bibinfo {author} {\bibfnamefont
				{P.-C.}\ \bibnamefont {Dai}}, \bibinfo {author} {\bibfnamefont {Z.-X.}\
				\bibnamefont {Shen}}, \bibinfo {author} {\bibfnamefont {S.-K.}\ \bibnamefont
				{Mo}}, \ and\ \bibinfo {author} {\bibfnamefont {D.}~\bibnamefont {Lu}},\
		}\bibfield  {title} {\enquote {\bibinfo {title} {{Quasiparticle coherence in
						the nematic state of FeSe}},}\ }\href {\doibase 10.1103/PhysRevB.104.L241101}
		{\bibfield  {journal} {\bibinfo  {journal} {Phys. Rev. B}\ }\textbf {\bibinfo
				{volume} {104}},\ \bibinfo {pages} {L241101} (\bibinfo {year}
			{2021})}\BibitemShut {NoStop}%
		\bibitem [{\citenamefont {Rhodes}\ \emph {et~al.}(2020)\citenamefont {Rhodes},
			\citenamefont {Watson}, \citenamefont {Haghighirad}, \citenamefont
			{Evtushinsky},\ and\ \citenamefont {Kim}}]{Rhodes2020}%
		\BibitemOpen
		\bibfield  {author} {\bibinfo {author} {\bibfnamefont {L.~C.}\ \bibnamefont
				{Rhodes}}, \bibinfo {author} {\bibfnamefont {M.~D.}\ \bibnamefont {Watson}},
			\bibinfo {author} {\bibfnamefont {A.~A.}\ \bibnamefont {Haghighirad}},
			\bibinfo {author} {\bibfnamefont {D.~V.}\ \bibnamefont {Evtushinsky}}, \ and\
			\bibinfo {author} {\bibfnamefont {T.~K.}\ \bibnamefont {Kim}},\ }\bibfield
		{title} {\enquote {\bibinfo {title} {{Revealing the single electron pocket of
						FeSe in a single orthorhombic domain}},}\ }\href {\doibase
			10.1103/PhysRevB.101.235128} {\bibfield  {journal} {\bibinfo  {journal}
				{Phys. Rev. B}\ }\textbf {\bibinfo {volume} {101}},\ \bibinfo {pages}
			{235128} (\bibinfo {year} {2020})}\BibitemShut {NoStop}%
		\bibitem [{\citenamefont {Kasahara}\ \emph {et~al.}(2014)\citenamefont
			{Kasahara}, \citenamefont {Watashige}, \citenamefont {Hanaguri},
			\citenamefont {Kohsaka}, \citenamefont {Yamashita}, \citenamefont
			{Shimoyama}, \citenamefont {Mizukami}, \citenamefont {Endo}, \citenamefont
			{Ikeda}, \citenamefont {Aoyama}, \citenamefont {Terashima}, \citenamefont
			{Uji}, \citenamefont {Wolf}, \citenamefont {von L{\"o}hneysen}, \citenamefont
			{Shibauchi},\ and\ \citenamefont {Matsuda}}]{Kasahara2014}%
		\BibitemOpen
		\bibfield  {author} {\bibinfo {author} {\bibfnamefont {S.}~\bibnamefont
				{Kasahara}}, \bibinfo {author} {\bibfnamefont {T.}~\bibnamefont {Watashige}},
			\bibinfo {author} {\bibfnamefont {T.}~\bibnamefont {Hanaguri}}, \bibinfo
			{author} {\bibfnamefont {Y.}~\bibnamefont {Kohsaka}}, \bibinfo {author}
			{\bibfnamefont {T.}~\bibnamefont {Yamashita}}, \bibinfo {author}
			{\bibfnamefont {Y.}~\bibnamefont {Shimoyama}}, \bibinfo {author}
			{\bibfnamefont {Y.}~\bibnamefont {Mizukami}}, \bibinfo {author}
			{\bibfnamefont {R.}~\bibnamefont {Endo}}, \bibinfo {author} {\bibfnamefont
				{H.}~\bibnamefont {Ikeda}}, \bibinfo {author} {\bibfnamefont
				{K.}~\bibnamefont {Aoyama}}, \bibinfo {author} {\bibfnamefont
				{T.}~\bibnamefont {Terashima}}, \bibinfo {author} {\bibfnamefont
				{S.}~\bibnamefont {Uji}}, \bibinfo {author} {\bibfnamefont {T.}~\bibnamefont
				{Wolf}}, \bibinfo {author} {\bibfnamefont {H.}~\bibnamefont {von
					L{\"o}hneysen}}, \bibinfo {author} {\bibfnamefont {T.}~\bibnamefont
				{Shibauchi}}, \ and\ \bibinfo {author} {\bibfnamefont {Y.}~\bibnamefont
				{Matsuda}},\ }\bibfield  {title} {\enquote {\bibinfo {title} {{Field-induced
						superconducting phase of FeSe in the BCS-BEC cross-over}},}\ }\href {\doibase
			10.1073/pnas.1413477111} {\bibfield  {journal} {\bibinfo  {journal}
				{Proceedings of the National Academy of Sciences}\ }\textbf {\bibinfo
				{volume} {111}},\ \bibinfo {pages} {16309--16313} (\bibinfo {year}
			{2014})}\BibitemShut {NoStop}%
		\bibitem [{\citenamefont {Hanaguri}\ \emph {et~al.}(2018)\citenamefont
			{Hanaguri}, \citenamefont {Iwaya}, \citenamefont {Kohsaka}, \citenamefont
			{Machida}, \citenamefont {Watashige}, \citenamefont {Kasahara}, \citenamefont
			{Shibauchi},\ and\ \citenamefont {Matsuda}}]{Hanaguri2018}%
		\BibitemOpen
		\bibfield  {author} {\bibinfo {author} {\bibfnamefont {T.}~\bibnamefont
				{Hanaguri}}, \bibinfo {author} {\bibfnamefont {K.}~\bibnamefont {Iwaya}},
			\bibinfo {author} {\bibfnamefont {Y.}~\bibnamefont {Kohsaka}}, \bibinfo
			{author} {\bibfnamefont {T.}~\bibnamefont {Machida}}, \bibinfo {author}
			{\bibfnamefont {T.}~\bibnamefont {Watashige}}, \bibinfo {author}
			{\bibfnamefont {S.}~\bibnamefont {Kasahara}}, \bibinfo {author}
			{\bibfnamefont {T.}~\bibnamefont {Shibauchi}}, \ and\ \bibinfo {author}
			{\bibfnamefont {Y.}~\bibnamefont {Matsuda}},\ }\bibfield  {title} {\enquote
			{\bibinfo {title} {{Two distinct superconducting pairing states divided by
						the nematic end point in
						${\mathrm{FeSe}}_{1\ensuremath{-}x}{\mathrm{S}}_{x}$}},}\ }\href
		{http://advances.sciencemag.org/content/4/5/eaar6419} {\bibfield  {journal}
			{\bibinfo  {journal} {Science Advances}\ }\textbf {\bibinfo {volume} {4}},\
			\bibinfo {pages} {eaar6419} (\bibinfo {year} {2018})}\BibitemShut {NoStop}%
		\bibitem [{\citenamefont {Kostin}\ \emph {et~al.}(2018)\citenamefont {Kostin},
			\citenamefont {Sprau}, \citenamefont {Kreisel}, \citenamefont {Chong},
			\citenamefont {B{\"{o}}hmer}, \citenamefont {Canfield}, \citenamefont
			{Hirschfeld}, \citenamefont {Andersen},\ and\ \citenamefont
			{Davis}}]{Kostin2018}%
		\BibitemOpen
		\bibfield  {author} {\bibinfo {author} {\bibfnamefont {A.}~\bibnamefont
				{Kostin}}, \bibinfo {author} {\bibfnamefont {P.~O.}\ \bibnamefont {Sprau}},
			\bibinfo {author} {\bibfnamefont {A.}~\bibnamefont {Kreisel}}, \bibinfo
			{author} {\bibfnamefont {Y-X.}\ \bibnamefont {Chong}}, \bibinfo {author}
			{\bibfnamefont {A.~E.}\ \bibnamefont {B{\"{o}}hmer}}, \bibinfo {author}
			{\bibfnamefont {P.~C.}\ \bibnamefont {Canfield}}, \bibinfo {author}
			{\bibfnamefont {P.~J.}\ \bibnamefont {Hirschfeld}}, \bibinfo {author}
			{\bibfnamefont {B.~M.}\ \bibnamefont {Andersen}}, \ and\ \bibinfo {author}
			{\bibfnamefont {J.~C.~S{\'{e}}amus}\ \bibnamefont {Davis}},\ }\bibfield
		{title} {\enquote {\bibinfo {title} {{Imaging orbital-selective
						quasiparticles in the Hund's metal state of FeSe}},}\ }\href
		{https://doi.org/10.1038/s41563-018-0151-0} {\bibfield  {journal} {\bibinfo
				{journal} {Nat. Mater.}\ }\textbf {\bibinfo {volume} {17}},\ \bibinfo {pages}
			{869--874} (\bibinfo {year} {2018})}\BibitemShut {NoStop}%
		\bibitem [{\citenamefont {Margadonna}\ \emph {et~al.}(2008)\citenamefont
			{Margadonna}, \citenamefont {Takabayashi}, \citenamefont {McDonald},
			\citenamefont {Kasperkiewicz}, \citenamefont {Mizuguchi}, \citenamefont
			{Takano}, \citenamefont {Fitch}, \citenamefont {Suard},\ and\ \citenamefont
			{Prassides}}]{Margadonna2008}%
		\BibitemOpen
		\bibfield  {author} {\bibinfo {author} {\bibfnamefont {S.}~\bibnamefont
				{Margadonna}}, \bibinfo {author} {\bibfnamefont {Y.}~\bibnamefont
				{Takabayashi}}, \bibinfo {author} {\bibfnamefont {M.}~\bibnamefont
				{McDonald}}, \bibinfo {author} {\bibfnamefont {K.}~\bibnamefont
				{Kasperkiewicz}}, \bibinfo {author} {\bibfnamefont {Y.}~\bibnamefont
				{Mizuguchi}}, \bibinfo {author} {\bibfnamefont {Y.}~\bibnamefont {Takano}},
			\bibinfo {author} {\bibfnamefont {A.}~\bibnamefont {Fitch}}, \bibinfo
			{author} {\bibfnamefont {E.}~\bibnamefont {Suard}}, \ and\ \bibinfo {author}
			{\bibfnamefont {K.}~\bibnamefont {Prassides}},\ }\bibfield  {title} {\enquote
			{\bibinfo {title} {{Crystal Structure of the new
						${\mathrm{FeSe}}_{1\ensuremath{-}x}$ Superconductor}},}\ }\href
		{https://pubs.rsc.org/en/content/articlelanding/2008/cc/b813076k} {\bibfield
			{journal} {\bibinfo  {journal} {Chem. Commun.}\ }\textbf {\bibinfo {volume}
				{4}},\ \bibinfo {pages} {5607--5609} (\bibinfo {year} {2008})}\BibitemShut
		{NoStop}%
		\bibitem [{\citenamefont {Andersen}\ and\ \citenamefont
			{Boeri}(2011)}]{AndersenBoeriGlideMirror}%
		\BibitemOpen
		\bibfield  {author} {\bibinfo {author} {\bibfnamefont {O.K.}\ \bibnamefont
				{Andersen}}\ and\ \bibinfo {author} {\bibfnamefont {L.}~\bibnamefont
				{Boeri}},\ }\bibfield  {title} {\enquote {\bibinfo {title} {On the
					multi-orbital band structure and itinerant magnetism of iron-based
					superconductors},}\ }\href {\doibase https://doi.org/10.1002/andp.201000149}
		{\bibfield  {journal} {\bibinfo  {journal} {Annalen der Physik}\ }\textbf
			{\bibinfo {volume} {523}},\ \bibinfo {pages} {8--50} (\bibinfo {year}
			{2011})}\BibitemShut {NoStop}%
		\bibitem [{\citenamefont {Eugenio}\ and\ \citenamefont
			{Vafek}(2018)}]{Eugenio2018}%
		\BibitemOpen
		\bibfield  {author} {\bibinfo {author} {\bibfnamefont {P.~M.}\ \bibnamefont
				{Eugenio}}\ and\ \bibinfo {author} {\bibfnamefont {O.}~\bibnamefont
				{Vafek}},\ }\bibfield  {title} {\enquote {\bibinfo {title} {{Classification
						of symmetry derived pairing at the $M$ point in FeSe}},}\ }\href {\doibase
			10.1103/PhysRevB.98.014503} {\bibfield  {journal} {\bibinfo  {journal} {Phys.
					Rev. B}\ }\textbf {\bibinfo {volume} {98}},\ \bibinfo {pages} {014503}
			(\bibinfo {year} {2018})}\BibitemShut {NoStop}%
		\bibitem [{\citenamefont {Borisenko}\ \emph {et~al.}(2016)\citenamefont
			{Borisenko}, \citenamefont {Evtushinsky}, \citenamefont {Liu}, \citenamefont
			{Morozov}, \citenamefont {Kappenberger}, \citenamefont {Wurmehl},
			\citenamefont {B{\"u}chner}, \citenamefont {Yaresko}, \citenamefont {Kim},
			\citenamefont {Hoesch}, \citenamefont {Wolf},\ and\ \citenamefont
			{Zhigadlo}}]{Borisenko2016}%
		\BibitemOpen
		\bibfield  {author} {\bibinfo {author} {\bibfnamefont {S.~V.}\ \bibnamefont
				{Borisenko}}, \bibinfo {author} {\bibfnamefont {D.~V.}\ \bibnamefont
				{Evtushinsky}}, \bibinfo {author} {\bibfnamefont {Z.-H.}\ \bibnamefont
				{Liu}}, \bibinfo {author} {\bibfnamefont {I.}~\bibnamefont {Morozov}},
			\bibinfo {author} {\bibfnamefont {R.}~\bibnamefont {Kappenberger}}, \bibinfo
			{author} {\bibfnamefont {S.}~\bibnamefont {Wurmehl}}, \bibinfo {author}
			{\bibfnamefont {B.}~\bibnamefont {B{\"u}chner}}, \bibinfo {author}
			{\bibfnamefont {A.~N.}\ \bibnamefont {Yaresko}}, \bibinfo {author}
			{\bibfnamefont {T.~K.}\ \bibnamefont {Kim}}, \bibinfo {author} {\bibfnamefont
				{M.}~\bibnamefont {Hoesch}}, \bibinfo {author} {\bibfnamefont
				{T.}~\bibnamefont {Wolf}}, \ and\ \bibinfo {author} {\bibfnamefont {N.~D.}\
				\bibnamefont {Zhigadlo}},\ }\bibfield  {title} {\enquote {\bibinfo {title}
				{{Direct observation of spin--orbit coupling in iron-based
						superconductors}},}\ }\href {\doibase 10.1038/nphys3594} {\bibfield
			{journal} {\bibinfo  {journal} {Nature Physics}\ }\textbf {\bibinfo {volume}
				{12}},\ \bibinfo {pages} {311--317} (\bibinfo {year} {2016})}\BibitemShut
		{NoStop}%
		\bibitem [{\citenamefont {Watson}\ \emph
			{et~al.}(2017{\natexlab{b}})\citenamefont {Watson}, \citenamefont
			{Haghighirad}, \citenamefont {Takita}, \citenamefont {Mansuer}, \citenamefont
			{Iwasawa}, \citenamefont {Schwier}, \citenamefont {Ino},\ and\ \citenamefont
			{Hoesch}}]{Watson2017b}%
		\BibitemOpen
		\bibfield  {author} {\bibinfo {author} {\bibfnamefont {M.~D.}\ \bibnamefont
				{Watson}}, \bibinfo {author} {\bibfnamefont {A.~A.}\ \bibnamefont
				{Haghighirad}}, \bibinfo {author} {\bibfnamefont {H.}~\bibnamefont {Takita}},
			\bibinfo {author} {\bibfnamefont {W.}~\bibnamefont {Mansuer}}, \bibinfo
			{author} {\bibfnamefont {H.}~\bibnamefont {Iwasawa}}, \bibinfo {author}
			{\bibfnamefont {E.~F.}\ \bibnamefont {Schwier}}, \bibinfo {author}
			{\bibfnamefont {A.}~\bibnamefont {Ino}}, \ and\ \bibinfo {author}
			{\bibfnamefont {M.}~\bibnamefont {Hoesch}},\ }\bibfield  {title} {\enquote
			{\bibinfo {title} {{Shifts and Splittings of the Hole Bands in the Nematic
						Phase of FeSe}},}\ }\href {\doibase 10.7566/JPSJ.86.053703} {\bibfield
			{journal} {\bibinfo  {journal} {Journal of the Physical Society of Japan}\
			}\textbf {\bibinfo {volume} {86}},\ \bibinfo {pages} {053703} (\bibinfo
			{year} {2017}{\natexlab{b}})}\BibitemShut {NoStop}%
		\bibitem [{\citenamefont {Day}\ \emph {et~al.}(2018)\citenamefont {Day},
			\citenamefont {Levy}, \citenamefont {Michiardi}, \citenamefont
			{Zwartsenberg}, \citenamefont {Zonno}, \citenamefont {Ji}, \citenamefont
			{Razzoli}, \citenamefont {Boschini}, \citenamefont {Chi}, \citenamefont
			{Liang}, \citenamefont {Das}, \citenamefont {Vobornik}, \citenamefont
			{Fujii}, \citenamefont {Hardy}, \citenamefont {Bonn}, \citenamefont
			{Elfimov},\ and\ \citenamefont {Damascelli}}]{Day2018}%
		\BibitemOpen
		\bibfield  {author} {\bibinfo {author} {\bibfnamefont {R.~P.}\ \bibnamefont
				{Day}}, \bibinfo {author} {\bibfnamefont {G.}~\bibnamefont {Levy}}, \bibinfo
			{author} {\bibfnamefont {M.}~\bibnamefont {Michiardi}}, \bibinfo {author}
			{\bibfnamefont {B.}~\bibnamefont {Zwartsenberg}}, \bibinfo {author}
			{\bibfnamefont {M.}~\bibnamefont {Zonno}}, \bibinfo {author} {\bibfnamefont
				{F.}~\bibnamefont {Ji}}, \bibinfo {author} {\bibfnamefont {E.}~\bibnamefont
				{Razzoli}}, \bibinfo {author} {\bibfnamefont {F.}~\bibnamefont {Boschini}},
			\bibinfo {author} {\bibfnamefont {S.}~\bibnamefont {Chi}}, \bibinfo {author}
			{\bibfnamefont {R.}~\bibnamefont {Liang}}, \bibinfo {author} {\bibfnamefont
				{P.~K.}\ \bibnamefont {Das}}, \bibinfo {author} {\bibfnamefont
				{I.}~\bibnamefont {Vobornik}}, \bibinfo {author} {\bibfnamefont
				{J.}~\bibnamefont {Fujii}}, \bibinfo {author} {\bibfnamefont {W.~N.}\
				\bibnamefont {Hardy}}, \bibinfo {author} {\bibfnamefont {D.~A.}\ \bibnamefont
				{Bonn}}, \bibinfo {author} {\bibfnamefont {I.~S.}\ \bibnamefont {Elfimov}}, \
			and\ \bibinfo {author} {\bibfnamefont {A.}~\bibnamefont {Damascelli}},\
		}\bibfield  {title} {\enquote {\bibinfo {title} {{Influence of Spin-Orbit
						Coupling in Iron-Based Superconductors}},}\ }\href {\doibase
			10.1103/PhysRevLett.121.076401} {\bibfield  {journal} {\bibinfo  {journal}
				{Phys. Rev. Lett.}\ }\textbf {\bibinfo {volume} {121}},\ \bibinfo {pages}
			{076401} (\bibinfo {year} {2018})}\BibitemShut {NoStop}%
		\bibitem [{\citenamefont {Rhodes}\ \emph {et~al.}(2017)\citenamefont {Rhodes},
			\citenamefont {Watson}, \citenamefont {Haghighirad}, \citenamefont
			{Eschrig},\ and\ \citenamefont {Kim}}]{Rhodes2017}%
		\BibitemOpen
		\bibfield  {author} {\bibinfo {author} {\bibfnamefont {L.~C.}\ \bibnamefont
				{Rhodes}}, \bibinfo {author} {\bibfnamefont {M.~D.}\ \bibnamefont {Watson}},
			\bibinfo {author} {\bibfnamefont {A.~A.}\ \bibnamefont {Haghighirad}},
			\bibinfo {author} {\bibfnamefont {M.}~\bibnamefont {Eschrig}}, \ and\
			\bibinfo {author} {\bibfnamefont {T.~K.}\ \bibnamefont {Kim}},\ }\bibfield
		{title} {\enquote {\bibinfo {title} {{Strongly enhanced temperature
						dependence of the chemical potential in FeSe}},}\ }\href {\doibase
			10.1103/PhysRevB.95.195111} {\bibfield  {journal} {\bibinfo  {journal} {Phys.
					Rev. B}\ }\textbf {\bibinfo {volume} {95}},\ \bibinfo {pages} {195111}
			(\bibinfo {year} {2017})}\BibitemShut {NoStop}%
		\bibitem [{\citenamefont {Zhang}\ \emph {et~al.}(2012)\citenamefont {Zhang},
			\citenamefont {He}, \citenamefont {Ye}, \citenamefont {Jiang}, \citenamefont
			{Chen}, \citenamefont {Xu}, \citenamefont {Ge}, \citenamefont {Xie},
			\citenamefont {Wei}, \citenamefont {Aeschlimann}, \citenamefont {Cui},
			\citenamefont {Shi}, \citenamefont {Hu},\ and\ \citenamefont
			{Feng}}]{Zhang2012}%
		\BibitemOpen
		\bibfield  {author} {\bibinfo {author} {\bibfnamefont {Y.}~\bibnamefont
				{Zhang}}, \bibinfo {author} {\bibfnamefont {C.}~\bibnamefont {He}}, \bibinfo
			{author} {\bibfnamefont {Z.~R.}\ \bibnamefont {Ye}}, \bibinfo {author}
			{\bibfnamefont {J.}~\bibnamefont {Jiang}}, \bibinfo {author} {\bibfnamefont
				{F.}~\bibnamefont {Chen}}, \bibinfo {author} {\bibfnamefont {M.}~\bibnamefont
				{Xu}}, \bibinfo {author} {\bibfnamefont {Q.~Q.}\ \bibnamefont {Ge}}, \bibinfo
			{author} {\bibfnamefont {B.~P.}\ \bibnamefont {Xie}}, \bibinfo {author}
			{\bibfnamefont {J.}~\bibnamefont {Wei}}, \bibinfo {author} {\bibfnamefont
				{M.}~\bibnamefont {Aeschlimann}}, \bibinfo {author} {\bibfnamefont {X.~Y.}\
				\bibnamefont {Cui}}, \bibinfo {author} {\bibfnamefont {M.}~\bibnamefont
				{Shi}}, \bibinfo {author} {\bibfnamefont {J.~P.}\ \bibnamefont {Hu}}, \ and\
			\bibinfo {author} {\bibfnamefont {D.~L.}\ \bibnamefont {Feng}},\ }\bibfield
		{title} {\enquote {\bibinfo {title} {{Symmetry breaking via orbital-dependent
						reconstruction of electronic structure in detwinned NaFeAs}},}\ }\href
		{\doibase 10.1103/PhysRevB.85.085121} {\bibfield  {journal} {\bibinfo
				{journal} {Phys. Rev. B}\ }\textbf {\bibinfo {volume} {85}},\ \bibinfo
			{pages} {085121} (\bibinfo {year} {2012})}\BibitemShut {NoStop}%
		\bibitem [{\citenamefont {Eschrig}\ and\ \citenamefont
			{Koepernik}(2009)}]{Eschrig2009}%
		\BibitemOpen
		\bibfield  {author} {\bibinfo {author} {\bibfnamefont {H.}~\bibnamefont
				{Eschrig}}\ and\ \bibinfo {author} {\bibfnamefont {K.}~\bibnamefont
				{Koepernik}},\ }\bibfield  {title} {\enquote {\bibinfo {title}
				{{Tight-binding models for the iron-based superconductors}},}\ }\href
		{\doibase 10.1103/PhysRevB.80.104503} {\bibfield  {journal} {\bibinfo
				{journal} {Phys. Rev. B}\ }\textbf {\bibinfo {volume} {80}},\ \bibinfo
			{pages} {104503} (\bibinfo {year} {2009})}\BibitemShut {NoStop}%
		\bibitem [{\citenamefont {Acharya}\ \emph {et~al.}(2021)\citenamefont
			{Acharya}, \citenamefont {Pashov}, \citenamefont {Jamet},\ and\ \citenamefont
			{van Schilfgaarde}}]{Acharya2021}%
		\BibitemOpen
		\bibfield  {author} {\bibinfo {author} {\bibfnamefont {S.}~\bibnamefont
				{Acharya}}, \bibinfo {author} {\bibfnamefont {D.}~\bibnamefont {Pashov}},
			\bibinfo {author} {\bibfnamefont {F.}~\bibnamefont {Jamet}}, \ and\ \bibinfo
			{author} {\bibfnamefont {M.}~\bibnamefont {van Schilfgaarde}},\ }\bibfield
		{title} {\enquote {\bibinfo {title} {{Electronic Origin of Tc in Bulk and
						Monolayer FeSe}},}\ }\href {\doibase 10.3390/sym13020169} {\bibfield
			{journal} {\bibinfo  {journal} {Symmetry}\ }\textbf {\bibinfo {volume} {13}}
			(\bibinfo {year} {2021}),\ 10.3390/sym13020169}\BibitemShut {NoStop}%
		\bibitem [{\citenamefont {Watson}\ \emph
			{et~al.}(2015{\natexlab{b}})\citenamefont {Watson}, \citenamefont {Kim},
			\citenamefont {Haghighirad}, \citenamefont {Blake}, \citenamefont {Davies},
			\citenamefont {Hoesch}, \citenamefont {Wolf},\ and\ \citenamefont
			{Coldea}}]{Watson2015b}%
		\BibitemOpen
		\bibfield  {author} {\bibinfo {author} {\bibfnamefont {M.~D.}\ \bibnamefont
				{Watson}}, \bibinfo {author} {\bibfnamefont {T.~K.}\ \bibnamefont {Kim}},
			\bibinfo {author} {\bibfnamefont {A.~A.}\ \bibnamefont {Haghighirad}},
			\bibinfo {author} {\bibfnamefont {S.~F.}\ \bibnamefont {Blake}}, \bibinfo
			{author} {\bibfnamefont {N.~R.}\ \bibnamefont {Davies}}, \bibinfo {author}
			{\bibfnamefont {M.}~\bibnamefont {Hoesch}}, \bibinfo {author} {\bibfnamefont
				{T.}~\bibnamefont {Wolf}}, \ and\ \bibinfo {author} {\bibfnamefont {A.~I.}\
				\bibnamefont {Coldea}},\ }\bibfield  {title} {\enquote {\bibinfo {title}
				{{Suppression of orbital ordering by chemical pressure in
						${\mathrm{FeSe}}_{1\ensuremath{-}x}{\mathrm{S}}_{x}$}},}\ }\href {\doibase
			10.1103/PhysRevB.92.121108} {\bibfield  {journal} {\bibinfo  {journal} {Phys.
					Rev. B}\ }\textbf {\bibinfo {volume} {92}},\ \bibinfo {pages} {121108}
			(\bibinfo {year} {2015}{\natexlab{b}})}\BibitemShut {NoStop}%
		\bibitem [{\citenamefont {Fernandes}\ and\ \citenamefont
			{Vafek}(2014)}]{Fernandes2014b}%
		\BibitemOpen
		\bibfield  {author} {\bibinfo {author} {\bibfnamefont {R.~M.}\ \bibnamefont
				{Fernandes}}\ and\ \bibinfo {author} {\bibfnamefont {O.}~\bibnamefont
				{Vafek}},\ }\bibfield  {title} {\enquote {\bibinfo {title} {{Distinguishing
						spin-orbit coupling and nematic order in the electronic spectrum of
						iron-based superconductors}},}\ }\href {\doibase 10.1103/PhysRevB.90.214514}
		{\bibfield  {journal} {\bibinfo  {journal} {Phys. Rev. B}\ }\textbf {\bibinfo
				{volume} {90}},\ \bibinfo {pages} {214514} (\bibinfo {year}
			{2014})}\BibitemShut {NoStop}%
		\bibitem [{\citenamefont {Yin}\ \emph {et~al.}(2011)\citenamefont {Yin},
			\citenamefont {Haule},\ and\ \citenamefont {Kotliar}}]{Yin2011}%
		\BibitemOpen
		\bibfield  {author} {\bibinfo {author} {\bibfnamefont {Z.~P.}\ \bibnamefont
				{Yin}}, \bibinfo {author} {\bibfnamefont {K.}~\bibnamefont {Haule}}, \ and\
			\bibinfo {author} {\bibfnamefont {G.}~\bibnamefont {Kotliar}},\ }\bibfield
		{title} {\enquote {\bibinfo {title} {{Kinetic frustration and the nature of
						the magnetic and paramagnetic states in iron pnictides and
						iron chalcogenides}},}\ }\href {\doibase 10.1038/nmat3120} {\bibfield
			{journal} {\bibinfo  {journal} {Nature Materials}\ }\textbf {\bibinfo
				{volume} {10}},\ \bibinfo {pages} {932--935} (\bibinfo {year}
			{2011})}\BibitemShut {NoStop}%
		\bibitem [{\citenamefont {Yi}\ \emph {et~al.}(2015)\citenamefont {Yi},
			\citenamefont {Liu}, \citenamefont {Zhang}, \citenamefont {Yu}, \citenamefont
			{Zhu}, \citenamefont {Lee}, \citenamefont {Moore}, \citenamefont {Schmitt},
			\citenamefont {Li}, \citenamefont {Riggs}, \citenamefont {Chu}, \citenamefont
			{Lv}, \citenamefont {Hu}, \citenamefont {Hashimoto}, \citenamefont {Mo},
			\citenamefont {Hussain}, \citenamefont {Mao}, \citenamefont {Chu},
			\citenamefont {Fisher}, \citenamefont {Si}, \citenamefont {Shen},\ and\
			\citenamefont {Lu}}]{Yi2015}%
		\BibitemOpen
		\bibfield  {author} {\bibinfo {author} {\bibfnamefont {M.}~\bibnamefont
				{Yi}}, \bibinfo {author} {\bibfnamefont {Z.-K.}\ \bibnamefont {Liu}},
			\bibinfo {author} {\bibfnamefont {Y.}~\bibnamefont {Zhang}}, \bibinfo
			{author} {\bibfnamefont {R.}~\bibnamefont {Yu}}, \bibinfo {author}
			{\bibfnamefont {J.-X.}\ \bibnamefont {Zhu}}, \bibinfo {author} {\bibfnamefont
				{J.~J.}\ \bibnamefont {Lee}}, \bibinfo {author} {\bibfnamefont {R.~G.}\
				\bibnamefont {Moore}}, \bibinfo {author} {\bibfnamefont {F.~T.}\ \bibnamefont
				{Schmitt}}, \bibinfo {author} {\bibfnamefont {W.}~\bibnamefont {Li}},
			\bibinfo {author} {\bibfnamefont {S.~C.}\ \bibnamefont {Riggs}}, \bibinfo
			{author} {\bibfnamefont {J.-H.}\ \bibnamefont {Chu}}, \bibinfo {author}
			{\bibfnamefont {B.}~\bibnamefont {Lv}}, \bibinfo {author} {\bibfnamefont
				{J.}~\bibnamefont {Hu}}, \bibinfo {author} {\bibfnamefont {M.}~\bibnamefont
				{Hashimoto}}, \bibinfo {author} {\bibfnamefont {S.-K.}\ \bibnamefont {Mo}},
			\bibinfo {author} {\bibfnamefont {Z.}~\bibnamefont {Hussain}}, \bibinfo
			{author} {\bibfnamefont {Z.~Q.}\ \bibnamefont {Mao}}, \bibinfo {author}
			{\bibfnamefont {C.~W.}\ \bibnamefont {Chu}}, \bibinfo {author} {\bibfnamefont
				{I.~R.}\ \bibnamefont {Fisher}}, \bibinfo {author} {\bibfnamefont
				{Q.}~\bibnamefont {Si}}, \bibinfo {author} {\bibfnamefont {Z.-X.}\
				\bibnamefont {Shen}}, \ and\ \bibinfo {author} {\bibfnamefont {D.~H.}\
				\bibnamefont {Lu}},\ }\bibfield  {title} {\enquote {\bibinfo {title}
				{{Observation of universal strong orbital-dependent correlation effects in
						iron chalcogenides}},}\ }\href {\doibase 10.1038/ncomms8777} {\bibfield
			{journal} {\bibinfo  {journal} {Nature Communications}\ }\textbf {\bibinfo
				{volume} {6}},\ \bibinfo {pages} {7777} (\bibinfo {year} {2015})}\BibitemShut
		{NoStop}%
		\bibitem [{\citenamefont {Watson}\ \emph
			{et~al.}(2017{\natexlab{c}})\citenamefont {Watson}, \citenamefont {Backes},
			\citenamefont {Haghighirad}, \citenamefont {Hoesch}, \citenamefont {Kim},
			\citenamefont {Coldea},\ and\ \citenamefont {Valent\'{\i}}}]{Watson2017c}%
		\BibitemOpen
		\bibfield  {author} {\bibinfo {author} {\bibfnamefont {M.~D.}\ \bibnamefont
				{Watson}}, \bibinfo {author} {\bibfnamefont {S.}~\bibnamefont {Backes}},
			\bibinfo {author} {\bibfnamefont {A.~A.}\ \bibnamefont {Haghighirad}},
			\bibinfo {author} {\bibfnamefont {M.}~\bibnamefont {Hoesch}}, \bibinfo
			{author} {\bibfnamefont {T.~K.}\ \bibnamefont {Kim}}, \bibinfo {author}
			{\bibfnamefont {A.~I.}\ \bibnamefont {Coldea}}, \ and\ \bibinfo {author}
			{\bibfnamefont {R.}~\bibnamefont {Valent\'{\i}}},\ }\bibfield  {title}
		{\enquote {\bibinfo {title} {{Formation of Hubbard-like bands as a
						fingerprint of strong electron-electron interactions in FeSe}},}\ }\href
		{\doibase 10.1103/PhysRevB.95.081106} {\bibfield  {journal} {\bibinfo
				{journal} {Phys. Rev. B}\ }\textbf {\bibinfo {volume} {95}},\ \bibinfo
			{pages} {081106} (\bibinfo {year} {2017}{\natexlab{c}})}\BibitemShut
		{NoStop}%
		\bibitem [{\citenamefont {Evtushinsky}\ \emph {et~al.}(2017)\citenamefont
			{Evtushinsky}, \citenamefont {Aichhorn}, \citenamefont {Sassa}, \citenamefont
			{Liu}, \citenamefont {Maletz}, \citenamefont {Wolf}, \citenamefont {Yaresko},
			\citenamefont {Biermann}, \citenamefont {Borisenko},\ and\ \citenamefont
			{Buchner}}]{Evtushinsky2017_arXiv}%
		\BibitemOpen
		\bibfield  {author} {\bibinfo {author} {\bibfnamefont {D.~V.}\ \bibnamefont
				{Evtushinsky}}, \bibinfo {author} {\bibfnamefont {M.}~\bibnamefont
				{Aichhorn}}, \bibinfo {author} {\bibfnamefont {Y.}~\bibnamefont {Sassa}},
			\bibinfo {author} {\bibfnamefont {Z.-H.}\ \bibnamefont {Liu}}, \bibinfo
			{author} {\bibfnamefont {J.}~\bibnamefont {Maletz}}, \bibinfo {author}
			{\bibfnamefont {T.}~\bibnamefont {Wolf}}, \bibinfo {author} {\bibfnamefont
				{A.~N.}\ \bibnamefont {Yaresko}}, \bibinfo {author} {\bibfnamefont
				{S.}~\bibnamefont {Biermann}}, \bibinfo {author} {\bibfnamefont {S.~V.}\
				\bibnamefont {Borisenko}}, \ and\ \bibinfo {author} {\bibfnamefont
				{B.}~\bibnamefont {Buchner}},\ }\bibfield  {title} {\enquote {\bibinfo
				{title} {{Direct observation of dispersive lower Hubbard band in iron-based
						superconductor FeSe}},}\ }\href {https://arxiv.org/abs/1612.02313} {\bibfield
			{journal} {\bibinfo  {journal} {arXiv:1612.02313}\ } (\bibinfo {year}
			{2017})}\BibitemShut {NoStop}%
		\bibitem [{\citenamefont {Glasbrenner}\ \emph {et~al.}(2015)\citenamefont
			{Glasbrenner}, \citenamefont {Mazin}, \citenamefont {Jeschke}, \citenamefont
			{Hirschfeld}, \citenamefont {Fernandes},\ and\ \citenamefont
			{Valenti}}]{Glasbrenner2015}%
		\BibitemOpen
		\bibfield  {author} {\bibinfo {author} {\bibfnamefont {J.~K.}\ \bibnamefont
				{Glasbrenner}}, \bibinfo {author} {\bibfnamefont {I.~I.}\ \bibnamefont
				{Mazin}}, \bibinfo {author} {\bibfnamefont {Harald~O.}\ \bibnamefont
				{Jeschke}}, \bibinfo {author} {\bibfnamefont {P.~J.}\ \bibnamefont
				{Hirschfeld}}, \bibinfo {author} {\bibfnamefont {R.~M.}\ \bibnamefont
				{Fernandes}}, \ and\ \bibinfo {author} {\bibfnamefont {R.}~\bibnamefont
				{Valent{\'i}}},\ }\bibfield  {title} {\enquote {\bibinfo {title} {{Effect of
						magnetic frustration on nematicity and superconductivity in iron
						chalcogenides}},}\ }\href {\doibase 10.1038/nphys3434} {\bibfield  {journal}
			{\bibinfo  {journal} {Nature Physics}\ }\textbf {\bibinfo {volume} {11}},\
			\bibinfo {pages} {953--958} (\bibinfo {year} {2015})}\BibitemShut {NoStop}%
		\bibitem [{\citenamefont {He}\ \emph {et~al.}(2018)\citenamefont {He},
			\citenamefont {Wang}, \citenamefont {Hardy}, \citenamefont {Xu},
			\citenamefont {Wolf}, \citenamefont {Adelmann},\ and\ \citenamefont
			{Meingast}}]{He2018}%
		\BibitemOpen
		\bibfield  {author} {\bibinfo {author} {\bibfnamefont {M.}~\bibnamefont
				{He}}, \bibinfo {author} {\bibfnamefont {L.}~\bibnamefont {Wang}}, \bibinfo
			{author} {\bibfnamefont {F.}~\bibnamefont {Hardy}}, \bibinfo {author}
			{\bibfnamefont {L.}~\bibnamefont {Xu}}, \bibinfo {author} {\bibfnamefont
				{T.}~\bibnamefont {Wolf}}, \bibinfo {author} {\bibfnamefont {P.}~\bibnamefont
				{Adelmann}}, \ and\ \bibinfo {author} {\bibfnamefont {C.}~\bibnamefont
				{Meingast}},\ }\bibfield  {title} {\enquote {\bibinfo {title} {{Evidence for
						short-range magnetic order in the nematic phase of FeSe from anisotropic
						in-plane magnetostriction and susceptibility measurements}},}\ }\href
		{\doibase 10.1103/PhysRevB.97.104107} {\bibfield  {journal} {\bibinfo
				{journal} {Phys. Rev. B}\ }\textbf {\bibinfo {volume} {97}},\ \bibinfo
			{pages} {104107} (\bibinfo {year} {2018})}\BibitemShut {NoStop}%
		\bibitem [{\citenamefont {Chen}\ \emph {et~al.}(2019)\citenamefont {Chen},
			\citenamefont {Chen}, \citenamefont {Kreisel}, \citenamefont {Lu},
			\citenamefont {Schneidewind}, \citenamefont {Qiu}, \citenamefont {Park},
			\citenamefont {Perring}, \citenamefont {Stewart}, \citenamefont {Cao},
			\citenamefont {Zhang}, \citenamefont {Li}, \citenamefont {Rong},
			\citenamefont {Wei}, \citenamefont {Andersen}, \citenamefont {Hirschfeld},
			\citenamefont {Broholm},\ and\ \citenamefont {Dai}}]{Chen2019}%
		\BibitemOpen
		\bibfield  {author} {\bibinfo {author} {\bibfnamefont {T.}~\bibnamefont
				{Chen}}, \bibinfo {author} {\bibfnamefont {Y.}~\bibnamefont {Chen}}, \bibinfo
			{author} {\bibfnamefont {A.}~\bibnamefont {Kreisel}}, \bibinfo {author}
			{\bibfnamefont {X.}~\bibnamefont {Lu}}, \bibinfo {author} {\bibfnamefont
				{A.}~\bibnamefont {Schneidewind}}, \bibinfo {author} {\bibfnamefont
				{Y.}~\bibnamefont {Qiu}}, \bibinfo {author} {\bibfnamefont {J.~T.}\
				\bibnamefont {Park}}, \bibinfo {author} {\bibfnamefont {T.~G.}\ \bibnamefont
				{Perring}}, \bibinfo {author} {\bibfnamefont {J.~R.}\ \bibnamefont
				{Stewart}}, \bibinfo {author} {\bibfnamefont {H.}~\bibnamefont {Cao}},
			\bibinfo {author} {\bibfnamefont {R.}~\bibnamefont {Zhang}}, \bibinfo
			{author} {\bibfnamefont {Y.}~\bibnamefont {Li}}, \bibinfo {author}
			{\bibfnamefont {Y.}~\bibnamefont {Rong}}, \bibinfo {author} {\bibfnamefont
				{Y.}~\bibnamefont {Wei}}, \bibinfo {author} {\bibfnamefont {B.~M.}\
				\bibnamefont {Andersen}}, \bibinfo {author} {\bibfnamefont {P.~J.}\
				\bibnamefont {Hirschfeld}}, \bibinfo {author} {\bibfnamefont
				{C.}~\bibnamefont {Broholm}}, \ and\ \bibinfo {author} {\bibfnamefont
				{P.}~\bibnamefont {Dai}},\ }\bibfield  {title} {\enquote {\bibinfo {title}
				{{Anisotropic spin fluctuations in detwinned FeSe}},}\ }\href {\doibase
			10.1038/s41563-019-0369-5} {\bibfield  {journal} {\bibinfo  {journal} {Nature
					Materials}\ }\textbf {\bibinfo {volume} {18}},\ \bibinfo {pages} {709--716}
			(\bibinfo {year} {2019})}\BibitemShut {NoStop}%
		\bibitem [{\citenamefont {Wang}\ \emph {et~al.}(2020)\citenamefont {Wang},
			\citenamefont {Zhao}, \citenamefont {Koch}, \citenamefont {Billinge},\ and\
			\citenamefont {Zunger}}]{Wang2020}%
		\BibitemOpen
		\bibfield  {author} {\bibinfo {author} {\bibfnamefont {Z.}~\bibnamefont
				{Wang}}, \bibinfo {author} {\bibfnamefont {X-G.}\ \bibnamefont {Zhao}},
			\bibinfo {author} {\bibfnamefont {R.}~\bibnamefont {Koch}}, \bibinfo {author}
			{\bibfnamefont {S.~J.~L.}\ \bibnamefont {Billinge}}, \ and\ \bibinfo {author}
			{\bibfnamefont {A.}~\bibnamefont {Zunger}},\ }\bibfield  {title} {\enquote
			{\bibinfo {title} {{Understanding electronic peculiarities in tetragonal FeSe
						as local structural symmetry breaking}},}\ }\href {\doibase
			10.1103/PhysRevB.102.235121} {\bibfield  {journal} {\bibinfo  {journal}
				{Phys. Rev. B}\ }\textbf {\bibinfo {volume} {102}},\ \bibinfo {pages}
			{235121} (\bibinfo {year} {2020})}\BibitemShut {NoStop}%
		\bibitem [{\citenamefont {Gorni}\ \emph {et~al.}(2021)\citenamefont {Gorni},
			\citenamefont {Villar~Arribi}, \citenamefont {Casula},\ and\ \citenamefont
			{de' Medici}}]{Gorni2021}%
		\BibitemOpen
		\bibfield  {author} {\bibinfo {author} {\bibfnamefont {T.}~\bibnamefont
				{Gorni}}, \bibinfo {author} {\bibfnamefont {P.}~\bibnamefont
				{Villar~Arribi}}, \bibinfo {author} {\bibfnamefont {M.}~\bibnamefont
				{Casula}}, \ and\ \bibinfo {author} {\bibfnamefont {L.}~\bibnamefont {de'
					Medici}},\ }\bibfield  {title} {\enquote {\bibinfo {title} {Accurate modeling
					of fese with screened fock exchange and hund metal correlations},}\ }\href
		{\doibase 10.1103/PhysRevB.104.014507} {\bibfield  {journal} {\bibinfo
				{journal} {Phys. Rev. B}\ }\textbf {\bibinfo {volume} {104}},\ \bibinfo
			{pages} {014507} (\bibinfo {year} {2021})}\BibitemShut {NoStop}%
		\bibitem [{\citenamefont {Mukherjee}\ \emph {et~al.}(2015)\citenamefont
			{Mukherjee}, \citenamefont {Kreisel}, \citenamefont {Hirschfeld},\ and\
			\citenamefont {Andersen}}]{Mukherjee2015}%
		\BibitemOpen
		\bibfield  {author} {\bibinfo {author} {\bibfnamefont {S.}~\bibnamefont
				{Mukherjee}}, \bibinfo {author} {\bibfnamefont {A.}~\bibnamefont {Kreisel}},
			\bibinfo {author} {\bibfnamefont {P. J.}\ \bibnamefont {Hirschfeld}}, \
			and\ \bibinfo {author} {\bibfnamefont {B.~M.}\ \bibnamefont {Andersen}},\
		}\bibfield  {title} {\enquote {\bibinfo {title} {{Model of Electronic
						Structure and Superconductivity in Orbitally Ordered FeSe}},}\ }\href
		{\doibase 10.1103/PhysRevLett.115.026402} {\bibfield  {journal} {\bibinfo
				{journal} {Phys. Rev. Lett.}\ }\textbf {\bibinfo {volume} {115}},\ \bibinfo
			{pages} {026402} (\bibinfo {year} {2015})}\BibitemShut {NoStop}%
		\bibitem [{\citenamefont {Rhodes}\ \emph {et~al.}(2021)\citenamefont {Rhodes},
			\citenamefont {B{\"o}ker}, \citenamefont {M{\"u}ller}, \citenamefont
			{Eschrig},\ and\ \citenamefont {Eremin}}]{Rhodes2021}%
		\BibitemOpen
		\bibfield  {author} {\bibinfo {author} {\bibfnamefont {L.~C.}\ \bibnamefont
				{Rhodes}}, \bibinfo {author} {\bibfnamefont {J.}~\bibnamefont {B{\"o}ker}},
			\bibinfo {author} {\bibfnamefont {M.~A.}\ \bibnamefont {M{\"u}ller}},
			\bibinfo {author} {\bibfnamefont {M.}~\bibnamefont {Eschrig}}, \ and\
			\bibinfo {author} {\bibfnamefont {I.~M.}\ \bibnamefont {Eremin}},\ }\bibfield
		{title} {\enquote {\bibinfo {title} {{Non-local dxy nematicity and the
						missing electron pocket in FeSe}},}\ }\href {\doibase
			10.1038/s41535-021-00341-6} {\bibfield  {journal} {\bibinfo  {journal} {npj
					Quantum Materials}\ }\textbf {\bibinfo {volume} {6}},\ \bibinfo {pages} {45}
			(\bibinfo {year} {2021})}\BibitemShut {NoStop}%
		\bibitem [{\citenamefont {Kushnirenko}\ \emph {et~al.}(2017)\citenamefont
			{Kushnirenko}, \citenamefont {Kordyuk}, \citenamefont {Fedorov},
			\citenamefont {Haubold}, \citenamefont {Wolf}, \citenamefont {B\"uchner},\
			and\ \citenamefont {Borisenko}}]{Kushnirenko2017}%
		\BibitemOpen
		\bibfield  {author} {\bibinfo {author} {\bibfnamefont {Y.~S.}\ \bibnamefont
				{Kushnirenko}}, \bibinfo {author} {\bibfnamefont {A.~A.}\ \bibnamefont
				{Kordyuk}}, \bibinfo {author} {\bibfnamefont {A.~V.}\ \bibnamefont
				{Fedorov}}, \bibinfo {author} {\bibfnamefont {E.}~\bibnamefont {Haubold}},
			\bibinfo {author} {\bibfnamefont {T.}~\bibnamefont {Wolf}}, \bibinfo {author}
			{\bibfnamefont {B.}~\bibnamefont {B\"uchner}}, \ and\ \bibinfo {author}
			{\bibfnamefont {S.~V.}\ \bibnamefont {Borisenko}},\ }\bibfield  {title}
		{\enquote {\bibinfo {title} {{Anomalous temperature evolution of the
						electronic structure of FeSe}},}\ }\href {\doibase
			10.1103/PhysRevB.96.100504} {\bibfield  {journal} {\bibinfo  {journal} {Phys.
					Rev. B}\ }\textbf {\bibinfo {volume} {96}},\ \bibinfo {pages} {100504}
			(\bibinfo {year} {2017})}\BibitemShut {NoStop}%
		\bibitem [{\citenamefont {Pustovit}\ \emph {et~al.}(2017)\citenamefont
			{Pustovit}, \citenamefont {Bezguba},\ and\ \citenamefont
			{Kordyuk}}]{Pustovit2017}%
		\BibitemOpen
		\bibfield  {author} {\bibinfo {author} {\bibfnamefont {Y.~V.}\ \bibnamefont
				{Pustovit}}, \bibinfo {author} {\bibfnamefont {V.~V.}\ \bibnamefont
				{Bezguba}}, \ and\ \bibinfo {author} {\bibfnamefont {A.~A.}\ \bibnamefont
				{Kordyuk}},\ }\bibfield  {title} {\enquote {\bibinfo {title} {{Temperature
						Dependence of the Electronic Structure of FeSe}},}\ }\href {\doibase
			10.15407/mfint.39.06.0709} {\bibfield  {journal} {\bibinfo  {journal}
				{Metallofiz. Noveishie Tekhnol.}\ }\textbf {\bibinfo {volume} {39}},\
			\bibinfo {pages} {709--718} (\bibinfo {year} {2017})}\BibitemShut {NoStop}%
		\bibitem [{\citenamefont {Pustovit}\ \emph {et~al.}(2018)\citenamefont
			{Pustovit}, \citenamefont {Brouet}, \citenamefont {Chareev},\ and\
			\citenamefont {Kordyuk}}]{Pustovit2018}%
		\BibitemOpen
		\bibfield  {author} {\bibinfo {author} {\bibfnamefont {Y..~V.}\ \bibnamefont
				{Pustovit}}, \bibinfo {author} {\bibfnamefont {V.}~\bibnamefont {Brouet}},
			\bibinfo {author} {\bibfnamefont {D.~A.}\ \bibnamefont {Chareev}}, \ and\
			\bibinfo {author} {\bibfnamefont {A.~A.}\ \bibnamefont {Kordyuk}},\
		}\bibfield  {title} {\enquote {\bibinfo {title} {{Temperature Evolution of
						Charge Carrier Density in the Centre of the Brillouin Zone of Fe(Se,Te)
						Superconductor}},}\ }\href {\doibase 10.15407/mfint.40.02.0139} {\bibfield
			{journal} {\bibinfo  {journal} {Metallofiz. Noveishie Tekhnol.}\ }\textbf
			{\bibinfo {volume} {40}},\ \bibinfo {pages} {139--146} (\bibinfo {year}
			{2018})}\BibitemShut {NoStop}%
		\bibitem [{\citenamefont {Schwier}\ \emph {et~al.}(2019)\citenamefont
			{Schwier}, \citenamefont {Takita}, \citenamefont {Mansur}, \citenamefont
			{Ino}, \citenamefont {Hoesch}, \citenamefont {Watson}, \citenamefont
			{Haghighirad},\ and\ \citenamefont {Shimada}}]{Schwier2019}%
		\BibitemOpen
		\bibfield  {author} {\bibinfo {author} {\bibfnamefont {E.~F.}\ \bibnamefont
				{Schwier}}, \bibinfo {author} {\bibfnamefont {H.}~\bibnamefont {Takita}},
			\bibinfo {author} {\bibfnamefont {W.}~\bibnamefont {Mansur}}, \bibinfo
			{author} {\bibfnamefont {A.}~\bibnamefont {Ino}}, \bibinfo {author}
			{\bibfnamefont {M.}~\bibnamefont {Hoesch}}, \bibinfo {author} {\bibfnamefont
				{M.~D.}\ \bibnamefont {Watson}}, \bibinfo {author} {\bibfnamefont {A.~A.}\
				\bibnamefont {Haghighirad}}, \ and\ \bibinfo {author} {\bibfnamefont
				{K.}~\bibnamefont {Shimada}},\ }\bibfield  {title} {\enquote {\bibinfo
				{title} {{Applications for ultimate spatial resolution in LASER based $\mu$ -
						ARPES: A FeSe case study}},}\ }\href {\doibase 10.1063/1.5084618} {\bibfield
			{journal} {\bibinfo  {journal} {AIP Conference Proceedings}\ }\textbf
			{\bibinfo {volume} {2054}},\ \bibinfo {pages} {040017} (\bibinfo {year}
			{2019})}\BibitemShut {NoStop}%
		\bibitem [{\citenamefont {Shimojima}\ \emph {et~al.}(2021)\citenamefont
			{Shimojima}, \citenamefont {Motoyui}, \citenamefont {Taniuchi}, \citenamefont
			{Bareille}, \citenamefont {Onari}, \citenamefont {Kontani}, \citenamefont
			{Nakajima}, \citenamefont {Kasahara}, \citenamefont {Shibauchi},
			\citenamefont {Matsuda},\ and\ \citenamefont {Shin}}]{Shimojima2021}%
		\BibitemOpen
		\bibfield  {author} {\bibinfo {author} {\bibfnamefont {T.}~\bibnamefont
				{Shimojima}}, \bibinfo {author} {\bibfnamefont {Y.}~\bibnamefont {Motoyui}},
			\bibinfo {author} {\bibfnamefont {T.}~\bibnamefont {Taniuchi}}, \bibinfo
			{author} {\bibfnamefont {C.}~\bibnamefont {Bareille}}, \bibinfo {author}
			{\bibfnamefont {S.}~\bibnamefont {Onari}}, \bibinfo {author} {\bibfnamefont
				{H.}~\bibnamefont {Kontani}}, \bibinfo {author} {\bibfnamefont
				{M.}~\bibnamefont {Nakajima}}, \bibinfo {author} {\bibfnamefont
				{S.}~\bibnamefont {Kasahara}}, \bibinfo {author} {\bibfnamefont
				{T.}~\bibnamefont {Shibauchi}}, \bibinfo {author} {\bibfnamefont
				{Y.}~\bibnamefont {Matsuda}}, \ and\ \bibinfo {author} {\bibfnamefont
				{S.}~\bibnamefont {Shin}},\ }\bibfield  {title} {\enquote {\bibinfo {title}
				{{Discovery of mesoscopic nematicity wave in iron-based superconductors}},}\
		}\href {\doibase 10.1126/science.abd6701} {\bibfield  {journal} {\bibinfo
				{journal} {Science}\ }\textbf {\bibinfo {volume} {373}},\ \bibinfo {pages}
			{1122--1125} (\bibinfo {year} {2021})}\BibitemShut {NoStop}%
		\bibitem [{\citenamefont {Tanatar}\ \emph {et~al.}(2016)\citenamefont
			{Tanatar}, \citenamefont {B\"ohmer}, \citenamefont {Timmons}, \citenamefont
			{Sch\"utt}, \citenamefont {Drachuck}, \citenamefont {Taufour}, \citenamefont
			{Kothapalli}, \citenamefont {Kreyssig}, \citenamefont {Bud'ko}, \citenamefont
			{Canfield}, \citenamefont {Fernandes},\ and\ \citenamefont
			{Prozorov}}]{Tanatar2016}%
		\BibitemOpen
		\bibfield  {author} {\bibinfo {author} {\bibfnamefont {M.~A.}\ \bibnamefont
				{Tanatar}}, \bibinfo {author} {\bibfnamefont {A.~E.}\ \bibnamefont
				{B\"ohmer}}, \bibinfo {author} {\bibfnamefont {E.~I.}\ \bibnamefont
				{Timmons}}, \bibinfo {author} {\bibfnamefont {M.}~\bibnamefont {Sch\"utt}},
			\bibinfo {author} {\bibfnamefont {G.}~\bibnamefont {Drachuck}}, \bibinfo
			{author} {\bibfnamefont {V.}~\bibnamefont {Taufour}}, \bibinfo {author}
			{\bibfnamefont {K.}~\bibnamefont {Kothapalli}}, \bibinfo {author}
			{\bibfnamefont {A.}~\bibnamefont {Kreyssig}}, \bibinfo {author}
			{\bibfnamefont {S.~L.}\ \bibnamefont {Bud'ko}}, \bibinfo {author}
			{\bibfnamefont {P.~C.}\ \bibnamefont {Canfield}}, \bibinfo {author}
			{\bibfnamefont {R.~M.}\ \bibnamefont {Fernandes}}, \ and\ \bibinfo {author}
			{\bibfnamefont {R.}~\bibnamefont {Prozorov}},\ }\bibfield  {title} {\enquote
			{\bibinfo {title} {{Origin of the Resistivity Anisotropy in the Nematic Phase
						of FeSe}},}\ }\href {\doibase 10.1103/PhysRevLett.117.127001} {\bibfield
			{journal} {\bibinfo  {journal} {Phys. Rev. Lett.}\ }\textbf {\bibinfo
				{volume} {117}},\ \bibinfo {pages} {127001} (\bibinfo {year}
			{2016})}\BibitemShut {NoStop}%
		\bibitem [{\citenamefont {Hoesch}\ \emph {et~al.}(2017)\citenamefont {Hoesch},
			\citenamefont {Kim}, \citenamefont {Dudin}, \citenamefont {Wang},
			\citenamefont {Scott}, \citenamefont {Harris}, \citenamefont {Patel},
			\citenamefont {Matthews}, \citenamefont {Hawkins}, \citenamefont {Alcock},
			\citenamefont {Richter}, \citenamefont {Mudd}, \citenamefont {Basham},
			\citenamefont {Pratt}, \citenamefont {Leicester}, \citenamefont {Longhi},
			\citenamefont {Tamai},\ and\ \citenamefont {Baumberger}}]{Hoesch2017}%
		\BibitemOpen
		\bibfield  {author} {\bibinfo {author} {\bibfnamefont {M.}~\bibnamefont
				{Hoesch}}, \bibinfo {author} {\bibfnamefont {T.~K.}\ \bibnamefont {Kim}},
			\bibinfo {author} {\bibfnamefont {P.}~\bibnamefont {Dudin}}, \bibinfo
			{author} {\bibfnamefont {H.}~\bibnamefont {Wang}}, \bibinfo {author}
			{\bibfnamefont {S.}~\bibnamefont {Scott}}, \bibinfo {author} {\bibfnamefont
				{P.}~\bibnamefont {Harris}}, \bibinfo {author} {\bibfnamefont
				{S.}~\bibnamefont {Patel}}, \bibinfo {author} {\bibfnamefont
				{M.}~\bibnamefont {Matthews}}, \bibinfo {author} {\bibfnamefont
				{D.}~\bibnamefont {Hawkins}}, \bibinfo {author} {\bibfnamefont {S.~G.}\
				\bibnamefont {Alcock}}, \bibinfo {author} {\bibfnamefont {T.}~\bibnamefont
				{Richter}}, \bibinfo {author} {\bibfnamefont {J.~J.}\ \bibnamefont {Mudd}},
			\bibinfo {author} {\bibfnamefont {M.}~\bibnamefont {Basham}}, \bibinfo
			{author} {\bibfnamefont {L.}~\bibnamefont {Pratt}}, \bibinfo {author}
			{\bibfnamefont {P.}~\bibnamefont {Leicester}}, \bibinfo {author}
			{\bibfnamefont {E.~C.}\ \bibnamefont {Longhi}}, \bibinfo {author}
			{\bibfnamefont {A.}~\bibnamefont {Tamai}}, \ and\ \bibinfo {author}
			{\bibfnamefont {F.}~\bibnamefont {Baumberger}},\ }\bibfield  {title}
		{\enquote {\bibinfo {title} {{A facility for the analysis of the electronic
						structures of solids and their surfaces by synchrotron radiation
						photoelectron spectroscopy}},}\ }\href {\doibase 10.1063/1.4973562}
		{\bibfield  {journal} {\bibinfo  {journal} {Rev. Sci. Instrum.}\ }\textbf
			{\bibinfo {volume} {88}},\ \bibinfo {pages} {013106} (\bibinfo {year}
			{2017})}\BibitemShut {NoStop}%
		\bibitem [{\citenamefont {Brouet}\ \emph {et~al.}(2012)\citenamefont {Brouet},
			\citenamefont {Jensen}, \citenamefont {Lin}, \citenamefont {T-I.},
			\citenamefont {Le~F\`evre}, \citenamefont {Bertran}, \citenamefont {Lin},
			\citenamefont {Ku}, \citenamefont {Forget},\ and\ \citenamefont
			{Colson}}]{Brouet2012}%
		\BibitemOpen
		\bibfield  {author} {\bibinfo {author} {\bibfnamefont {V.}~\bibnamefont
				{Brouet}}, \bibinfo {author} {\bibfnamefont {M.~Fuglsang}\ \bibnamefont
				{Jensen}}, \bibinfo {author} {\bibfnamefont {P-H.}\ \bibnamefont {Lin}},
			\bibinfo {author} {\bibfnamefont {A.}~\bibnamefont {T-I.}}, \bibinfo {author}
			{\bibfnamefont {P.}~\bibnamefont {Le~F\`evre}}, \bibinfo {author}
			{\bibfnamefont {F.}~\bibnamefont {Bertran}}, \bibinfo {author} {\bibfnamefont
				{C-H.}\ \bibnamefont {Lin}}, \bibinfo {author} {\bibfnamefont
				{W.}~\bibnamefont {Ku}}, \bibinfo {author} {\bibfnamefont {A.}~\bibnamefont
				{Forget}}, \ and\ \bibinfo {author} {\bibfnamefont {D.}~\bibnamefont
				{Colson}},\ }\bibfield  {title} {\enquote {\bibinfo {title} {{Impact of the
						two Fe unit cell on the electronic structure measured by ARPES in iron
						pnictides}},}\ }\href {\doibase 10.1103/PhysRevB.86.075123} {\bibfield
			{journal} {\bibinfo  {journal} {Phys. Rev. B}\ }\textbf {\bibinfo {volume}
				{86}},\ \bibinfo {pages} {075123} (\bibinfo {year} {2012})}\BibitemShut
		{NoStop}%
		\bibitem [{\citenamefont {Day}\ \emph {et~al.}(2019)\citenamefont {Day},
			\citenamefont {Zwartsenberg}, \citenamefont {Elfimov},\ and\ \citenamefont
			{Damascelli}}]{Day2019}%
		\BibitemOpen
		\bibfield  {author} {\bibinfo {author} {\bibfnamefont {R.~P.}\ \bibnamefont
				{Day}}, \bibinfo {author} {\bibfnamefont {B.}~\bibnamefont {Zwartsenberg}},
			\bibinfo {author} {\bibfnamefont {I.~S.}\ \bibnamefont {Elfimov}}, \ and\
			\bibinfo {author} {\bibfnamefont {A.}~\bibnamefont {Damascelli}},\ }\bibfield
		{title} {\enquote {\bibinfo {title} {{Computational framework chinook for
						angle-resolved photoemission spectroscopy}},}\ }\href {\doibase
			10.1038/s41535-019-0194-8} {\bibfield  {journal} {\bibinfo  {journal} {npj
					Quantum Materials}\ }\textbf {\bibinfo {volume} {4}},\ \bibinfo {pages} {54}
			(\bibinfo {year} {2019})}\BibitemShut {NoStop}%
		\bibitem [{\citenamefont {Reiss}\ \emph {et~al.}(2017)\citenamefont {Reiss},
			\citenamefont {Watson}, \citenamefont {Kim}, \citenamefont {Haghighirad},
			\citenamefont {Woodruff}, \citenamefont {Bruma}, \citenamefont {Clarke},\
			and\ \citenamefont {Coldea}}]{Reiss2017}%
		\BibitemOpen
		\bibfield  {author} {\bibinfo {author} {\bibfnamefont {P.}~\bibnamefont
				{Reiss}}, \bibinfo {author} {\bibfnamefont {M.~D.}\ \bibnamefont {Watson}},
			\bibinfo {author} {\bibfnamefont {T.~K.}\ \bibnamefont {Kim}}, \bibinfo
			{author} {\bibfnamefont {A.~A.}\ \bibnamefont {Haghighirad}}, \bibinfo
			{author} {\bibfnamefont {D.~N.}\ \bibnamefont {Woodruff}}, \bibinfo {author}
			{\bibfnamefont {M.}~\bibnamefont {Bruma}}, \bibinfo {author} {\bibfnamefont
				{S.~J.}\ \bibnamefont {Clarke}}, \ and\ \bibinfo {author} {\bibfnamefont
				{A.~I.}\ \bibnamefont {Coldea}},\ }\bibfield  {title} {\enquote {\bibinfo
				{title} {{Suppression of electronic correlations by chemical pressure from
						FeSe to FeS}},}\ }\href {\doibase 10.1103/PhysRevB.96.121103} {\bibfield
			{journal} {\bibinfo  {journal} {Phys. Rev. B}\ }\textbf {\bibinfo {volume}
				{96}},\ \bibinfo {pages} {121103} (\bibinfo {year} {2017})}\BibitemShut
		{NoStop}%
		\bibitem [{\citenamefont {Pustovit}\ and\ \citenamefont
			{Kordyuk}(2016)}]{Pustovit2016}%
		\BibitemOpen
		\bibfield  {author} {\bibinfo {author} {\bibfnamefont {Y..~V.}\ \bibnamefont
				{Pustovit}}\ and\ \bibinfo {author} {\bibfnamefont {A.~A.}\ \bibnamefont
				{Kordyuk}},\ }\bibfield  {title} {\enquote {\bibinfo {title} {{Metamorphoses
						of electronic structure of FeSe-based superconductors}},}\ }\href {\doibase
			10.1063/1.4969896} {\bibfield  {journal} {\bibinfo  {journal} {Low
					Temperature Physics}\ }\textbf {\bibinfo {volume} {42}},\ \bibinfo {pages}
			{995--1007} (\bibinfo {year} {2016})}\BibitemShut {NoStop}%
		\bibitem [{\citenamefont {Fisher}\ \emph {et~al.}(2011)\citenamefont {Fisher},
			\citenamefont {Degiorgi},\ and\ \citenamefont {Shen}}]{Fisher2011}%
		\BibitemOpen
		\bibfield  {author} {\bibinfo {author} {\bibfnamefont {I.~R.}\ \bibnamefont
				{Fisher}}, \bibinfo {author} {\bibfnamefont {L.}~\bibnamefont {Degiorgi}}, \
			and\ \bibinfo {author} {\bibfnamefont {Z.~X.}\ \bibnamefont {Shen}},\
		}\bibfield  {title} {\enquote {\bibinfo {title} {{In-plane electronic
						anisotropy of underdoped `122' Fe-arsenide superconductors revealed by
						measurements of detwinned single crystals}},}\ }\href {\doibase
			10.1088/0034-4885/74/12/124506} {\bibfield  {journal} {\bibinfo  {journal}
				{Rep. Prog. Phys.}\ }\textbf {\bibinfo {volume} {74}},\ \bibinfo {pages}
			{124506} (\bibinfo {year} {2011})}\BibitemShut {NoStop}%
		\bibitem [{\citenamefont {Christensen}\ \emph {et~al.}(2020)\citenamefont
			{Christensen}, \citenamefont {Fernandes},\ and\ \citenamefont
			{Chubukov}}]{Christensen2020}%
		\BibitemOpen
		\bibfield  {author} {\bibinfo {author} {\bibfnamefont {M.~H.}\ \bibnamefont
				{Christensen}}, \bibinfo {author} {\bibfnamefont {R.~M.}\ \bibnamefont
				{Fernandes}}, \ and\ \bibinfo {author} {\bibfnamefont {A.~V.}\ \bibnamefont
				{Chubukov}},\ }\bibfield  {title} {\enquote {\bibinfo {title} {{Orbital
						transmutation and the electronic spectrum of FeSe in the nematic phase}},}\
		}\href {\doibase 10.1103/PhysRevResearch.2.013015} {\bibfield  {journal}
			{\bibinfo  {journal} {Phys. Rev. Research}\ }\textbf {\bibinfo {volume}
				{2}},\ \bibinfo {pages} {013015} (\bibinfo {year} {2020})}\BibitemShut
		{NoStop}%
		\bibitem [{\citenamefont {Rhodes}\ \emph {et~al.}(2019)\citenamefont {Rhodes},
			\citenamefont {Watson}, \citenamefont {Kim},\ and\ \citenamefont
			{Eschrig}}]{Rhodes2019}%
		\BibitemOpen
		\bibfield  {author} {\bibinfo {author} {\bibfnamefont {L.~C.}\ \bibnamefont
				{Rhodes}}, \bibinfo {author} {\bibfnamefont {M.~D.}\ \bibnamefont {Watson}},
			\bibinfo {author} {\bibfnamefont {T.~K.}\ \bibnamefont {Kim}}, \ and\
			\bibinfo {author} {\bibfnamefont {M.}~\bibnamefont {Eschrig}},\ }\bibfield
		{title} {\enquote {\bibinfo {title} {{${k}_{z}$ Selective Scattering within
						Quasiparticle Interference Measurements of FeSe}},}\ }\href {\doibase
			10.1103/PhysRevLett.123.216404} {\bibfield  {journal} {\bibinfo  {journal}
				{Phys. Rev. Lett.}\ }\textbf {\bibinfo {volume} {123}},\ \bibinfo {pages}
			{216404} (\bibinfo {year} {2019})}\BibitemShut {NoStop}%
		\bibitem [{\citenamefont {Sunko}\ \emph {et~al.}(2019)\citenamefont {Sunko},
			\citenamefont {Abarca~Morales}, \citenamefont {Markovi{\'{c}}}, \citenamefont
			{Barber}, \citenamefont {Milosavljevi{\'{c}}}, \citenamefont {Mazzola},
			\citenamefont {Sokolov}, \citenamefont {Kikugawa}, \citenamefont {Cacho},
			\citenamefont {Dudin}, \citenamefont {Rosner}, \citenamefont {Hicks},
			\citenamefont {King},\ and\ \citenamefont {Mackenzie}}]{Sunko2019}%
		\BibitemOpen
		\bibfield  {author} {\bibinfo {author} {\bibfnamefont {V.}~\bibnamefont
				{Sunko}}, \bibinfo {author} {\bibfnamefont {E.}~\bibnamefont
				{Abarca~Morales}}, \bibinfo {author} {\bibfnamefont {I.}~\bibnamefont
				{Markovi{\'{c}}}}, \bibinfo {author} {\bibfnamefont {M.~E.}\ \bibnamefont
				{Barber}}, \bibinfo {author} {\bibfnamefont {D.}~\bibnamefont
				{Milosavljevi{\'{c}}}}, \bibinfo {author} {\bibfnamefont {F.}~\bibnamefont
				{Mazzola}}, \bibinfo {author} {\bibfnamefont {D.~A.}\ \bibnamefont
				{Sokolov}}, \bibinfo {author} {\bibfnamefont {N.}~\bibnamefont {Kikugawa}},
			\bibinfo {author} {\bibfnamefont {C.}~\bibnamefont {Cacho}}, \bibinfo
			{author} {\bibfnamefont {P.}~\bibnamefont {Dudin}}, \bibinfo {author}
			{\bibfnamefont {H.}~\bibnamefont {Rosner}}, \bibinfo {author} {\bibfnamefont
				{C.~W.}\ \bibnamefont {Hicks}}, \bibinfo {author} {\bibfnamefont {P.~D.~C.}\
				\bibnamefont {King}}, \ and\ \bibinfo {author} {\bibfnamefont {A.~P.}\
				\bibnamefont {Mackenzie}},\ }\bibfield  {title} {\enquote {\bibinfo {title}
				{{Direct observation of a uniaxial stress-driven Lifshitz transition in
						Sr$_{2}$RuO$_{4}$}},}\ }\href {\doibase 10.1038/s41535-019-0185-9} {\bibfield
			{journal} {\bibinfo  {journal} {npj Quantum Materials}\ }\textbf {\bibinfo
				{volume} {4}},\ \bibinfo {pages} {46} (\bibinfo {year} {2019})}\BibitemShut
		{NoStop}%
		\bibitem [{\citenamefont {Iwasawa}(2020)}]{Iwasawa_2020}%
		\BibitemOpen
		\bibfield  {author} {\bibinfo {author} {\bibfnamefont {H.}~\bibnamefont
				{Iwasawa}},\ }\bibfield  {title} {\enquote {\bibinfo {title}
				{{High-resolution angle-resolved photoemission spectroscopy and
						microscopy}},}\ }\href {\doibase 10.1088/2516-1075/abb379} {\bibfield
			{journal} {\bibinfo  {journal} {Electronic Structure}\ }\textbf {\bibinfo
				{volume} {2}},\ \bibinfo {pages} {043001} (\bibinfo {year}
			{2020})}\BibitemShut {NoStop}%
		\bibitem [{\citenamefont {Watson}\ \emph {et~al.}(2019)\citenamefont {Watson},
			\citenamefont {Dudin}, \citenamefont {Rhodes}, \citenamefont {Evtushinsky},
			\citenamefont {Iwasawa}, \citenamefont {Aswartham}, \citenamefont {Wurmehl},
			\citenamefont {B{\"u}chner}, \citenamefont {Hoesch},\ and\ \citenamefont
			{Kim}}]{Watson2019}%
		\BibitemOpen
		\bibfield  {author} {\bibinfo {author} {\bibfnamefont {M.~D.}\ \bibnamefont
				{Watson}}, \bibinfo {author} {\bibfnamefont {P.}~\bibnamefont {Dudin}},
			\bibinfo {author} {\bibfnamefont {L.~C.}\ \bibnamefont {Rhodes}}, \bibinfo
			{author} {\bibfnamefont {D.~V.}\ \bibnamefont {Evtushinsky}}, \bibinfo
			{author} {\bibfnamefont {H.}~\bibnamefont {Iwasawa}}, \bibinfo {author}
			{\bibfnamefont {S.}~\bibnamefont {Aswartham}}, \bibinfo {author}
			{\bibfnamefont {S.}~\bibnamefont {Wurmehl}}, \bibinfo {author} {\bibfnamefont
				{B.}~\bibnamefont {B{\"u}chner}}, \bibinfo {author} {\bibfnamefont
				{M.}~\bibnamefont {Hoesch}}, \ and\ \bibinfo {author} {\bibfnamefont {T.~K.}\
				\bibnamefont {Kim}},\ }\bibfield  {title} {\enquote {\bibinfo {title}
				{{Probing the reconstructed Fermi surface of antiferromagnetic BaFe$_2$As$_2$
						in one domain}},}\ }\href {\doibase 10.1038/s41535-019-0174-z} {\bibfield
			{journal} {\bibinfo  {journal} {npj Quantum Materials}\ }\textbf {\bibinfo
				{volume} {4}},\ \bibinfo {pages} {36} (\bibinfo {year} {2019})}\BibitemShut
		{NoStop}%
		\bibitem [{\citenamefont {Jiao}\ \emph {et~al.}(2017)\citenamefont {Jiao},
			\citenamefont {Huang}, \citenamefont {R{\"o}{\ss}ler}, \citenamefont {Koz},
			\citenamefont {R{\"o}{\ss}ler}, \citenamefont {Schwarz},\ and\ \citenamefont
			{Wirth}}]{Jiao2017}%
		\BibitemOpen
		\bibfield  {author} {\bibinfo {author} {\bibfnamefont {L.}~\bibnamefont
				{Jiao}}, \bibinfo {author} {\bibfnamefont {C-L.}\ \bibnamefont {Huang}},
			\bibinfo {author} {\bibfnamefont {S}~\bibnamefont {R{\"o}{\ss}ler}}, \bibinfo
			{author} {\bibfnamefont {C.}~\bibnamefont {Koz}}, \bibinfo {author}
			{\bibfnamefont {U.~K.}\ \bibnamefont {R{\"o}{\ss}ler}}, \bibinfo {author}
			{\bibfnamefont {U.}~\bibnamefont {Schwarz}}, \ and\ \bibinfo {author}
			{\bibfnamefont {S.}~\bibnamefont {Wirth}},\ }\bibfield  {title} {\enquote
			{\bibinfo {title} {Superconducting gap structure of fese},}\ }\href {\doibase
			10.1038/srep44024} {\bibfield  {journal} {\bibinfo  {journal} {Scientific
					Reports}\ }\textbf {\bibinfo {volume} {7}},\ \bibinfo {pages} {44024}
			(\bibinfo {year} {2017})}\BibitemShut {NoStop}%
		\bibitem [{\citenamefont {Choubey}\ \emph {et~al.}(2014)\citenamefont
			{Choubey}, \citenamefont {Berlijn}, \citenamefont {Kreisel}, \citenamefont
			{Cao},\ and\ \citenamefont {Hirschfeld}}]{Choubey2014}%
		\BibitemOpen
		\bibfield  {author} {\bibinfo {author} {\bibfnamefont {P.}~\bibnamefont
				{Choubey}}, \bibinfo {author} {\bibfnamefont {T.}~\bibnamefont {Berlijn}},
			\bibinfo {author} {\bibfnamefont {A.}~\bibnamefont {Kreisel}}, \bibinfo
			{author} {\bibfnamefont {C.}~\bibnamefont {Cao}}, \ and\ \bibinfo {author}
			{\bibfnamefont {P.~J.}\ \bibnamefont {Hirschfeld}},\ }\bibfield  {title}
		{\enquote {\bibinfo {title} {{Visualization of atomic-scale phenomena in
						superconductors: Application to FeSe}},}\ }\href {\doibase
			10.1103/PhysRevB.90.134520} {\bibfield  {journal} {\bibinfo  {journal} {Phys.
					Rev. B}\ }\textbf {\bibinfo {volume} {90}},\ \bibinfo {pages} {134520}
			(\bibinfo {year} {2014})}\BibitemShut {NoStop}%
		\bibitem [{\citenamefont {Bu}\ \emph {et~al.}(2019)\citenamefont {Bu},
			\citenamefont {Wang}, \citenamefont {Zhang}, \citenamefont {Fei},
			\citenamefont {Zheng}, \citenamefont {Ai}, \citenamefont {Wu}, \citenamefont
			{Wang}, \citenamefont {Wo}, \citenamefont {Zhao}, \citenamefont {Jin},\ and\
			\citenamefont {Yin}}]{Bu2019}%
		\BibitemOpen
		\bibfield  {author} {\bibinfo {author} {\bibfnamefont {K.}~\bibnamefont
				{Bu}}, \bibinfo {author} {\bibfnamefont {B.}~\bibnamefont {Wang}}, \bibinfo
			{author} {\bibfnamefont {W.}~\bibnamefont {Zhang}}, \bibinfo {author}
			{\bibfnamefont {Y.}~\bibnamefont {Fei}}, \bibinfo {author} {\bibfnamefont
				{Y.}~\bibnamefont {Zheng}}, \bibinfo {author} {\bibfnamefont
				{F.}~\bibnamefont {Ai}}, \bibinfo {author} {\bibfnamefont {Z.}~\bibnamefont
				{Wu}}, \bibinfo {author} {\bibfnamefont {Q.}~\bibnamefont {Wang}}, \bibinfo
			{author} {\bibfnamefont {H.}~\bibnamefont {Wo}}, \bibinfo {author}
			{\bibfnamefont {J.}~\bibnamefont {Zhao}}, \bibinfo {author} {\bibfnamefont
				{C.}~\bibnamefont {Jin}}, \ and\ \bibinfo {author} {\bibfnamefont
				{Y.}~\bibnamefont {Yin}},\ }\bibfield  {title} {\enquote {\bibinfo {title}
				{{Study of intrinsic defect states of FeSe with scanning tunneling
						microscopy}},}\ }\href {\doibase 10.1103/PhysRevB.100.155127} {\bibfield
			{journal} {\bibinfo  {journal} {Phys. Rev. B}\ }\textbf {\bibinfo {volume}
				{100}},\ \bibinfo {pages} {155127} (\bibinfo {year} {2019})}\BibitemShut
		{NoStop}%
		\bibitem [{\citenamefont {Macdonald}\ \emph {et~al.}(2016)\citenamefont
			{Macdonald}, \citenamefont {Tremblay-Johnston}, \citenamefont {Grothe},
			\citenamefont {Chi}, \citenamefont {Dosanjh}, \citenamefont {Johnston},\ and\
			\citenamefont {Burke}}]{Macdonald2016}%
		\BibitemOpen
		\bibfield  {author} {\bibinfo {author} {\bibfnamefont {A.~J.}\ \bibnamefont
				{Macdonald}}, \bibinfo {author} {\bibfnamefont {Y-S.}\ \bibnamefont
				{Tremblay-Johnston}}, \bibinfo {author} {\bibfnamefont {S.}~\bibnamefont
				{Grothe}}, \bibinfo {author} {\bibfnamefont {S.}~\bibnamefont {Chi}},
			\bibinfo {author} {\bibfnamefont {P.}~\bibnamefont {Dosanjh}}, \bibinfo
			{author} {\bibfnamefont {S.}~\bibnamefont {Johnston}}, \ and\ \bibinfo
			{author} {\bibfnamefont {S.~A.}\ \bibnamefont {Burke}},\ }\bibfield  {title}
		{\enquote {\bibinfo {title} {{Dispersing artifacts in {FT}-{STS}: a
						comparison of set point effects across acquisition modes}},}\ }\href
		{\doibase 10.1088/0957-4484/27/41/414004} {\bibfield  {journal} {\bibinfo
				{journal} {Nanotechnology}\ }\textbf {\bibinfo {volume} {27}},\ \bibinfo
			{pages} {414004} (\bibinfo {year} {2016})}\BibitemShut {NoStop}%
		\bibitem [{\citenamefont {Weismann}\ \emph {et~al.}(2009)\citenamefont
			{Weismann}, \citenamefont {Wenderoth}, \citenamefont {Lounis}, \citenamefont
			{Zahn}, \citenamefont {Quaas}, \citenamefont {Ulbrich}, \citenamefont
			{Dederichs},\ and\ \citenamefont {Blügel}}]{Weismann2009}%
		\BibitemOpen
		\bibfield  {author} {\bibinfo {author} {\bibfnamefont {A.}~\bibnamefont
				{Weismann}}, \bibinfo {author} {\bibfnamefont {M.}~\bibnamefont {Wenderoth}},
			\bibinfo {author} {\bibfnamefont {S.}~\bibnamefont {Lounis}}, \bibinfo
			{author} {\bibfnamefont {P.}~\bibnamefont {Zahn}}, \bibinfo {author}
			{\bibfnamefont {N.}~\bibnamefont {Quaas}}, \bibinfo {author} {\bibfnamefont
				{R.~G.}\ \bibnamefont {Ulbrich}}, \bibinfo {author} {\bibfnamefont {P.~H.}\
				\bibnamefont {Dederichs}}, \ and\ \bibinfo {author} {\bibfnamefont
				{S.}~\bibnamefont {Blügel}},\ }\bibfield  {title} {\enquote {\bibinfo
				{title} {{Seeing the Fermi Surface in Real Space by Nanoscale Electron
						Focusing}},}\ }\href {\doibase 10.1126/science.1168738} {\bibfield  {journal}
			{\bibinfo  {journal} {Science}\ }\textbf {\bibinfo {volume} {323}},\ \bibinfo
			{pages} {1190--1193} (\bibinfo {year} {2009})}\BibitemShut {NoStop}%
		\bibitem [{\citenamefont {Lounis}\ \emph {et~al.}(2011)\citenamefont {Lounis},
			\citenamefont {Zahn}, \citenamefont {Weismann}, \citenamefont {Wenderoth},
			\citenamefont {Ulbrich}, \citenamefont {Mertig}, \citenamefont {Dederichs},\
			and\ \citenamefont {Bl\"ugel}}]{Lounis2011}%
		\BibitemOpen
		\bibfield  {author} {\bibinfo {author} {\bibfnamefont {S.}~\bibnamefont
				{Lounis}}, \bibinfo {author} {\bibfnamefont {P.}~\bibnamefont {Zahn}},
			\bibinfo {author} {\bibfnamefont {A.}~\bibnamefont {Weismann}}, \bibinfo
			{author} {\bibfnamefont {M.}~\bibnamefont {Wenderoth}}, \bibinfo {author}
			{\bibfnamefont {R.~G.}\ \bibnamefont {Ulbrich}}, \bibinfo {author}
			{\bibfnamefont {I.}~\bibnamefont {Mertig}}, \bibinfo {author} {\bibfnamefont
				{P.~H.}\ \bibnamefont {Dederichs}}, \ and\ \bibinfo {author} {\bibfnamefont
				{S.}~\bibnamefont {Bl\"ugel}},\ }\bibfield  {title} {\enquote {\bibinfo
				{title} {{Theory of real space imaging of Fermi surface parts}},}\ }\href
		{\doibase 10.1103/PhysRevB.83.035427} {\bibfield  {journal} {\bibinfo
				{journal} {Phys. Rev. B}\ }\textbf {\bibinfo {volume} {83}},\ \bibinfo
			{pages} {035427} (\bibinfo {year} {2011})}\BibitemShut {NoStop}%
		\bibitem [{\citenamefont {Marques}\ \emph {et~al.}(2021)\citenamefont
			{Marques}, \citenamefont {Bahramy}, \citenamefont {Trainer}, \citenamefont
			{Markovi{\'{c}}}, \citenamefont {Watson}, \citenamefont {Mazzola},
			\citenamefont {Rajan}, \citenamefont {Raub}, \citenamefont {King},\ and\
			\citenamefont {Wahl}}]{Marques2021}%
		\BibitemOpen
		\bibfield  {author} {\bibinfo {author} {\bibfnamefont {C.~A.}\ \bibnamefont
				{Marques}}, \bibinfo {author} {\bibfnamefont {M.~S.}\ \bibnamefont
				{Bahramy}}, \bibinfo {author} {\bibfnamefont {C.}~\bibnamefont {Trainer}},
			\bibinfo {author} {\bibfnamefont {I.}~\bibnamefont {Markovi{\'{c}}}},
			\bibinfo {author} {\bibfnamefont {M.~D.}\ \bibnamefont {Watson}}, \bibinfo
			{author} {\bibfnamefont {F.}~\bibnamefont {Mazzola}}, \bibinfo {author}
			{\bibfnamefont {A.}~\bibnamefont {Rajan}}, \bibinfo {author} {\bibfnamefont
				{T.~D.}\ \bibnamefont {Raub}}, \bibinfo {author} {\bibfnamefont {P.~D.~C.}\
				\bibnamefont {King}}, \ and\ \bibinfo {author} {\bibfnamefont
				{P.}~\bibnamefont {Wahl}},\ }\bibfield  {title} {\enquote {\bibinfo {title}
				{{Tomographic mapping of the hidden dimension in quasi-particle
						interference}},}\ }\href {\doibase 10.1038/s41467-021-27082-1} {\bibfield
			{journal} {\bibinfo  {journal} {Nat. Comm.}\ }\textbf {\bibinfo {volume}
				{12}},\ \bibinfo {pages} {6739} (\bibinfo {year} {2021})}\BibitemShut
		{NoStop}%
		\bibitem [{\citenamefont {Li}\ \emph {et~al.}(2020{\natexlab{a}})\citenamefont
			{Li}, \citenamefont {Wu}, \citenamefont {Wang}, \citenamefont {Liu},
			\citenamefont {Cai}, \citenamefont {Wang}, \citenamefont {Gao}, \citenamefont
			{Song}, \citenamefont {Huang}, \citenamefont {Dong}, \citenamefont {Liu},
			\citenamefont {Ai}, \citenamefont {Luo}, \citenamefont {Yin}, \citenamefont
			{Liu}, \citenamefont {Huang}, \citenamefont {Wang}, \citenamefont {Jia},
			\citenamefont {Zhang}, \citenamefont {Zhang}, \citenamefont {Yang},
			\citenamefont {Wang}, \citenamefont {Peng}, \citenamefont {Xu}, \citenamefont
			{Shi}, \citenamefont {Hu}, \citenamefont {Xiang}, \citenamefont {Zhao},\ and\
			\citenamefont {Zhou}}]{Li2020PRXARPES}%
		\BibitemOpen
		\bibfield  {author} {\bibinfo {author} {\bibfnamefont {C.}~\bibnamefont
				{Li}}, \bibinfo {author} {\bibfnamefont {X.}~\bibnamefont {Wu}}, \bibinfo
			{author} {\bibfnamefont {L.}~\bibnamefont {Wang}}, \bibinfo {author}
			{\bibfnamefont {D.}~\bibnamefont {Liu}}, \bibinfo {author} {\bibfnamefont
				{Y.}~\bibnamefont {Cai}}, \bibinfo {author} {\bibfnamefont {Y.}~\bibnamefont
				{Wang}}, \bibinfo {author} {\bibfnamefont {Q.}~\bibnamefont {Gao}}, \bibinfo
			{author} {\bibfnamefont {C.}~\bibnamefont {Song}}, \bibinfo {author}
			{\bibfnamefont {J.}~\bibnamefont {Huang}}, \bibinfo {author} {\bibfnamefont
				{C.}~\bibnamefont {Dong}}, \bibinfo {author} {\bibfnamefont {J.}~\bibnamefont
				{Liu}}, \bibinfo {author} {\bibfnamefont {P.}~\bibnamefont {Ai}}, \bibinfo
			{author} {\bibfnamefont {H.}~\bibnamefont {Luo}}, \bibinfo {author}
			{\bibfnamefont {C-H.}\ \bibnamefont {Yin}}, \bibinfo {author} {\bibfnamefont
				{G.}~\bibnamefont {Liu}}, \bibinfo {author} {\bibfnamefont {Y.}~\bibnamefont
				{Huang}}, \bibinfo {author} {\bibfnamefont {Q.}~\bibnamefont {Wang}},
			\bibinfo {author} {\bibfnamefont {X.}~\bibnamefont {Jia}}, \bibinfo {author}
			{\bibfnamefont {F.}~\bibnamefont {Zhang}}, \bibinfo {author} {\bibfnamefont
				{S.}~\bibnamefont {Zhang}}, \bibinfo {author} {\bibfnamefont
				{F.}~\bibnamefont {Yang}}, \bibinfo {author} {\bibfnamefont {Z.}~\bibnamefont
				{Wang}}, \bibinfo {author} {\bibfnamefont {Q.}~\bibnamefont {Peng}}, \bibinfo
			{author} {\bibfnamefont {Z.}~\bibnamefont {Xu}}, \bibinfo {author}
			{\bibfnamefont {Y.}~\bibnamefont {Shi}}, \bibinfo {author} {\bibfnamefont
				{J.}~\bibnamefont {Hu}}, \bibinfo {author} {\bibfnamefont {T.}~\bibnamefont
				{Xiang}}, \bibinfo {author} {\bibfnamefont {L.}~\bibnamefont {Zhao}}, \ and\
			\bibinfo {author} {\bibfnamefont {X.~J.}\ \bibnamefont {Zhou}},\ }\bibfield
		{title} {\enquote {\bibinfo {title} {{Spectroscopic Evidence for an
						Additional Symmetry Breaking in the Nematic State of FeSe Superconductor}},}\
		}\href {\doibase 10.1103/PhysRevX.10.031033} {\bibfield  {journal} {\bibinfo
				{journal} {Phys. Rev. X}\ }\textbf {\bibinfo {volume} {10}},\ \bibinfo
			{pages} {031033} (\bibinfo {year} {2020}{\natexlab{a}})}\BibitemShut
		{NoStop}%
		\bibitem [{\citenamefont {Lee}\ \emph {et~al.}(2009)\citenamefont {Lee},
			\citenamefont {Yin},\ and\ \citenamefont {Ku}}]{Lee2009}%
		\BibitemOpen
		\bibfield  {author} {\bibinfo {author} {\bibfnamefont {C-C.}\ \bibnamefont
				{Lee}}, \bibinfo {author} {\bibfnamefont {W-G.}\ \bibnamefont {Yin}}, \ and\
			\bibinfo {author} {\bibfnamefont {W.}~\bibnamefont {Ku}},\ }\bibfield
		{title} {\enquote {\bibinfo {title} {{Ferro-Orbital Order and Strong Magnetic
						Anisotropy in the Parent Compounds of Iron-Pnictide Superconductors }},}\
		}\href {\doibase 10.1103/PhysRevLett.103.267001} {\bibfield  {journal}
			{\bibinfo  {journal} {Phys. Rev. Lett.}\ }\textbf {\bibinfo {volume} {103}},\
			\bibinfo {pages} {267001} (\bibinfo {year} {2009})}\BibitemShut {NoStop}%
		\bibitem [{\citenamefont {Pradhan}\ \emph {et~al.}(2021)\citenamefont
			{Pradhan}, \citenamefont {Parida},\ and\ \citenamefont
			{Sahoo}}]{Pradhan2021}%
		\BibitemOpen
		\bibfield  {author} {\bibinfo {author} {\bibfnamefont {B.}~\bibnamefont
				{Pradhan}}, \bibinfo {author} {\bibfnamefont {P.~K.}\ \bibnamefont {Parida}},
			\ and\ \bibinfo {author} {\bibfnamefont {S.}~\bibnamefont {Sahoo}},\
		}\bibfield  {title} {\enquote {\bibinfo {title} {{Superconductivity and
						Jahn-Teller Distortion in s{\textpm}-Wave Iron-Based Superconductors}},}\
		}\href {\doibase 10.1007/s13538-020-00827-x} {\bibfield  {journal} {\bibinfo
				{journal} {Brazilian Journal of Physics}\ }\textbf {\bibinfo {volume} {51}},\
			\bibinfo {pages} {393--400} (\bibinfo {year} {2021})}\BibitemShut {NoStop}%
		\bibitem [{\citenamefont {Kreisel}\ \emph {et~al.}(2015)\citenamefont
			{Kreisel}, \citenamefont {Mukherjee}, \citenamefont {Hirschfeld},\ and\
			\citenamefont {Andersen}}]{Kreisel2015}%
		\BibitemOpen
		\bibfield  {author} {\bibinfo {author} {\bibfnamefont {A.}~\bibnamefont
				{Kreisel}}, \bibinfo {author} {\bibfnamefont {S.}~\bibnamefont {Mukherjee}},
			\bibinfo {author} {\bibfnamefont {P.~J.}\ \bibnamefont {Hirschfeld}}, \ and\
			\bibinfo {author} {\bibfnamefont {Brian~M.}\ \bibnamefont {Andersen}},\
		}\bibfield  {title} {\enquote {\bibinfo {title} {{Spin excitations in a model
						of FeSe with orbital ordering}},}\ }\href {\doibase
			10.1103/PhysRevB.92.224515} {\bibfield  {journal} {\bibinfo  {journal} {Phys.
					Rev. B}\ }\textbf {\bibinfo {volume} {92}},\ \bibinfo {pages} {224515}
			(\bibinfo {year} {2015})}\BibitemShut {NoStop}%
		\bibitem [{\citenamefont {Kreisel}\ \emph {et~al.}(2018)\citenamefont
			{Kreisel}, \citenamefont {Andersen},\ and\ \citenamefont
			{Hirschfeld}}]{Kreisel2018}%
		\BibitemOpen
		\bibfield  {author} {\bibinfo {author} {\bibfnamefont {A.}~\bibnamefont
				{Kreisel}}, \bibinfo {author} {\bibfnamefont {B.~M.}\ \bibnamefont
				{Andersen}}, \ and\ \bibinfo {author} {\bibfnamefont {P.~J.}\ \bibnamefont
				{Hirschfeld}},\ }\bibfield  {title} {\enquote {\bibinfo {title} {{Itinerant
						approach to magnetic neutron scattering of FeSe: Effect of orbital
						selectivity}},}\ }\href {\doibase 10.1103/PhysRevB.98.214518} {\bibfield
			{journal} {\bibinfo  {journal} {Phys. Rev. B}\ }\textbf {\bibinfo {volume}
				{98}},\ \bibinfo {pages} {214518} (\bibinfo {year} {2018})}\BibitemShut
		{NoStop}%
		\bibitem [{\citenamefont {Yu}\ \emph {et~al.}(2021)\citenamefont {Yu},
			\citenamefont {Hu}, \citenamefont {Nica}, \citenamefont {Zhu},\ and\
			\citenamefont {Si}}]{Yu2021}%
		\BibitemOpen
		\bibfield  {author} {\bibinfo {author} {\bibfnamefont {R.}~\bibnamefont
				{Yu}}, \bibinfo {author} {\bibfnamefont {H.}~\bibnamefont {Hu}}, \bibinfo
			{author} {\bibfnamefont {E.~M.}\ \bibnamefont {Nica}}, \bibinfo {author}
			{\bibfnamefont {J-X.}\ \bibnamefont {Zhu}}, \ and\ \bibinfo {author}
			{\bibfnamefont {Q.}~\bibnamefont {Si}},\ }\bibfield  {title} {\enquote
			{\bibinfo {title} {{Orbital Selectivity in Electron Correlations and
						Superconducting Pairing of Iron-Based Superconductors}},}\ }\href {\doibase
			10.3389/fphy.2021.578347} {\bibfield  {journal} {\bibinfo  {journal}
				{Frontiers in Physics}\ }\textbf {\bibinfo {volume} {9}},\ \bibinfo {pages}
			{92} (\bibinfo {year} {2021})}\BibitemShut {NoStop}%
		\bibitem [{\citenamefont {Cercellier}\ \emph {et~al.}(2019)\citenamefont
			{Cercellier}, \citenamefont {Rodi\`ere}, \citenamefont {Toulemonde},
			\citenamefont {Marcenat},\ and\ \citenamefont {Klein}}]{Cercellier2019}%
		\BibitemOpen
		\bibfield  {author} {\bibinfo {author} {\bibfnamefont {H.}~\bibnamefont
				{Cercellier}}, \bibinfo {author} {\bibfnamefont {P.}~\bibnamefont
				{Rodi\`ere}}, \bibinfo {author} {\bibfnamefont {P.}~\bibnamefont
				{Toulemonde}}, \bibinfo {author} {\bibfnamefont {C.}~\bibnamefont
				{Marcenat}}, \ and\ \bibinfo {author} {\bibfnamefont {T.}~\bibnamefont
				{Klein}},\ }\bibfield  {title} {\enquote {\bibinfo {title} {{Influence of the
						quasiparticle spectral weight in FeSe on spectroscopic, magnetic, and
						thermodynamic properties}},}\ }\href {\doibase 10.1103/PhysRevB.100.104516}
		{\bibfield  {journal} {\bibinfo  {journal} {Phys. Rev. B}\ }\textbf {\bibinfo
				{volume} {100}},\ \bibinfo {pages} {104516} (\bibinfo {year}
			{2019})}\BibitemShut {NoStop}%
		\bibitem [{\citenamefont {Jiang}\ \emph {et~al.}(2016)\citenamefont {Jiang},
			\citenamefont {Hu}, \citenamefont {Ding},\ and\ \citenamefont
			{Wang}}]{Jiang2016}%
		\BibitemOpen
		\bibfield  {author} {\bibinfo {author} {\bibfnamefont {K.}~\bibnamefont
				{Jiang}}, \bibinfo {author} {\bibfnamefont {J.}~\bibnamefont {Hu}}, \bibinfo
			{author} {\bibfnamefont {H.}~\bibnamefont {Ding}}, \ and\ \bibinfo {author}
			{\bibfnamefont {Z.}~\bibnamefont {Wang}},\ }\bibfield  {title} {\enquote
			{\bibinfo {title} {{Interatomic Coulomb interaction and electron nematic bond
						order in FeSe}},}\ }\href {\doibase 10.1103/PhysRevB.93.115138} {\bibfield
			{journal} {\bibinfo  {journal} {Phys. Rev. B}\ }\textbf {\bibinfo {volume}
				{93}},\ \bibinfo {pages} {115138} (\bibinfo {year} {2016})}\BibitemShut
		{NoStop}%
		\bibitem [{\citenamefont {Biswas}\ \emph {et~al.}(2018)\citenamefont {Biswas},
			\citenamefont {Kreisel}, \citenamefont {Wang}, \citenamefont {Adroja},
			\citenamefont {Hillier}, \citenamefont {Zhao}, \citenamefont {Khasanov},
			\citenamefont {Orain}, \citenamefont {Amato},\ and\ \citenamefont
			{Morenzoni}}]{Biswas2018}%
		\BibitemOpen
		\bibfield  {author} {\bibinfo {author} {\bibfnamefont {P.~K.}\ \bibnamefont
				{Biswas}}, \bibinfo {author} {\bibfnamefont {A.}~\bibnamefont {Kreisel}},
			\bibinfo {author} {\bibfnamefont {Q.}~\bibnamefont {Wang}}, \bibinfo {author}
			{\bibfnamefont {D.~T.}\ \bibnamefont {Adroja}}, \bibinfo {author}
			{\bibfnamefont {A.~D.}\ \bibnamefont {Hillier}}, \bibinfo {author}
			{\bibfnamefont {J.}~\bibnamefont {Zhao}}, \bibinfo {author} {\bibfnamefont
				{R.}~\bibnamefont {Khasanov}}, \bibinfo {author} {\bibfnamefont {J-C.}\
				\bibnamefont {Orain}}, \bibinfo {author} {\bibfnamefont {A.}~\bibnamefont
				{Amato}}, \ and\ \bibinfo {author} {\bibfnamefont {E.}~\bibnamefont
				{Morenzoni}},\ }\bibfield  {title} {\enquote {\bibinfo {title} {{Evidence of
						nodal gap structure in the basal plane of the FeSe superconductor}},}\ }\href
		{\doibase 10.1103/PhysRevB.98.180501} {\bibfield  {journal} {\bibinfo
				{journal} {Phys. Rev. B}\ }\textbf {\bibinfo {volume} {98}},\ \bibinfo
			{pages} {180501} (\bibinfo {year} {2018})}\BibitemShut {NoStop}%
		\bibitem [{\citenamefont {Xing}\ \emph {et~al.}(2018)\citenamefont {Xing},
			\citenamefont {Classen},\ and\ \citenamefont {Chubukov}}]{Xing2018}%
		\BibitemOpen
		\bibfield  {author} {\bibinfo {author} {\bibfnamefont {R-Q.}\ \bibnamefont
				{Xing}}, \bibinfo {author} {\bibfnamefont {L.}~\bibnamefont {Classen}}, \
			and\ \bibinfo {author} {\bibfnamefont {A.~V.}\ \bibnamefont {Chubukov}},\
		}\bibfield  {title} {\enquote {\bibinfo {title} {{Orbital order in FeSe: The
						case for vertex renormalization}},}\ }\href {\doibase
			10.1103/PhysRevB.98.041108} {\bibfield  {journal} {\bibinfo  {journal} {Phys.
					Rev. B}\ }\textbf {\bibinfo {volume} {98}},\ \bibinfo {pages} {041108}
			(\bibinfo {year} {2018})}\BibitemShut {NoStop}%
		\bibitem [{\citenamefont {Kang}\ \emph
			{et~al.}(2018{\natexlab{b}})\citenamefont {Kang}, \citenamefont {Chubukov},\
			and\ \citenamefont {Fernandes}}]{Kang2018b}%
		\BibitemOpen
		\bibfield  {author} {\bibinfo {author} {\bibfnamefont {J.}~\bibnamefont
				{Kang}}, \bibinfo {author} {\bibfnamefont {A.~V.}\ \bibnamefont {Chubukov}},
			\ and\ \bibinfo {author} {\bibfnamefont {R.~M.}\ \bibnamefont {Fernandes}},\
		}\bibfield  {title} {\enquote {\bibinfo {title} {{Time-reversal
						symmetry-breaking nematic superconductivity in FeSe}},}\ }\href {\doibase
			10.1103/PhysRevB.98.064508} {\bibfield  {journal} {\bibinfo  {journal} {Phys.
					Rev. B}\ }\textbf {\bibinfo {volume} {98}},\ \bibinfo {pages} {064508}
			(\bibinfo {year} {2018}{\natexlab{b}})}\BibitemShut {NoStop}%
		\bibitem [{\citenamefont {Mandal}\ \emph {et~al.}(2017)\citenamefont {Mandal},
			\citenamefont {Zhang}, \citenamefont {Ismail-Beigi},\ and\ \citenamefont
			{Haule}}]{Mandal2017}%
		\BibitemOpen
		\bibfield  {author} {\bibinfo {author} {\bibfnamefont {S.}~\bibnamefont
				{Mandal}}, \bibinfo {author} {\bibfnamefont {P.}~\bibnamefont {Zhang}},
			\bibinfo {author} {\bibfnamefont {S.}~\bibnamefont {Ismail-Beigi}}, \ and\
			\bibinfo {author} {\bibfnamefont {K.}~\bibnamefont {Haule}},\ }\bibfield
		{title} {\enquote {\bibinfo {title} {{How Correlated is the
						$\mathrm{FeSe}/{\mathrm{SrTiO}}_{3}$ System?}}}\ }\href {\doibase
			10.1103/PhysRevLett.119.067004} {\bibfield  {journal} {\bibinfo  {journal}
				{Phys. Rev. Lett.}\ }\textbf {\bibinfo {volume} {119}},\ \bibinfo {pages}
			{067004} (\bibinfo {year} {2017})}\BibitemShut {NoStop}%
		\bibitem [{\citenamefont {Long}\ \emph {et~al.}(2020)\citenamefont {Long},
			\citenamefont {Zhang}, \citenamefont {Wang},\ and\ \citenamefont
			{Liu}}]{Long2020}%
		\BibitemOpen
		\bibfield  {author} {\bibinfo {author} {\bibfnamefont {X.}~\bibnamefont
				{Long}}, \bibinfo {author} {\bibfnamefont {S.}~\bibnamefont {Zhang}},
			\bibinfo {author} {\bibfnamefont {F.}~\bibnamefont {Wang}}, \ and\ \bibinfo
			{author} {\bibfnamefont {Z.}~\bibnamefont {Liu}},\ }\bibfield  {title}
		{\enquote {\bibinfo {title} {{A first-principle perspective on electronic
						nematicity in FeSe}},}\ }\href {\doibase 10.1038/s41535-020-00253-x}
		{\bibfield  {journal} {\bibinfo  {journal} {npj Quantum Materials}\ }\textbf
			{\bibinfo {volume} {5}},\ \bibinfo {pages} {50} (\bibinfo {year}
			{2020})}\BibitemShut {NoStop}%
		\bibitem [{\citenamefont {Yamada}\ and\ \citenamefont
			{Tohyama}(2021)}]{Yamada2021}%
		\BibitemOpen
		\bibfield  {author} {\bibinfo {author} {\bibfnamefont {T.}~\bibnamefont
				{Yamada}}\ and\ \bibinfo {author} {\bibfnamefont {T.}~\bibnamefont
				{Tohyama}},\ }\bibfield  {title} {\enquote {\bibinfo {title} {{Multipolar
						nematic state of nonmagnetic FeSe based on $\mathrm{DFT+U}$}},}\ }\href
		{\doibase 10.1103/PhysRevB.104.L161110} {\bibfield  {journal} {\bibinfo
				{journal} {Phys. Rev. B}\ }\textbf {\bibinfo {volume} {104}},\ \bibinfo
			{pages} {L161110} (\bibinfo {year} {2021})}\BibitemShut {NoStop}%
		\bibitem [{\citenamefont {Steffensen}\ \emph {et~al.}(2021)\citenamefont
			{Steffensen}, \citenamefont {Kreisel}, \citenamefont {Hirschfeld},\ and\
			\citenamefont {Andersen}}]{Steffensen2021}%
		\BibitemOpen
		\bibfield  {author} {\bibinfo {author} {\bibfnamefont {D}~\bibnamefont
				{Steffensen}}, \bibinfo {author} {\bibfnamefont {A.}~\bibnamefont {Kreisel}},
			\bibinfo {author} {\bibfnamefont {P.~J.}\ \bibnamefont {Hirschfeld}}, \ and\
			\bibinfo {author} {\bibfnamefont {B.~M.}\ \bibnamefont {Andersen}},\
		}\bibfield  {title} {\enquote {\bibinfo {title} {{Interorbital nematicity and
						the origin of a single electron Fermi pocket in FeSe}},}\ }\href {\doibase
			10.1103/PhysRevB.103.054505} {\bibfield  {journal} {\bibinfo  {journal}
				{Phys. Rev. B}\ }\textbf {\bibinfo {volume} {103}},\ \bibinfo {pages}
			{054505} (\bibinfo {year} {2021})}\BibitemShut {NoStop}%
		\bibitem [{\citenamefont {Rodriguez}\ \emph {et~al.}(2011)\citenamefont
			{Rodriguez}, \citenamefont {Stock}, \citenamefont {Zajdel}, \citenamefont
			{Krycka}, \citenamefont {Majkrzak}, \citenamefont {Zavalij},\ and\
			\citenamefont {Green}}]{Rodriquez2011}%
		\BibitemOpen
		\bibfield  {author} {\bibinfo {author} {\bibfnamefont {E.~E.}\ \bibnamefont
				{Rodriguez}}, \bibinfo {author} {\bibfnamefont {C.}~\bibnamefont {Stock}},
			\bibinfo {author} {\bibfnamefont {P.}~\bibnamefont {Zajdel}}, \bibinfo
			{author} {\bibfnamefont {K.~L.}\ \bibnamefont {Krycka}}, \bibinfo {author}
			{\bibfnamefont {C.~F.}\ \bibnamefont {Majkrzak}}, \bibinfo {author}
			{\bibfnamefont {P.}~\bibnamefont {Zavalij}}, \ and\ \bibinfo {author}
			{\bibfnamefont {M.~A.}\ \bibnamefont {Green}},\ }\bibfield  {title} {\enquote
			{\bibinfo {title} {{Magnetic-crystallographic phase diagram of the
						superconducting parent compound Fe${}_{1+x}$Te}},}\ }\href {\doibase
			10.1103/PhysRevB.84.064403} {\bibfield  {journal} {\bibinfo  {journal} {Phys.
					Rev. B}\ }\textbf {\bibinfo {volume} {84}},\ \bibinfo {pages} {064403}
			(\bibinfo {year} {2011})}\BibitemShut {NoStop}%
		\bibitem [{\citenamefont {Trainer}\ \emph {et~al.}(2019)\citenamefont
			{Trainer}, \citenamefont {Yim}, \citenamefont {Heil}, \citenamefont
			{Giustino}, \citenamefont {Croitori}, \citenamefont {Tsurkan}, \citenamefont
			{Loidl}, \citenamefont {Rodriguez}, \citenamefont {Stock},\ and\
			\citenamefont {Wahl}}]{Trainer2019}%
		\BibitemOpen
		\bibfield  {author} {\bibinfo {author} {\bibfnamefont {C.}~\bibnamefont
				{Trainer}}, \bibinfo {author} {\bibfnamefont {C-M.}\ \bibnamefont {Yim}},
			\bibinfo {author} {\bibfnamefont {C.}~\bibnamefont {Heil}}, \bibinfo {author}
			{\bibfnamefont {F.}~\bibnamefont {Giustino}}, \bibinfo {author}
			{\bibfnamefont {D.}~\bibnamefont {Croitori}}, \bibinfo {author}
			{\bibfnamefont {V.}~\bibnamefont {Tsurkan}}, \bibinfo {author} {\bibfnamefont
				{A.}~\bibnamefont {Loidl}}, \bibinfo {author} {\bibfnamefont {E.~E.}\
				\bibnamefont {Rodriguez}}, \bibinfo {author} {\bibfnamefont {C.}~\bibnamefont
				{Stock}}, \ and\ \bibinfo {author} {\bibfnamefont {P.}~\bibnamefont {Wahl}},\
		}\bibfield  {title} {\enquote {\bibinfo {title} {{Manipulating surface
						magnetic order in iron telluride}},}\ }\href {\doibase
			10.1126/sciadv.aav3478} {\bibfield  {journal} {\bibinfo  {journal} {Science
					Advances}\ }\textbf {\bibinfo {volume} {5}},\ \bibinfo {pages} {eaav3478}
			(\bibinfo {year} {2019})}\BibitemShut {NoStop}%
		\bibitem [{\citenamefont {Chubukov}\ \emph {et~al.}(2016)\citenamefont
			{Chubukov}, \citenamefont {Khodas},\ and\ \citenamefont
			{Fernandes}}]{Chubukov2016}%
		\BibitemOpen
		\bibfield  {author} {\bibinfo {author} {\bibfnamefont {A.~V.}\ \bibnamefont
				{Chubukov}}, \bibinfo {author} {\bibfnamefont {M.}~\bibnamefont {Khodas}}, \
			and\ \bibinfo {author} {\bibfnamefont {R.~M.}\ \bibnamefont {Fernandes}},\
		}\bibfield  {title} {\enquote {\bibinfo {title} {{Magnetism,
						Superconductivity, and Spontaneous Orbital Order in Iron-Based
						Superconductors: Which Comes First and Why?}}}\ }\href {\doibase
			10.1103/PhysRevX.6.041045} {\bibfield  {journal} {\bibinfo  {journal} {Phys.
					Rev. X}\ }\textbf {\bibinfo {volume} {6}},\ \bibinfo {pages} {041045}
			(\bibinfo {year} {2016})}\BibitemShut {NoStop}%
		\bibitem [{\citenamefont {Xing}\ \emph {et~al.}(2017)\citenamefont {Xing},
			\citenamefont {Classen}, \citenamefont {Khodas},\ and\ \citenamefont
			{Chubukov}}]{Xing2017}%
		\BibitemOpen
		\bibfield  {author} {\bibinfo {author} {\bibfnamefont {R.-Q.}\ \bibnamefont
				{Xing}}, \bibinfo {author} {\bibfnamefont {L.}~\bibnamefont {Classen}},
			\bibinfo {author} {\bibfnamefont {M.}~\bibnamefont {Khodas}}, \ and\ \bibinfo
			{author} {\bibfnamefont {A.~V.}\ \bibnamefont {Chubukov}},\ }\bibfield
		{title} {\enquote {\bibinfo {title} {{Competing instabilities, orbital
						ordering, and splitting of band degeneracies from a parquet renormalization
						group analysis of a four-pocket model for iron-based superconductors:
						Application to FeSe}},}\ }\href {\doibase 10.1103/PhysRevB.95.085108}
		{\bibfield  {journal} {\bibinfo  {journal} {Phys. Rev. B}\ }\textbf {\bibinfo
				{volume} {95}},\ \bibinfo {pages} {085108} (\bibinfo {year}
			{2017})}\BibitemShut {NoStop}%
		\bibitem [{\citenamefont {Classen}\ \emph {et~al.}(2017)\citenamefont
			{Classen}, \citenamefont {Xing}, \citenamefont {Khodas},\ and\ \citenamefont
			{Chubukov}}]{Classen2018}%
		\BibitemOpen
		\bibfield  {author} {\bibinfo {author} {\bibfnamefont {L.}~\bibnamefont
				{Classen}}, \bibinfo {author} {\bibfnamefont {R.-Q.}\ \bibnamefont {Xing}},
			\bibinfo {author} {\bibfnamefont {M.}~\bibnamefont {Khodas}}, \ and\ \bibinfo
			{author} {\bibfnamefont {A.~V.}\ \bibnamefont {Chubukov}},\ }\bibfield
		{title} {\enquote {\bibinfo {title} {{Interplay between Magnetism,
						Superconductivity, and Orbital Order in 5-Pocket Model for Iron-Based
						Superconductors: Parquet Renormalization Group Study}},}\ }\href {\doibase
			10.1103/PhysRevLett.118.037001} {\bibfield  {journal} {\bibinfo  {journal}
				{Phys. Rev. Lett.}\ }\textbf {\bibinfo {volume} {118}},\ \bibinfo {pages}
			{037001} (\bibinfo {year} {2017})}\BibitemShut {NoStop}%
		\bibitem [{\citenamefont {Islam}\ \emph {et~al.}(2021)\citenamefont {Islam},
			\citenamefont {B\"oker}, \citenamefont {Eremin},\ and\ \citenamefont
			{Chubukov}}]{Islam2021}%
		\BibitemOpen
		\bibfield  {author} {\bibinfo {author} {\bibfnamefont {K.~R.}\ \bibnamefont
				{Islam}}, \bibinfo {author} {\bibfnamefont {J.}~\bibnamefont {B\"oker}},
			\bibinfo {author} {\bibfnamefont {I.~M.}\ \bibnamefont {Eremin}}, \ and\
			\bibinfo {author} {\bibfnamefont {A.~V.}\ \bibnamefont {Chubukov}},\
		}\bibfield  {title} {\enquote {\bibinfo {title} {{Specific heat and gap
						structure of a nematic superconductor: Application to FeSe}},}\ }\href
		{\doibase 10.1103/PhysRevB.104.094522} {\bibfield  {journal} {\bibinfo
				{journal} {Phys. Rev. B}\ }\textbf {\bibinfo {volume} {104}},\ \bibinfo
			{pages} {094522} (\bibinfo {year} {2021})}\BibitemShut {NoStop}%
		\bibitem [{\citenamefont {Li}\ \emph {et~al.}(2020{\natexlab{b}})\citenamefont
			{Li}, \citenamefont {Lei}, \citenamefont {Zhao}, \citenamefont {Nie},
			\citenamefont {Song}, \citenamefont {Zheng}, \citenamefont {Li},
			\citenamefont {Kang}, \citenamefont {Luo}, \citenamefont {Wu},\ and\
			\citenamefont {Chen}}]{Li2020NMR}%
		\BibitemOpen
		\bibfield  {author} {\bibinfo {author} {\bibfnamefont {J.}~\bibnamefont
				{Li}}, \bibinfo {author} {\bibfnamefont {B.}~\bibnamefont {Lei}}, \bibinfo
			{author} {\bibfnamefont {D.}~\bibnamefont {Zhao}}, \bibinfo {author}
			{\bibfnamefont {L.~P.}\ \bibnamefont {Nie}}, \bibinfo {author} {\bibfnamefont
				{D.~W.}\ \bibnamefont {Song}}, \bibinfo {author} {\bibfnamefont {L.~X.}\
				\bibnamefont {Zheng}}, \bibinfo {author} {\bibfnamefont {S.~J.}\ \bibnamefont
				{Li}}, \bibinfo {author} {\bibfnamefont {B.~L.}\ \bibnamefont {Kang}},
			\bibinfo {author} {\bibfnamefont {X.~G.}\ \bibnamefont {Luo}}, \bibinfo
			{author} {\bibfnamefont {T.}~\bibnamefont {Wu}}, \ and\ \bibinfo {author}
			{\bibfnamefont {X.~H.}\ \bibnamefont {Chen}},\ }\bibfield  {title} {\enquote
			{\bibinfo {title} {{Spin-Orbital-Intertwined Nematic State in FeSe}},}\
		}\href {\doibase 10.1103/PhysRevX.10.011034} {\bibfield  {journal} {\bibinfo
				{journal} {Phys. Rev. X}\ }\textbf {\bibinfo {volume} {10}},\ \bibinfo
			{pages} {011034} (\bibinfo {year} {2020}{\natexlab{b}})}\BibitemShut
		{NoStop}%
		\bibitem [{\citenamefont {Liu}\ \emph {et~al.}(2021)\citenamefont {Liu},
			\citenamefont {Yuan}, \citenamefont {Ma}, \citenamefont {Lu}, \citenamefont
			{Zhang}, \citenamefont {Ma}, \citenamefont {Zhang}, \citenamefont {Jin},
			\citenamefont {Yu}, \citenamefont {Zhou}, \citenamefont {Dong},\ and\
			\citenamefont {Zhao}}]{Liu2021}%
		\BibitemOpen
		\bibfield  {author} {\bibinfo {author} {\bibfnamefont {S.}~\bibnamefont
				{Liu}}, \bibinfo {author} {\bibfnamefont {J.}~\bibnamefont {Yuan}}, \bibinfo
			{author} {\bibfnamefont {S.}~\bibnamefont {Ma}}, \bibinfo {author}
			{\bibfnamefont {Z.}~\bibnamefont {Lu}}, \bibinfo {author} {\bibfnamefont
				{Y.}~\bibnamefont {Zhang}}, \bibinfo {author} {\bibfnamefont
				{M.}~\bibnamefont {Ma}}, \bibinfo {author} {\bibfnamefont {H.}~\bibnamefont
				{Zhang}}, \bibinfo {author} {\bibfnamefont {K.}~\bibnamefont {Jin}}, \bibinfo
			{author} {\bibfnamefont {L.}~\bibnamefont {Yu}}, \bibinfo {author}
			{\bibfnamefont {F.}~\bibnamefont {Zhou}}, \bibinfo {author} {\bibfnamefont
				{X.}~\bibnamefont {Dong}}, \ and\ \bibinfo {author} {\bibfnamefont
				{Z.}~\bibnamefont {Zhao}},\ }\bibfield  {title} {\enquote {\bibinfo {title}
				{{Magnetic-Field-Induced Spin Nematicity in
						${\mathrm{FeSe}}_{1\ensuremath{-}x}{\mathrm{S}}_{x}$ and
						${\mathrm{FeSe}}_{1\ensuremath{-}y}{\mathrm{Te}}_{y}$ Superconductor
						Systems}},}\ }\href
		{https://iopscience.iop.org/article/10.1088/0256-307X/38/8/087401/pdf}
		{\bibfield  {journal} {\bibinfo  {journal} {Chinese Phys. Lett.}\ }\textbf
			{\bibinfo {volume} {38}},\ \bibinfo {pages} {087401} (\bibinfo {year}
			{2021})}\BibitemShut {NoStop}%
		\bibitem [{\citenamefont {Medvedev}\ \emph {et~al.}(2009)\citenamefont
			{Medvedev}, \citenamefont {McQueen}, \citenamefont {Troyan}, \citenamefont
			{Palasyuk}, \citenamefont {Eremets}, \citenamefont {Cava}, \citenamefont
			{Naghavi}, \citenamefont {Casper}, \citenamefont {Ksenofontov}, \citenamefont
			{Wortmann},\ and\ \citenamefont {Felser}}]{Medvedev2009}%
		\BibitemOpen
		\bibfield  {author} {\bibinfo {author} {\bibfnamefont {S.}~\bibnamefont
				{Medvedev}}, \bibinfo {author} {\bibfnamefont {T.~M.}\ \bibnamefont
				{McQueen}}, \bibinfo {author} {\bibfnamefont {I.~A.}\ \bibnamefont {Troyan}},
			\bibinfo {author} {\bibfnamefont {T.}~\bibnamefont {Palasyuk}}, \bibinfo
			{author} {\bibfnamefont {M.~I.}\ \bibnamefont {Eremets}}, \bibinfo {author}
			{\bibfnamefont {R.~J.}\ \bibnamefont {Cava}}, \bibinfo {author}
			{\bibfnamefont {S.}~\bibnamefont {Naghavi}}, \bibinfo {author} {\bibfnamefont
				{F.}~\bibnamefont {Casper}}, \bibinfo {author} {\bibfnamefont
				{V.}~\bibnamefont {Ksenofontov}}, \bibinfo {author} {\bibfnamefont
				{G.}~\bibnamefont {Wortmann}}, \ and\ \bibinfo {author} {\bibfnamefont
				{C.}~\bibnamefont {Felser}},\ }\bibfield  {title} {\enquote {\bibinfo {title}
				{{Electronic and magnetic phase diagram of $\beta$-Fe1.01Se with
						superconductivity at 36.7 K under pressure}},}\ }\href {\doibase
			10.1038/nmat2491} {\bibfield  {journal} {\bibinfo  {journal} {Nat. Mater.}\
			}\textbf {\bibinfo {volume} {8}},\ \bibinfo {pages} {630--633} (\bibinfo
			{year} {2009})}\BibitemShut {NoStop}%
		\bibitem [{\citenamefont {Huang}\ and\ \citenamefont
			{Hoffman}(2017)}]{Huang2017}%
		\BibitemOpen
		\bibfield  {author} {\bibinfo {author} {\bibfnamefont {D.}~\bibnamefont
				{Huang}}\ and\ \bibinfo {author} {\bibfnamefont {J.~E.}\ \bibnamefont
				{Hoffman}},\ }\bibfield  {title} {\enquote {\bibinfo {title} {{Monolayer FeSe
						on ${\mathrm{SrTiO}}_{3}$}},}\ }\href {\doibase
			10.1146/annurev-conmatphys-031016-025242} {\bibfield  {journal} {\bibinfo
				{journal} {Annu. Rev. Condens. Matter Phys.}\ }\textbf {\bibinfo {volume}
				{8}},\ \bibinfo {pages} {311--336} (\bibinfo {year} {2017})}\BibitemShut
		{NoStop}%
		\bibitem [{\citenamefont {Hardy}\ \emph {et~al.}(2019)\citenamefont {Hardy},
			\citenamefont {He}, \citenamefont {Wang}, \citenamefont {Wolf}, \citenamefont
			{Schweiss}, \citenamefont {Merz}, \citenamefont {Barth}, \citenamefont
			{Adelmann}, \citenamefont {Eder}, \citenamefont {Haghighirad},\ and\
			\citenamefont {Meingast}}]{Hardy2019}%
		\BibitemOpen
		\bibfield  {author} {\bibinfo {author} {\bibfnamefont {F.}~\bibnamefont
				{Hardy}}, \bibinfo {author} {\bibfnamefont {M.}~\bibnamefont {He}}, \bibinfo
			{author} {\bibfnamefont {L.}~\bibnamefont {Wang}}, \bibinfo {author}
			{\bibfnamefont {T.}~\bibnamefont {Wolf}}, \bibinfo {author} {\bibfnamefont
				{P.}~\bibnamefont {Schweiss}}, \bibinfo {author} {\bibfnamefont
				{M.}~\bibnamefont {Merz}}, \bibinfo {author} {\bibfnamefont {M.}~\bibnamefont
				{Barth}}, \bibinfo {author} {\bibfnamefont {P.}~\bibnamefont {Adelmann}},
			\bibinfo {author} {\bibfnamefont {R.}~\bibnamefont {Eder}}, \bibinfo {author}
			{\bibfnamefont {A.~A.}\ \bibnamefont {Haghighirad}}, \ and\ \bibinfo {author}
			{\bibfnamefont {C.}~\bibnamefont {Meingast}},\ }\bibfield  {title} {\enquote
			{\bibinfo {title} {{Calorimetric evidence of nodal gaps in the nematic
						superconductor FeSe}},}\ }\href {\doibase 10.1103/PhysRevB.99.035157}
		{\bibfield  {journal} {\bibinfo  {journal} {Phys. Rev. B}\ }\textbf {\bibinfo
				{volume} {99}},\ \bibinfo {pages} {035157} (\bibinfo {year}
			{2019})}\BibitemShut {NoStop}%
		\bibitem [{\citenamefont {Sun}\ \emph {et~al.}(2017)\citenamefont {Sun},
			\citenamefont {Kittaka}, \citenamefont {Nakamura}, \citenamefont
			{Sakakibara}, \citenamefont {Irie}, \citenamefont {Nomoto}, \citenamefont
			{Machida}, \citenamefont {Chen},\ and\ \citenamefont {Tamegai}}]{Sun2017}%
		\BibitemOpen
		\bibfield  {author} {\bibinfo {author} {\bibfnamefont {Y.}~\bibnamefont
				{Sun}}, \bibinfo {author} {\bibfnamefont {S.}~\bibnamefont {Kittaka}},
			\bibinfo {author} {\bibfnamefont {S.}~\bibnamefont {Nakamura}}, \bibinfo
			{author} {\bibfnamefont {T.}~\bibnamefont {Sakakibara}}, \bibinfo {author}
			{\bibfnamefont {K.}~\bibnamefont {Irie}}, \bibinfo {author} {\bibfnamefont
				{T.}~\bibnamefont {Nomoto}}, \bibinfo {author} {\bibfnamefont
				{K.}~\bibnamefont {Machida}}, \bibinfo {author} {\bibfnamefont
				{J.}~\bibnamefont {Chen}}, \ and\ \bibinfo {author} {\bibfnamefont
				{T.}~\bibnamefont {Tamegai}},\ }\bibfield  {title} {\enquote {\bibinfo
				{title} {{Gap structure of FeSe determined by angle-resolved specific heat
						measurements in applied rotating magnetic field}},}\ }\href {\doibase
			10.1103/PhysRevB.96.220505} {\bibfield  {journal} {\bibinfo  {journal} {Phys.
					Rev. B}\ }\textbf {\bibinfo {volume} {96}},\ \bibinfo {pages} {220505}
			(\bibinfo {year} {2017})}\BibitemShut {NoStop}%
		\bibitem [{\citenamefont {Matsuura}\ \emph {et~al.}(2017)\citenamefont
			{Matsuura}, \citenamefont {Mizukami}, \citenamefont {Arai}, \citenamefont
			{Sugimura}, \citenamefont {Maejima}, \citenamefont {Machida}, \citenamefont
			{Watanuki}, \citenamefont {Fukuda}, \citenamefont {Yajima}, \citenamefont
			{Hiroi}, \citenamefont {Yip}, \citenamefont {Chan}, \citenamefont {Niu},
			\citenamefont {Hosoi}, \citenamefont {Ishida}, \citenamefont {Mukasa},
			\citenamefont {Kasahara}, \citenamefont {Cheng}, \citenamefont {Goh},
			\citenamefont {Matsuda}, \citenamefont {Uwatoko},\ and\ \citenamefont
			{Shibauchi}}]{Matsuura2017}%
		\BibitemOpen
		\bibfield  {author} {\bibinfo {author} {\bibfnamefont {K.}~\bibnamefont
				{Matsuura}}, \bibinfo {author} {\bibfnamefont {Y.}~\bibnamefont {Mizukami}},
			\bibinfo {author} {\bibfnamefont {Y.}~\bibnamefont {Arai}}, \bibinfo {author}
			{\bibfnamefont {Y.}~\bibnamefont {Sugimura}}, \bibinfo {author}
			{\bibfnamefont {N.}~\bibnamefont {Maejima}}, \bibinfo {author} {\bibfnamefont
				{A.}~\bibnamefont {Machida}}, \bibinfo {author} {\bibfnamefont
				{T.}~\bibnamefont {Watanuki}}, \bibinfo {author} {\bibfnamefont
				{T.}~\bibnamefont {Fukuda}}, \bibinfo {author} {\bibfnamefont
				{T.}~\bibnamefont {Yajima}}, \bibinfo {author} {\bibfnamefont
				{Z.}~\bibnamefont {Hiroi}}, \bibinfo {author} {\bibfnamefont {K.~Y.}\
				\bibnamefont {Yip}}, \bibinfo {author} {\bibfnamefont {Y.~C.}\ \bibnamefont
				{Chan}}, \bibinfo {author} {\bibfnamefont {Q.}~\bibnamefont {Niu}}, \bibinfo
			{author} {\bibfnamefont {S.}~\bibnamefont {Hosoi}}, \bibinfo {author}
			{\bibfnamefont {K.}~\bibnamefont {Ishida}}, \bibinfo {author} {\bibfnamefont
				{K.}~\bibnamefont {Mukasa}}, \bibinfo {author} {\bibfnamefont
				{S.}~\bibnamefont {Kasahara}}, \bibinfo {author} {\bibfnamefont {J.-G.}\
				\bibnamefont {Cheng}}, \bibinfo {author} {\bibfnamefont {S.~K.}\ \bibnamefont
				{Goh}}, \bibinfo {author} {\bibfnamefont {Y.}~\bibnamefont {Matsuda}},
			\bibinfo {author} {\bibfnamefont {Y.}~\bibnamefont {Uwatoko}}, \ and\
			\bibinfo {author} {\bibfnamefont {T.}~\bibnamefont {Shibauchi}},\ }\bibfield
		{title} {\enquote {\bibinfo {title} {{Maximizing Tc by tuning nematicity and
						magnetism in ${\mathrm{FeSe}}_{1-x}{\mathrm{S}}_{x}$ superconductors}},}\
		}\href {\doibase 10.1038/s41467-017-01277-x} {\bibfield  {journal} {\bibinfo
				{journal} {Nature Communications}\ }\textbf {\bibinfo {volume} {8}},\
			\bibinfo {pages} {1143} (\bibinfo {year} {2017})}\BibitemShut {NoStop}%
		\bibitem [{\citenamefont {Wang}\ \emph
			{et~al.}(2016{\natexlab{a}})\citenamefont {Wang}, \citenamefont {Shen},
			\citenamefont {Pan}, \citenamefont {Zhang}, \citenamefont {Ikeuchi},
			\citenamefont {Iida}, \citenamefont {Christianson}, \citenamefont {Walker},
			\citenamefont {Adroja}, \citenamefont {Abdel-Hafiez}, \citenamefont {Chen},
			\citenamefont {Chareev}, \citenamefont {Vasiliev},\ and\ \citenamefont
			{Zhao}}]{Wang2016}%
		\BibitemOpen
		\bibfield  {author} {\bibinfo {author} {\bibfnamefont {Q.}~\bibnamefont
				{Wang}}, \bibinfo {author} {\bibfnamefont {Y.}~\bibnamefont {Shen}}, \bibinfo
			{author} {\bibfnamefont {B.}~\bibnamefont {Pan}}, \bibinfo {author}
			{\bibfnamefont {X.}~\bibnamefont {Zhang}}, \bibinfo {author} {\bibfnamefont
				{K.}~\bibnamefont {Ikeuchi}}, \bibinfo {author} {\bibfnamefont
				{K.}~\bibnamefont {Iida}}, \bibinfo {author} {\bibfnamefont {A.~D.}\
				\bibnamefont {Christianson}}, \bibinfo {author} {\bibfnamefont {H.~C.}\
				\bibnamefont {Walker}}, \bibinfo {author} {\bibfnamefont {D.~T.}\
				\bibnamefont {Adroja}}, \bibinfo {author} {\bibfnamefont {M.}~\bibnamefont
				{Abdel-Hafiez}}, \bibinfo {author} {\bibfnamefont {Xiaojia}\ \bibnamefont
				{Chen}}, \bibinfo {author} {\bibfnamefont {D.~A.}\ \bibnamefont {Chareev}},
			\bibinfo {author} {\bibfnamefont {A.~N.}\ \bibnamefont {Vasiliev}}, \ and\
			\bibinfo {author} {\bibfnamefont {J.}~\bibnamefont {Zhao}},\ }\bibfield
		{title} {\enquote {\bibinfo {title} {{Magnetic ground state of FeSe}},}\
		}\href {\doibase 10.1038/ncomms12182} {\bibfield  {journal} {\bibinfo
				{journal} {Nat. Comm.}\ }\textbf {\bibinfo {volume} {7}},\ \bibinfo {pages}
			{12182} (\bibinfo {year} {2016}{\natexlab{a}})}\BibitemShut {NoStop}%
		\bibitem [{\citenamefont {Wang}\ \emph
			{et~al.}(2016{\natexlab{b}})\citenamefont {Wang}, \citenamefont {Shen},
			\citenamefont {Pan}, \citenamefont {Hao}, \citenamefont {Ma}, \citenamefont
			{Zhou}, \citenamefont {Steffens}, \citenamefont {Schmalzl}, \citenamefont
			{Forrest}, \citenamefont {Abdel-Hafiez}, \citenamefont {Chen}, \citenamefont
			{Chareev}, \citenamefont {Vasiliev}, \citenamefont {Bourges}, \citenamefont
			{Sidis}, \citenamefont {Cao},\ and\ \citenamefont {Zhao}}]{Wang2016b}%
		\BibitemOpen
		\bibfield  {author} {\bibinfo {author} {\bibfnamefont {Q.}~\bibnamefont
				{Wang}}, \bibinfo {author} {\bibfnamefont {Y.}~\bibnamefont {Shen}}, \bibinfo
			{author} {\bibfnamefont {B.}~\bibnamefont {Pan}}, \bibinfo {author}
			{\bibfnamefont {Y.}~\bibnamefont {Hao}}, \bibinfo {author} {\bibfnamefont
				{M.}~\bibnamefont {Ma}}, \bibinfo {author} {\bibfnamefont {F.}~\bibnamefont
				{Zhou}}, \bibinfo {author} {\bibfnamefont {P.}~\bibnamefont {Steffens}},
			\bibinfo {author} {\bibfnamefont {K.}~\bibnamefont {Schmalzl}}, \bibinfo
			{author} {\bibfnamefont {T.~R.}\ \bibnamefont {Forrest}}, \bibinfo {author}
			{\bibfnamefont {M.}~\bibnamefont {Abdel-Hafiez}}, \bibinfo {author}
			{\bibfnamefont {Xiaojia}\ \bibnamefont {Chen}}, \bibinfo {author}
			{\bibfnamefont {D.~A.}\ \bibnamefont {Chareev}}, \bibinfo {author}
			{\bibfnamefont {A.~N.}\ \bibnamefont {Vasiliev}}, \bibinfo {author}
			{\bibfnamefont {P.}~\bibnamefont {Bourges}}, \bibinfo {author} {\bibfnamefont
				{Y.}~\bibnamefont {Sidis}}, \bibinfo {author} {\bibfnamefont
				{H.}~\bibnamefont {Cao}}, \ and\ \bibinfo {author} {\bibfnamefont
				{J.}~\bibnamefont {Zhao}},\ }\bibfield  {title} {\enquote {\bibinfo {title}
				{{Strong interplay between stripe spin fluctuations, nematicity and
						superconductivity in FeSe}},}\ }\href {\doibase 10.1038/nmat4492} {\bibfield
			{journal} {\bibinfo  {journal} {Nat. Mater.}\ }\textbf {\bibinfo {volume}
				{15}},\ \bibinfo {pages} {159--163} (\bibinfo {year}
			{2016}{\natexlab{b}})}\BibitemShut {NoStop}%
		\bibitem [{\citenamefont {Watson}\ \emph {et~al.}(2018)\citenamefont {Watson},
			\citenamefont {Aswartham}, \citenamefont {Rhodes}, \citenamefont {Parrett},
			\citenamefont {Iwasawa}, \citenamefont {Hoesch}, \citenamefont {Morozov},
			\citenamefont {B\"uchner},\ and\ \citenamefont {Kim}}]{Watson2018}%
		\BibitemOpen
		\bibfield  {author} {\bibinfo {author} {\bibfnamefont {M.~D.}\ \bibnamefont
				{Watson}}, \bibinfo {author} {\bibfnamefont {S.}~\bibnamefont {Aswartham}},
			\bibinfo {author} {\bibfnamefont {L.~C.}\ \bibnamefont {Rhodes}}, \bibinfo
			{author} {\bibfnamefont {B.}~\bibnamefont {Parrett}}, \bibinfo {author}
			{\bibfnamefont {H.}~\bibnamefont {Iwasawa}}, \bibinfo {author} {\bibfnamefont
				{M.}~\bibnamefont {Hoesch}}, \bibinfo {author} {\bibfnamefont
				{I.}~\bibnamefont {Morozov}}, \bibinfo {author} {\bibfnamefont
				{B.}~\bibnamefont {B\"uchner}}, \ and\ \bibinfo {author} {\bibfnamefont
				{T.~K.}\ \bibnamefont {Kim}},\ }\bibfield  {title} {\enquote {\bibinfo
				{title} {{Three-dimensional electronic structure of the nematic and
						antiferromagnetic phases of NaFeAs from detwinned angle-resolved
						photoemission spectroscopy}},}\ }\href {\doibase 10.1103/PhysRevB.97.035134}
		{\bibfield  {journal} {\bibinfo  {journal} {Phys. Rev. B}\ }\textbf {\bibinfo
				{volume} {97}},\ \bibinfo {pages} {035134} (\bibinfo {year}
			{2018})}\BibitemShut {NoStop}%
		\bibitem [{\citenamefont {Coldea}\ \emph {et~al.}(2019)\citenamefont {Coldea},
			\citenamefont {Blake}, \citenamefont {Kasahara}, \citenamefont {Haghighirad},
			\citenamefont {Watson}, \citenamefont {Knafo}, \citenamefont {Choi},
			\citenamefont {McCollam}, \citenamefont {Reiss}, \citenamefont {Yamashita},
			\citenamefont {Bruma}, \citenamefont {Speller}, \citenamefont {Matsuda},
			\citenamefont {Wolf}, \citenamefont {Shibauchi},\ and\ \citenamefont
			{Schofield}}]{Coldea2019}%
		\BibitemOpen
		\bibfield  {author} {\bibinfo {author} {\bibfnamefont {A.~I.}\ \bibnamefont
				{Coldea}}, \bibinfo {author} {\bibfnamefont {S.~F.}\ \bibnamefont {Blake}},
			\bibinfo {author} {\bibfnamefont {S.}~\bibnamefont {Kasahara}}, \bibinfo
			{author} {\bibfnamefont {A.~A.}\ \bibnamefont {Haghighirad}}, \bibinfo
			{author} {\bibfnamefont {M.~D.}\ \bibnamefont {Watson}}, \bibinfo {author}
			{\bibfnamefont {W.}~\bibnamefont {Knafo}}, \bibinfo {author} {\bibfnamefont
				{E-S.}\ \bibnamefont {Choi}}, \bibinfo {author} {\bibfnamefont
				{A.}~\bibnamefont {McCollam}}, \bibinfo {author} {\bibfnamefont
				{P.}~\bibnamefont {Reiss}}, \bibinfo {author} {\bibfnamefont
				{T.}~\bibnamefont {Yamashita}}, \bibinfo {author} {\bibfnamefont
				{M.}~\bibnamefont {Bruma}}, \bibinfo {author} {\bibfnamefont {S.~C.}\
				\bibnamefont {Speller}}, \bibinfo {author} {\bibfnamefont {Y.}~\bibnamefont
				{Matsuda}}, \bibinfo {author} {\bibfnamefont {T.}~\bibnamefont {Wolf}},
			\bibinfo {author} {\bibfnamefont {T.}~\bibnamefont {Shibauchi}}, \ and\
			\bibinfo {author} {\bibfnamefont {A.~J.}\ \bibnamefont {Schofield}},\
		}\bibfield  {title} {\enquote {\bibinfo {title} {{Evolution of the
						low-temperature Fermi surface of superconducting
						${\mathrm{FeSe}}_{1-x}{\mathrm{S}}_{x}$ across a nematic phase
						transition}},}\ }\href {\doibase 10.1038/s41535-018-0141-0} {\bibfield
			{journal} {\bibinfo  {journal} {npj Quantum Materials}\ }\textbf {\bibinfo
				{volume} {4}},\ \bibinfo {pages} {2} (\bibinfo {year} {2019})}\BibitemShut
		{NoStop}%
	\end{thebibliography}
	
	%

\end{document}